\documentclass[12pt]{article}
\input xy
\xyoption{all}
\input epsf.tex
\usepackage{pstricks}

\usepackage{graphicx}
\usepackage{mathrsfs}
\usepackage{cancel}
\usepackage{amssymb,amsmath}
\usepackage{slashed}

\def\harr#1#2{\smash{\mathop{\hbox to .3in{\rightarrowfill}}
 \limits^{\scriptstyle#1}_{\scriptstyle#2}}}

\def\s2{\frac{1}{\sqrt2}}

\def\be{\begin{equation}}
\def\ee{\end{equation}}
\def\beqa{\begin{eqnarray}}
\def\eeqa{\end{eqnarray}}

\def\Dsl{\,\raise.15ex\hbox{/}\mkern-13.5mu D} 
\def\d3{d^3}

\begin{document}

\begin{center}
\Large{\bf Radiation Damping of a Yang-Mills Particle Revisited}\\
\end{center}

\vspace{1cm}

\begin{center}
{\large{Sair Arquez}\footnote{e-mail address: {\tt
sarquez@fis.cinvestav.mx}}} \\
{\it Departamento de F\'{\i}sica, Escuela Superior de F\'{\i}sica y
Matem\'aticas del Instituto Polit\'ecnico Nacional\\ Unidad Adolfo
L\'opez Mateos, Edificio 9,
07738, Ciudad de M\'exico, M\'exico} \\
\vskip .2truecm
{\it Departamento de Ciencias Naturales y Exactas, Universidad de la Costa,\\ Calle 58 num. 55-66, Barranquilla, Colombia.}\\
\vspace{.5cm} {\large{Rub\'en Cordero}\footnote{e-mail address: {\tt
cordero@esfm.ipn.mx}}} \\
{\it Departamento de F\'{\i}sica, Escuela Superior de F\'{\i}sica y
Matem\'aticas del Instituto Polit\'ecnico Nacional \\ Unidad Adolfo
L\'opez Mateos, Edificio 9,
07738, Ciudad de M\'exico, M\'exico}\\
\vspace{.5cm} {\large{Hugo Garc\'{\i}a-Compe\'an}\footnote{e-mail
address:
{\tt compean@fis.cinvestav.mx}}} \\
{\it Departamento de F\'{\i}sica, Centro de Investigaci\'on y de
Estudios Avanzados del IPN\\ P.O. Box 14-740, 07000, Ciudad de M\'exico, M\'exico}\\
\end{center}
\vspace{1cm}
\begin{abstract}
The problem of a color-charged point particle interacting with a
four dimensional Yang-Mills gauge theory is revisited. The radiation
damping is obtained inspired in the Dirac's computation. The
difficulties in the non-abelian case were solved by using an ansatz
for the Li\'enard-Wiechert potentials, already used in the
literature \cite{Sarioglu:2002qb} for finding solutions to the
Yang-Mills equations. Three non-trivial examples of radiation
damping for the non-abelian particle are discussed in detail.

\end{abstract}


\vspace{5cm}

\leftline{September 13, 2018}
\newpage

\section{Introduction}

The problem of motion for a classical radiating particles  has been
studied for a long time in the non-relativistic and relativistic
contexts \cite{Abraham,Dirac:1938nz}. It is well known that this
particle suffers a radiation damping due the self-force. In the
Dirac's correction there is a third-order time derivative of the
position of the particle. Thus there are an unphysical phenomena
such as runaway's behavior and preaccelerating
\cite{Teitelboim,bonnor,AresdeParga:1998ti}. It is right now known
that this bad behavior is a consequence of the pointlike nature of
the particle. An exhaustive study of the mentioned phenomena was
performed by many authors and the systematic description of some
divergences require to renormalize the electron charge and mass. For
some recent overviews, see
\cite{Jackson:1998nia,FRohrlich,RohrlichAJP2000,Rohrlich:1999gd}.
The solution of the mentioned problem has been of great importance
to understand the relativistic dynamics of the particles interacting
with radiation. Particularly in the Ref. \cite{Dirac:1938nz}, Dirac
calculated the self-force of the emitted electromagnetic radiation
(or radiation damping) from an electron. He considered expressions
for retarded and advanced potentials, in terms of a parameter
$\sigma$ that up on regularization is function of the electron size,
which obviously must be taking to zero at the end of the
computation.

The equations of motion for a particle carrying isotopic-spin in a
non-abelian Yang-Mills field has been originally worked out in the
70s by S.K. Wong \cite{Wong:1970fu}.  Moreover the study of
non-abelian particles and their equations of motion have been of
great importance along the years. Later important developments were
given in Refs. \cite{Balachandran:1976ya,Balachandran:1977ub}. In
these papers Lagrangian and Hamiltonian for the charged particle in
a non-abelian gauge field are proposed such that Wong's equations
are obtained from them. The canonical quantization and some detailed
cases were worked out in the second paper
\cite{Balachandran:1977ub}. Further developments using Wong's
equations and incorporating spin were described in
\cite{Arodz:1982fm}. The non-abelian particle is also described by
using the world-line formalism in Ref. \cite{Bastianelli:2013pta}.
The application to the transport properties of non-abelian fluids or
plasmas and in general to non-abelian hydrodynamics were discussed
in Refs. \cite{Heinz:1983nx,Heinz:1985vf,Kelly:1994dh,
Gyulassy:1994ug,Nayak:1996ex,Litim:1999id,Bistrovic:2002jx,Arnold:2005vb,Dumitru:2007rp,PeraltaRamos:2012er,Fernandez-Melgarejo:2016xiv}.
In particular the importance of Wong's equations in the problem of
the quark-gluon plasma has been considered in Refs.
\cite{Poschl:1998fr,JalilianMarian:2000ad,Voronyuk:2015ita,Mrowczynski:2016etf,Dumitru:2018vpr}.
More recently, the consideration of the Wong's equations of motion
has been discussed in the context of the double copy approach
\cite{Goldberger:2016iau} and also as a possible candidate of dark
matter \cite{Dzhunushaliev:2018hui}.

As a next logical step, to extend the ideas in \cite{Dirac:1938nz}
to the case of non-abelian theories, however this simple idea has
strong complications. The problem of motion for particles in a
Yang-Mills fields have been discussed in a subsequent series of
papers
\cite{Drechsler:1981nc,Kates:1984cn,Trautman:1981qd,Oh:1985gj}. In
\cite{Drechsler:1981nc} the equations of motion were derived from
the energy and momentum conservation. Also some solutions were
obtained in terms of the first order abelian L\'ienard-Wiechert
solution. In \cite{Kates:1984cn} it was studied the radiation
damping of color in Yang-Mills theory.  In Ref.
\cite{Trautman:1981qd} it was argued that the color charge can be
changed for a suitable gauge group. If the gauge group is compact
and semisimple then the color charge remains constant, however there
is a transfer of energy described by the L\'ienard-Wiechert (LW)
potentials. Moreover in Ref. \cite{Oh:1985gj} it was studied the
variation of the color charges with respect to the choice of the
non-abelian waves.

One of the most important problems in this context is to find
suitable L\'ienard-Wiechert's potentials as solutions to the
Yang-Mills equations. However, it is possible to construct these
potentials satisfying physical conditions and being also compatible
with the Yang-Mills equations. The simplest case is to consider an
expansion in the form ${1\over R}$, where $R$ gives the retarded
distance, for a retarded potential such that it vanishes at
infinity. However is possible to make an ansatz and write the LW
potentials as superposition of terms with different negative powers
of $R$. The non-abelian dependence lies on certain $\alpha_{i}$
functions that take values in the Lie algebra of the Lie group and
also depend on the proper-time \cite{Sarioglu:2002qb}. In the
present paper we use this ansatz and the Dirac's procedure
\cite{Dirac:1938nz} in order to obtain the self-force and
consequently the radiation damping of a Yang-Mills particle
following Wong's equations.

This paper is organized as follows. In Section 2 we introduce the
notation and conventions we will follow along the article. In
particular we give a brief review of the L\'ienard-Wiechert's
potentials and the ansatz \cite{Sarioglu:2002qb} that we will use to
compute the radiation damping in subsequent sections. In this
Section 2 we compute the retarded Yang-Mills field strength and the
associated radiated field strength. Section 3 is devoted to obtain
the self-force correction to Wong's equations. We follow Dirac's
method of computing the flux in the volume of the energy-momentum
tensor. In this same section we find the equations of motion for
different cases. In the first case we recover the abelian Dirac's
modified equations of motion.  Later we present three non-trivial
cases of the radiation damping of the non-abelian particle. Finally
in Section 4 we present our final comments. Three appendices were
included in order to write down some important details of the very
long computations relevant to Sections 2 and 3.

\section{Li\'enard-Wiechert Potentials for a Non-Abelian Yang-Mills Charge}\label{S2b2}

In the present section we will give some preliminaries and the set
up of the next sections. We also introduce the notation and
conventions we will follow along the article.

Consider the Yang-Mills potential $A_{\mu}(x) = A^{a}_{\mu}(x)
T_{a}$, where $T_{a}$ $a=1,\dots , $dim$(G)$ are the generators of
the compact and simple gauge group $G$, and $x$ labels the points in
the Minkowski space-time with signature $(+,-,-,-)$. These fields
have a field strength given by
\begin{equation}\label{k1}
F_{\mu\nu}=\partial_{\mu} A_{\nu}-\partial_{\nu}
A_{\mu}+g\left[A_{\mu},A_{\nu}\right],
\end{equation}
where $g$ is the coupling constant. The action of the gauge field
$A_\mu$ in the presence of an external source $J_\mu$ reads
\begin{equation}
S=-\frac{1}{4}\int d^{4}x {\rm Tr} \left[ F_{\mu \nu}(x)F^{\mu
\nu}(x)- J_\mu(x) A^\mu(x)\right].
\end{equation}
The corresponding Euler-Lagrange equations for $A_{\mu}$ are the
Yang-Mills equations
\begin{equation}
D_{\mu}F^{\mu\nu}=-J^{\nu}.
\end{equation}
The classical interpretation of $J_{\nu}$ is known as the isotopic
spin current \cite{Wong:1970fu}, and expressed in the simple form
\begin{equation}
J_{\nu}=\int g \textit{I}(s) \dot{z}_{\mu}(s)\delta^{(4)}\left(
x_{\mu}-z_{\mu}(s)\right)ds,
\end{equation}
which is the current associated to particles moving along the
worldline $z_{\mu}(s)$. Here $I(s)$ is a generic element of the Lie
algebra of $G$ and  $z_\mu(s)$ stands for the position of the
particle describing a smooth curve $\Gamma$ in spacetime and $x_\mu$
is any point in Minkowski spacetime.

Now let $(x_{\mu}-z_{\mu}(s))(x^{\mu}-z^{\mu}(s))=0$ be the
light-cone condition, where $s$ denotes the proper time and
$R=\dot{z}^{\mu}(x_{\mu}-z_{\mu}(s))$ is the retarded distance. In
Ref. \cite{Sarioglu:2002qb}, the author proposed an ansatz in order
to obtain some interesting classical solutions to the Yang-Mills
equations in terms of the LW potentials. In the present article we
will use this ansatz in order to study the radiation damping of an
accelerated Yang-Mills particle.

Consider the ansatz for the LW potential worked out in
\cite{Sarioglu:2002qb}
\begin{equation}
A_{\mu}^{(ret)}= H(R) \dot{z}_{\mu} + G(R) \lambda_\mu,
\label{ansatz}
\end{equation}
where $H$ and $G$ are smooth functions of $R$. $\lambda_\mu$ is
given by $\lambda^\mu = {(x^\mu - z^\mu(s)) \over R}$, such that
$\lambda^\mu \dot{z}_\mu =1$ and $\lambda^\mu R_{,\mu}=1$ and
$\lambda^\mu a_{, \mu}=0$ ($a_{, \mu} \equiv {\partial a \over
\partial x^\mu}$).  In the latter condition for $\lambda_\mu$, $a$
represents the acceleration of the particle which is defined by
$a={1 \over R}\ddot{z}^\mu(x_\mu - z_\mu(s))$. Moreover one can
introduce additional scalars $a_k$ defined by $a_k\equiv \lambda_\mu
{d^k \ddot{z}^\mu \over ds^k}$, $k=0,1,2,3$ with $a_0=a$. These
scalars $a_k$ satisfy the following relation
\begin{equation}
\frac{(x^{\mu}-z^{\mu}(s))}{R}\partial_{\mu}a_{k}=0.
\end{equation}

Taking account these considerations the ansatz (\ref{ansatz}) is
given by
\begin{equation}\label{k10}
A_{\mu}^{(ret)}=\left(
\frac{\alpha_{1}}{R}+\frac{\alpha_{2}}{R^{2}}\right)\dot{z}_{\mu}+\left(
\frac{\alpha_{3}}{R}+\frac{\alpha_{4}}{R^{2}}+\frac{\alpha_{5}}{R^{3}}\right)\left(
x_{\mu}-z_{\mu}(s)\right),
\end{equation}
where $\alpha_{i}$ with $i=1,\dots,5$ are Lie algebra-valued
functions of the proper time and they are $R$-independent functions.

For a smooth curve $\Gamma$ in space-time one has that the velocity
and its respective derivatives in the proper time satisfies the
following relations
\begin{equation}
\nonumber
\textbf{v}^{2}=1\;\;\;\;\:\;\textbf{v}\dot{\textbf{v}}=0,\;\;\;\;\:\;\textbf{v}\ddot{\textbf{v}}=-\dot{\textbf{v}}^{2},\;\;\;\;\:\;\textbf{v}\dddot{\textbf{v}}=-3\dot{\textbf{v}}\ddot{\textbf{v}},
\end{equation}
\begin{equation}
\nonumber
\textbf{v}\textbf{v}^{(4)}=-3\ddot{\textbf{v}}^{2}-4\dot{\textbf{v}}\dddot{\textbf{v}},\;\;\;\;\:\;\textbf{v}\textbf{v}^{(5)}=-5\dot{\textbf{v}}\textbf{v}^{(4)}-10
\ddot{\textbf{v}}\dddot{\textbf{v}},
\end{equation}
\begin{equation}
\nonumber
\textbf{v}\textbf{v}^{(6)}=-10\dddot{\textbf{v}}^{2}-6\dot{\textbf{v}}\textbf{v}^{(5)}-15\ddot{\textbf{v}}\textbf{v}^{(4)},\;\;\;\;\:\;\textbf{v}\textbf{v}^{(7)}=-7\dot{\textbf{v}}\textbf{v}^{(6)}-21\ddot{\textbf{v}}\textbf{v}^{(5)}-35\dddot{\textbf{v}}\textbf{v}^{(4)},
\end{equation}
\begin{equation}
\textbf{v}\textbf{v}^{(8)}=-8\dot{\textbf{v}}\textbf{v}^{(7)}-28\ddot{\textbf{v}}\textbf{v}^{(6)}-56\dddot{\textbf{v}}\textbf{v}^{(5)}-35\left(
\textbf{v}^{(4)}\right)^{2},
\end{equation}
where we adopted Dirac's notation \cite{Dirac:1938nz} for the scalar
product: ${\bf v} {\bf v} \equiv {\bf v}^2 \equiv v^{\mu} v_{\mu}$,
and similarly for the other products, for instance ${\bf v} \dot{\bf
v} \equiv v^{\mu} \dot{v}_{\mu}$ with $v_{\mu}=\dot{z}_{\mu}(s)$.
Moreover the velocity of the particle ${\bf v}^{(n)}$ in components
reads as $v_{\mu}^{(n)}$ for $n \geq 4$ and they stand for the
$n$-th derivative of $v_{\mu}$ with respect to parameter $s$. Notice
that in general $\textbf{v}^{2}=\varepsilon$ with $\varepsilon=0,\pm
1$. Within the adopted signature $\varepsilon=0$ for null curves,
$\varepsilon=1$ for time-like curves, and $\varepsilon=-1$ for
space-like curves.

We can use the properties of the Dirac delta functions to express
Eq. (\ref{k10}) in the following form and to localize the LW
potentials along the path of the particle. This procedure allow us
to establish a relationship between the trajectory of the particle
and the process of emission and absorbtion of radiation. Moreover it
is easier to work in this representation given the possible
divergences of the potential at $R = 0$. The expansion (\ref{k10})
in $1/R$ is finite for $R\neq 0$. This physical condition is also
valid for non-abelian gauge theories, in which the potentials and
the interaction is finite at long distances, at least within the
classical description.

We follow the procedure introduced by Dirac in \cite{Dirac:1938nz}.
Thus we can express $A_{\mu}$ in the following form according to the
ansatz (\ref{ansatz}) adapted for the retarded gauge field
$$
A_{\mu}^{(ret)}=2\int \bigg[\alpha_{1}\dot{z}_{\mu}+\alpha_{3}\left(
x_{\mu}-z_{\mu}(s)\right)\bigg]\delta\left( \Omega\right)ds + 4 \int
\bigg[\alpha_{2}\dot{\overline{z}}_{\mu}+\alpha_{4}\left(
x_{\mu}-\overline{z}_{\mu}(s)\right)\bigg]
 \delta\left( \Omega\right)\delta\left( \overline{\Omega}\right)dsd\overline{s}
$$
\begin{equation}
+8\int \bar{\alpha}_{5}\left(
x_{\mu}-\widetilde{z}_{\mu}(s)\right)\delta\left(
\Omega\right)\delta\left( \overline{\Omega}\right)\delta\left(
\widetilde{\Omega}\right)ds d\overline{s} d\widetilde{s},
\label{fieldone}
\end{equation}
where $\Omega:= (x_{\mu}-z_{\mu}(s))(x^{\mu}-z^{\mu}(s))$ vanishes
as soon as $x^\mu$ coincides with the positions of the worldline of
the particle $z^{\mu}(s)$.

In order to calculate the corresponding retarded field strength we
compute (\ref{fieldone}) and plugging it into the Eq. (\ref{k1}),
then we obtain
\begin{align}
\label{efemunuret} \nonumber F_{\mu \nu}^{
(ret)}=&-\frac{\alpha_{1}}{R}\dfrac{d}{ds}\left[
\frac{1}{R}\dot{z}_{\mu}\left(
x_{\nu}-z_{\nu}\right)-\frac{1}{R}\dot{z}_{\nu}\left(
x_{\mu}-z_{\mu}\right)\right] \\ \nonumber
  &-\frac{2\alpha_{2}}{R^{2}} \dfrac{d}{ds}\left[ \frac{1}{R}\dot{z}_{\mu}\left( x_{\nu}-z_{\nu}\right)-\frac{1}{R}\dot{z}_{\nu}\left( x_{\mu}-z_{\mu}\right)\right]
  \\ \nonumber
 &+\left\lbrace \dfrac{-2\dot{\alpha}_{2}+\left[ \alpha_{1},\alpha_{4}\right]+\left[ \alpha_{2},\alpha_{3}\right] }{R^{3}}\right\rbrace
 \bigg[\dot{z}_{\mu}\left( x_{\nu}-z_{\nu}\right)-\dot{z}_{\nu}\left( x_{\mu}-z_{\mu}\right)\bigg]\\
   &+\left\lbrace \dfrac{\left[\alpha_{1},\alpha_{5}\right]+\left[\alpha_{2},\alpha_{4}\right]  }{R^{4}}+\dfrac{\left[ \alpha_{2} ,\alpha_{5}\right] }{R^{5}}\right\rbrace
   \bigg[\dot{z}_{\mu}\left( x_{\nu}-z_{\nu}\right)-\dot{z}_{\nu}\left( x_{\mu}-z_{\mu}\right)\bigg].\\
\nonumber
\end{align}
In the process of getting Eq. (\ref{efemunuret}) we used some
relations between the $\alpha_{i}$'s found in
\cite{Sarioglu:2002qb}.

To calculate the Yang-Mills field strength in the neighborhood of
the particle along its trajectory,  we need to perform a Taylor
expansion for $x_{\mu}=z_{\mu}(s_{0})+\gamma_{\mu}$, such that
$\gamma_{\mu}$ is a small function of the parameter $\sigma$. Here
$\sigma$ is the correction to the proper-time starting from its
initial value at $s\to s_{0}-\sigma$. This difference establishes to
first order the retarded-time, which is delayed from the observer's
time $s$ and gives the time when the field begun to propagate, then
we have the Taylor expansion in the form:
\begin{equation}\label{k3}
x_{\mu}-z_{\mu}(s_{0}-\sigma)=\gamma_{\mu}+\sum_{k}\frac{(-1)^{k-1}}{k!}\sigma^{k}v_{\mu}^{(k-1)}
\end{equation}
and
\begin{equation}\label{k4}
\dot{z}_{\mu}(s_{0}-\sigma)=-\sum_{k}\frac{(-1)^{k}}{(k-1)!}\sigma^{k-1}v_{\mu}^{(k-1)},
\end{equation}
where  $k = 1,2, \dots, 9$. The possible values of $k$ is due the
last term in $F_{\mu\nu}$ is of order $\sigma^{-8}$, thus it is
reasonable to take expansions up to order $\sigma^{9}$ to have all
terms independent of $\sigma$, analogously to the process carried
out by Dirac in Ref. \cite{Dirac:1938nz}. Using equations (\ref{k3})
and (\ref{k4}) we can calculate the expansions for $R$ and the
products $\dot{z}^{\mu}(x_{\nu}-z_{\nu}(s))$. This is done in
appendix A.

Yang-Mills field is singular at $R=0$, thus in order to avoid these
divergences it is convenient to introduce a regularization parameter
$\varepsilon$ which can be interpreted as the size of the particle.
Equations of motion cannot depend on $\varepsilon$, thus at the end
of the computation we will have the finite and the divergent parts.
By taking the limit $\varepsilon$ tends to zero we will take zero
all the positive powers of $\varepsilon$ and the remaining few
negative powers of $\varepsilon$ will be renormalized.

To calculate the radiated field strength of the radiating particle
$F_{\mu \nu}^{(rad)}$, which can be determined by the equation:
\begin{equation}\label{k30}
F^{(rad)}_{\mu \nu}= F^{(ret)}_{\mu \nu} - F^{(adv)}_{\mu \nu} =
F^{(out)}_{\mu \nu} - F^{(in)}_{\mu \nu}.
\end{equation}
In analogy to the procedure performed by Dirac in Ref.
\cite{Dirac:1938nz} we can define the following relations
\begin{equation}\label{Frad}
f_{\mu \nu}=F_{\mu \nu}^{(act)}-\frac{1}{2}\left( F_{\mu
\nu}^{(ret)}+F_{\mu \nu}^{(adv)}\right),  \;\;\; f_{\mu \nu}=F_{\mu
\nu}^{(in)}+\frac{1}{2}F_{\mu \nu}^{(rad)}.
\end{equation}
The possibility to define the field in this way will allow to
present the equations of motion in a compact and familiar form.

At this point it is worth highlighting the fact that in general
these definitions were set out from first principles under the
conditions of conservation of the moment and energy and in addition
to the relativistic invariance of the Lagrangian. In our case all
these conditions are satisfied classically, for this reason we have
used these same tools. From Eq. (\ref{k30}) we can calculate
$F_{\mu\nu}^{(rad)}$ (for the details of the computation, see
appendix B) which are expressed in the form
\begin{align}\label{a}
\nonumber F_{\mu \nu }^{(rad)}=&\left(
-\frac{3}{4}\dot{\textbf{v}}^2A_2-\frac{7}{12}\dot{\textbf{v}}^2A_4-\frac{11}{12}\dot{\textbf{v}}\ddot{\textbf{v}}
A_5\right) \left(v_{\mu } \dot{v}_{\text{$\nu $}}- v_{\nu
}\dot{v}_{\text{$\mu $}}\right)  \\  \nonumber
+&\frac{1}{3}\left(\frac{3}{2}\dot{\textbf{v}}^{2} A_5 +2 A_3 +4 A_1\right)
\left(\ddot{v}_{\text{$\mu $}} v_{\nu }-\ddot{v}_{\text{$\nu
$}}v_{\mu }\right)\\  \nonumber
+&\frac{1}{4}\left(A_4+3 A_2\right)  \left( v_{\mu } \dddot{v}_{\text{$\nu $}}+\frac{2}{3}\dot{v}_{\text{$\mu $}} \ddot{v}_{\text{$\nu $}}-\frac{2}{3}\ddot{v}_{\text{$\mu $}} \dot{v}_{\text{$\nu $}}-\dddot{v}_{\text{$\mu $}} v_{\nu } \right)\\
+&2 A_5 \left( -\frac{v_{\mu } v^{(4)}_{\text{$\nu
$}}}{30}-\frac{\dot{v}_{\text{$\mu $}} \dddot{v}_{\text{$\nu
$}}}{24}+\frac{\dddot{v}_{\text{$\mu $}} \dot{v}_{\text{$\nu
$}}}{24}+\frac{v^{(4)}_{\text{$\mu $}} v_{\nu }}{30}\right),
\end{align}
where $A_{1}=-\alpha_{1}, A_{2}=-2\alpha_{2},
A_{3}=-2\dot{\alpha}_{2}+\left[ \alpha_{1},\alpha_{4}\right]+\left[
\alpha_{2},\alpha_{3}\right]$,
$A_{4}=\left[\alpha_{1},\alpha_{5}\right]+\left[\alpha_{2},\alpha_{4}\right]$,
$A_{5}=\left[ \alpha_{2},\alpha_{5}\right]$.

To calculate the advance field strength $F_{\mu\nu }^{(adv)}$ from
$F_{\mu\nu }^{(ret)}$ it is only required to change $\varepsilon$ by
$-\varepsilon$, however unlike the Dirac case \cite{Dirac:1938nz} in
the computation of $F_{\mu \nu }^{(rad)}$ there will survive the
even powers terms in $F_{\mu \nu }^{(rad)}$. This result is
interesting because physically the radiated field has a subtle
dependence on these terms which would involve divergences of the
radiated field. However without these divergences the equations of
motion would not be obtained correctly.

An important point is that in the second term from (\ref{a}) there
is a term that corresponds exactly to Dirac's radiative field
\begin{equation}
F_{\mu \nu \ D}^{(rad)}=\frac{4}{3} A_1 \left( \ddot{v}_{\text{$\mu
$}} v_{\nu }-\ddot{v}_{\text{$\nu $}}v_{\mu }\right),
\end{equation}
which is the usual one for the radiation produced by an accelerated
abelian charge.

\section{The Equations of Motion}\label{Ka}

If a charged classical particle moves along its worldline it
generates a flow on its trajectory. This flow can be calculated
through the energy-momentum tensor of the field in the neighborhood
of a point along the trajectory of the particle. The energy-momentum
tensor is given by
\begin{equation}\label{T1}
T^{F}_{\mu\nu}= {\rm Tr}\bigg(F^{\rho}_{\mu}
F_{\rho\nu}-\frac{1}{4}g_{\mu\nu}F^{\alpha\beta}
F_{\alpha\beta}\bigg).
\end{equation}
We can consider a spherical hyper-surface of radius $\varepsilon$
covering the path. The integral on this surface is
$4\pi\varepsilon^{2}$, and the expression between the braces is
precisely the equation of motion outlined in appendix C
\begin{align}
\label{eq0}
\nonumber
&\int \varepsilon^{-1}T_{\mu\rho}\gamma^{\rho}ds\left|dx_{\parallel}\right|=\int {\rm Tr}\left\lbrace \frac{A_5 f_{\mu \nu } v^{\nu }+Q_{14} v_{\mu }+Q_{11} \dot{v}_{\text{$\mu $}}+Q_{17} \dot{v}_{\text{$\mu $}}+Q_{16} \dddot{v}_{\text{$\mu $}}}{\varepsilon ^2}\right.  \nonumber\\
+&\frac{A_5 f_{\mu \nu } \dot{v}^{\nu }+Q_{36} f_{\mu \nu } v^{\nu }+Q_9 v_{\mu }+Q_3 \dot{v}_{\text{$\mu $}}+Q_{10} \ddot{v}_{\text{$\mu $}}+Q_{15} \dddot{v}_{\text{$\mu $}}}{\varepsilon } \nonumber\\
+&Q_{33} f_{\mu \nu } \ddot{v}^{\nu }+Q_8 v_{\mu }+Q_6 \dot{v}_{\text{$\mu $}}+Q_2 \ddot{v}_{\text{$\mu $}}+Q_{12} \dddot{v}_{\text{$\mu $}}-A_4 f_{\mu \nu } \dot{v}^{\nu }+Q_{35} f_{\mu \nu } v^{\nu } \nonumber\\
+&\left. \frac{Q_{24} \dot{v}_{\text{$\mu $}}}{\varepsilon
^5}+\frac{Q_{23} v_{\mu }+Q_{22} \dot{v}_{\text{$\mu $}}+Q_{20}
\ddot{v}_{\text{$\mu $}}}{\varepsilon ^4}+\frac{Q_{13} v_{\mu
}+Q_{21} \dot{v}_{\text{$\mu $}}+Q_{19} \ddot{v}_{\text{$\mu
$}}+Q_{18} \dddot{v}_{\text{$\mu $}}}{\varepsilon ^3}\right\rbrace
ds,
\end{align}
where the values of the functions $Q_{i}$'s are given in appendix C.

All terms depending on $\varepsilon$ must disappear when we take the
limit $\varepsilon \to 0$. This is an impediment to obtain the
equations of motion directly from the calculation, due to the terms
that arise as negative powers of $\varepsilon$. These terms induce
divergences in the equation of motion that should be removed. One
strategy is to use the fact that the braces is an exact differential
and match to a vector $\dot{B}_{\mu}$. However there is an
obstruction to find the Dirac's constraint in which
$\textbf{v}\dot{\textbf{B}} \equiv v_{\mu}\dot{B}^{\mu}=0 $. A
useful procedure is to consider
\begin{equation}\label{eq1}
\begin{split}
\dot{B}_{\mu}=&\left(\frac{Q_{23} }{\varepsilon^{4}}+\frac{Q_{13} }{\varepsilon^{3}}+\frac{Q_{14} }{\varepsilon^{2}}+\frac{Q_{9} }{\varepsilon}+Q_{8}\right) v_{\mu}+\left(\frac{Q_{24} }{\varepsilon^{5}}+\frac{Q_{22} }{\varepsilon^{4}}+\frac{Q_{21} }{\varepsilon^{3}}+\frac{Q_{11} }{\varepsilon^{2}}+\frac{Q_{3} }{\varepsilon}+Q_{6}\right) \dot{v}_{\mu}\\
& + \left(\frac{Q_{20} }{\varepsilon^{4}}+\frac{Q_{19} }{\varepsilon^{3}}+\frac{Q_{17} }{\varepsilon^{2}}+\frac{Q_{10} }{\varepsilon}+Q_{2}\right) \ddot{v}_{\mu}+\left(\frac{Q_{18} }{\varepsilon^{3}}+\frac{Q_{16} }{\varepsilon^{2}}+\frac{Q_{15} }{\varepsilon}+Q_{12}\right) \dddot{v}_{\mu}\\
& + \left(\frac{A_{5} }{\varepsilon^{2}}+\frac{Q_{36}
}{\varepsilon}+Q_{35}\right) f_{\mu\nu}v_{\nu}+\left(\frac{A_{5}
}{\varepsilon}-A_{4}\right)
f_{\mu\nu}\dot{v}_{\nu}+Q_{33}f_{\mu\nu}\ddot{v}^{\nu}.
\end{split}
\end{equation}
We will determine a solution to the above expression in such a way
that we can obtain the value of  $B_{\mu}$. In the present case the
product $\textbf{v}\dot{\textbf{B}}\neq 0$ which complicates the
calculation of the equations of motion. However a good strategy is
to find, from our original set of vectors $\dot{B}_{\mu}, v_{\mu},
\dot{v}_{\mu}, \ddot{v}_{\mu}, \dddot{v}_{\mu}$,  a new set of
normalized vectors $\dot{\widehat{B}}_{\mu}, \widehat{u}_{1\mu},
\widehat{u}_{2\mu}, \widehat{u}_{3\mu}, \widehat{u}_{4\mu}$ (see
appendix C). It is now easy to prove that
$\widehat{\textbf{u}}_{i}\dot{\textbf{B}}= 0$, $i=1, \dots ,4$. Here
$\widehat{u}_{i\mu}=u_{i\mu}/ \left|u_{i}\right|$ and we have
$\widehat{\textbf{u}}_{i}\dot{\widehat{\textbf{u}}}_{i}=0$, due
$\widehat{\textbf{u}}_{i}\widehat{\textbf{u}}_{i}=1$. Thus a simple
solution to the equation $\widehat{\textbf{u}}_{i}\dot{\textbf{B}}=
0$ is given by
\begin{equation}\label{solut}
B_{\mu}=k\widehat{u}_{4\mu}.
\end{equation}

However, by replacing the value of $\widehat{u}_{4\mu}$ in Eq.
(\ref{solut}), we have a non-homogeneous first-order differential
equation, whose coefficients are function of the norm $|\dot{\bf
B}|=\dot{B}$, that involves the limit $\varepsilon \to  0$ and which
does not exactly cancel the divergences. A solution to this problem
is to choose $\dot{B}_{\mu}$ so that it respects the general form of
$\dot{B}_{\mu}$, i.e. $\widehat{\textbf{u}}_{i}\dot{\textbf{B}}= 0$
and also eliminates the divergences. Therefore the best way to do
this is to take the linear combination
\begin{equation}\label{sol}
\dot{B}_{\mu}=f_{0}(\varepsilon)v_{\mu}+f_{1}(\varepsilon)\dot{v}_{\text{$\mu
$}}+f_{2}(\varepsilon)\ddot{v}_{\text{$\mu $}},
\end{equation}
for some suitable functions $f_i$. Moreover the choice of
(\ref{sol}) is not arbitrary because
$\widehat{\textbf{u}}_{i}\dot{\textbf{B}}= 0$ requires the
additional condition: $Q_{18}=0$ (since
$\widehat{\textbf{u}}_{i}\textbf{v}^{(j)}=0$ except for $j=3$). This
implies that there is an abelian subgroup of the Lie's algebra
satisfying  $[\alpha_{2},\alpha_{5}]^{2} = 0$. Consequently it
reduces the equations to
\begin{align}
\label{eq1}
\nonumber
\dot{B}_{\mu}=&\left[ \frac{-\frac{2}{3}\left(  A_4^2 + A_2^2 + A_2 A_4 \right) \dot{\textbf{v}}^2}{\varepsilon^{2}}+\frac{\left(  \frac{2}{3} A_1 A_2 +\frac{1}{3} A_2 A_3 +\frac{5}{6} A_1 A_4+\frac{7}{6} A_3 A_4\right) \dot{\textbf{v}}^{2}  }{\varepsilon}\right.  \nonumber\\
+&\left.  \frac{\left( -\frac{1}{2} A_2^2-\frac{5}{6} A_4^2 -\frac{5}{3} A_2 A_4\right)  \dot{\textbf{v}}\ddot{\textbf{v}}}{\varepsilon}-\frac{1}{24} \left( 8A_2^2+9A_4^2 +13 A_2 A_4 \right) \dot{\textbf{v}}^4\right. \nonumber\\
-&\frac{1}{6} \left( 2A_2^2 + A_4^2+3 A_2 A_4\right)  \dot{\textbf{v}} \dddot{\textbf{v}}-\frac{1}{6}\left( 2 A_2^2 +A_4^2 +3A_2 A_4\right)  \ddot{\textbf{v}}^2 \nonumber\\
 -&\left. \frac{1}{2} \left( A_3^2 + A_1 A_3\right)  \dot{\textbf{v}}^{2}+\frac{1}{12}\left(  9A_1 A_2 +17A_2 A_3+9 A_1 A_4+13A_3 A_4
\right) \dot{\textbf{v}}\ddot{\textbf{v}}\right]  v_{\mu} \nonumber\\
+&\left[ \frac{-A_2^2+A_4^2 }{\varepsilon^{3}}+\frac{3 A_1 A_2+A_1 A_4-A_2 A_3-3 A_3 A_4}{2\varepsilon^{2}}+\frac{- A_2^2 \dot{\textbf{v}}^2- A_2 A_4 \dot{\textbf{v}}^2+A_3^2-A_1^2}{2\varepsilon}\right. \nonumber\\
+&\left. \frac{1}{8}\left(  5A_1 A_2+3 A_2 A_3+3 A_1 A_4 +A_3 A_4\right)  \dot{\textbf{v}}^2+\frac{1}{6}\left( A_4^2-5A_2^2 -4A_2 A_4 \right)  \dot{\textbf{v}}\ddot{\textbf{v}} \right]  \dot{v}_{\mu} \nonumber\\
+&\left[ \frac{- 2 A_2^2+A_4 A_2+A_4^2 }{3\varepsilon^{2}}+\frac{2
A_1 A_2+A_1 A_4-2 A_2
A_3-A_3 A_4}{3\varepsilon}-\frac{1}{6} \left( 2A_2^2 +A_2 A_4\right)  \dot{\textbf{v}}^2\right]  \ddot{v}_{\mu} \nonumber \\
+&\left(\frac{A_2-A_4 }{\varepsilon}-A_1+A_3\right)
f_{\mu\nu}v^{\nu}-A_{4} f_{\mu\nu}\dot{v}^{\nu}.
\end{align}

We now can take $\dot{B}_{\mu}$ of the general form:
\begin{equation}\label{alef}
\dot{B}_{\mu}=f_{0}(\varepsilon)v_{\mu}+f_{1}(\varepsilon)\dot{v}_{\text{$\mu
$}}+f_{2}(\varepsilon)\ddot{v}_{\text{$\mu
$}}+k_{3}f_{\mu\nu}v^{\nu},
\end{equation}
where the last term in (\ref{alef}) is not consistent with
(\ref{sol}). However, it can be proved that this term satisfies the
condition $\dot{\textbf{B}}\widehat{\textbf{u}}_{i}=0 $ but this is
possible only if $ L_{\nu}\dot{v}^{\nu}:=0 $ where
$L_{\nu}:=f_{\mu\nu}\widehat{u}_{4}^{\mu}$ is a null vector. The
physical meaning of this is to have a privileged direction of the
Yang-Mills field, see Eq. (\ref{UL}) in appendix C, for which it is
allowed to add it to  Eq. (\ref{alef}). Then the equations of motion
can be expressed in the form
\begin{align}\label{eq1b}
\nonumber
&\left[ \rho_{1}\dot{\textbf{v}}^{2} +\rho_{2}\dot{\textbf{v}}\ddot{\textbf{v}}-\frac{1}{24} \left( 8A_2^2+9A_4^2 +13 A_2 A_4 \right) \dot{\textbf{v}}^4\right. \nonumber\\
-&\frac{1}{6} \left( 2A_2^2 + A_4^2+3 A_2 A_4\right)  \dot{\textbf{v}}\dddot{\textbf{v}}-\frac{1}{6}\left( 2 A_2^2 +A_4^2 +3A_2 A_4\right)  \ddot{\textbf{v}}^2 \nonumber\\
 -&\left. \frac{1}{2} \left( A_3^2 + A_1 A_3\right)  \dot{\textbf{v}}^{2}+\frac{1}{12}\left(  9A_1 A_2 +17A_2 A_3+9 A_1 A_4+13A_3 A_4
\right) \dot{\textbf{v}}\ddot{\textbf{v}}\right]  v_{\mu} \nonumber\\
+&\left[m+\mu_{1}\dot{\textbf{v}}^{2}+\frac{1}{8}\left(  5A_1 A_2+3 A_2 A_3+3 A_1 A_4 +A_3 A_4\right)  \dot{\textbf{v}}^2+\frac{1}{6}\left( A_4^2-5A_2^2 -4A_2 A_4 \right)  \dot{\textbf{v}}\ddot{\textbf{v}} \right]  \dot{v}_{\mu} \nonumber\\
+&\left[\mu_{3}-\frac{1}{6} \left( 2A_2^2 +A_2 A_4\right)
\dot{\textbf{v}}^2\right]  \ddot{v}^{\mu}+\left(q-A_1+A_3\right)
f_{\mu\nu}v_{\nu}-A_{4} f_{\mu\nu}\dot{v}^{\nu}=0,
\end{align}
where $rho_1$, $\rho_2$, $\mu_1$, $\mu_3$, $m$ and $q$ are suitable
renormalization constants. Furthermore the form of the functions
$f_{i}(\varepsilon)$ that cancels the divergences in Eq. (\ref{eq1})
is of the form
$$
f_{0}=\rho_{1}\dot{\textbf{v}}^{2} +\frac{-\frac{2}{3}\left(  A_4^2
+ A_2^2 + A_2 A_4 \right)
\dot{\textbf{v}}^2}{\varepsilon^{2}}+\frac{\left(  \frac{2}{3} A_1
A_2 +\frac{1}{3} A_2 A_3 +\frac{5}{6} A_1 A_4+\frac{7}{6} A_3
A_4\right) \dot{\textbf{v}}^{2}  }{\varepsilon}
$$
\begin{equation}
+\rho_{2}\dot{\textbf{v}}\ddot{\textbf{v}}+ \frac{\left(
-\frac{1}{2} A_2^2-\frac{5}{6} A_4^2 -\frac{5}{3} A_2 A_4\right)
\dot{\textbf{v}}\ddot{\textbf{v}}}{\varepsilon},
\end{equation}
$$
f_{1}=m+\mu_{1}\dot{\textbf{v}}^{2}+ \frac{-A_2^2+A_4^2
}{\varepsilon^{3}}+\frac{3 A_1 A_2+A_1 A_4-A_2 A_3-3 A_3
A_4}{2\varepsilon^{2}}
$$
\begin{equation}
+\frac{- A_2^2 \dot{\textbf{v}}^2- A_2 A_4
\dot{\textbf{v}}^2+A_3^2-A_1^2}{2\varepsilon},
\end{equation}
\begin{equation}
f_{2}=\mu_{3}+\frac{- 2 A_2^2+A_4 A_2+A_4^2
}{3\varepsilon^{2}}+\frac{2 A_1 A_2+A_1 A_4-2 A_2 A_3-A_3
A_4}{3\varepsilon}.
\end{equation}
Here $k_{3}$ is a correction to the charge of the non-abelian
particle \cite{bonnor,AresdeParga:1998ti}
\begin{equation}
k_{3}=-q+\frac{A_2-A_4 }{\varepsilon}.
\end{equation}

\subsection{The Dirac's radiation damping effect}

A particular case of Eq. (\ref{eq1}) is when one considers
$\alpha_{2},\alpha_{3},\alpha_{4},\alpha_{5}=0$, which is justly the
abelian case according to Eq. (\ref{efemunuret}). For this case we
have that Eq. (\ref{eq1}) is of the form
\begin{equation}\label{20}
\dot{B}_{\mu}=\frac{1}{2}\alpha_{1}^{2}\varepsilon^{-1}\dot{v}_{\mu}-\alpha_{1}f_{\mu\nu}v^{\nu},
\end{equation}
where $f_{\mu \nu}$ is determined by Eq. (\ref{Frad}).

In this case $B_{\mu}=kv_{\mu}$ is solution to Eq. (\ref{20}), where
$k=1/2\alpha_{1}^{2}\varepsilon^{-1}+ \beta$. Thus we have the
traditional equation of a charged particle in an electromagnetic
field with corrections given the emission/absorption of radiation or
the Abraham-Lorentz-Dirac equation \cite{Dirac:1938nz}
\begin{equation}
m\dot{v}_{\mu}=- {\rm Tr} \big( \alpha_1f_{\mu\nu}v^{\nu}\big),
\label{ALD}
\end{equation}
where $m={\rm Tr}(\beta)$.

\subsection{The First non-abelian radiation damping effect}

Now we consider the case in which $\alpha_{2}=\alpha_{5}=0 $. For
this situation we find that the $A_{1}=-\alpha_{1}$, and
$A_{3}=\left[ \alpha_{1},\alpha_{4}\right]$. Plugging these
expressions into Eq. (\ref{eq1}) we have that the equations of
motion take the simple form
\begin{equation}
\dot{B}_{\mu}=\frac{1}{2}\left( A_3^2-A_1^2 \right)\varepsilon^{-1}
\dot{v}_{\mu}-\frac{1}{2}\left(  A_3^2+A_1 A_3\right)  \dot{\bf v}^2
v_{\mu }+\left( A_3 -A_1\right)  f_{\mu \nu } v^{\nu }.
\end{equation}
In order to obtain the actual equations of motion we have to take
the trace in both sides of the previous equation
\begin{equation}
{\rm Tr}(\dot{B}_{\mu})= {\rm Tr} \bigg(\frac{1}{2}\left(
A_3^2-A_1^2 \right)\varepsilon^{-1} \dot{v}_{\mu}-\frac{1}{2}\left(
A_3^2+A_1 A_3\right)  \dot{\bf v}^2 v_{\mu }+\left( A_3 -A_1\right)
f_{\mu \nu } v^{\nu }\bigg).
\end{equation}
The first term on the right-hand side can be renormalized by taking
$B_{\mu}=kv_{\mu}$ and $k={1 \over
2}(\left[\alpha_{1},\alpha_{4}\right]^{2}-\alpha_{1}^{2})\varepsilon^{-1}+
\beta.$ Then we find the equation of motion
\begin{equation}
m\dot{v}_{\mu}+\mu_{1} \dot{v}^2 v_{\mu }= {\rm
Tr}\big(\widetilde{\mu} f_{\mu \nu } v^{\nu}\big).
\end{equation}
Thus we have the simplest non-abelian equation with $m= {\rm
Tr}(\beta)$ and $\mu_1={\rm Tr}(\widetilde{\mu}_1)$, where
$\widetilde{\mu}_{1}=\frac{1}{2}\left( \left[
\alpha_{1},\alpha_{4}\right]^2-\alpha_1 \left[
\alpha_{1},\alpha_{4}\right]\right)$ and  $\widetilde{\mu}= \left[
\alpha_{1},\alpha_{4}\right] +\alpha_1$, respectively. The field
$f_{\mu\nu}$ in (\ref{Frad}) is modified only in the non-abelian
charges of the radiation term which has the form
\begin{equation}
F_{\mu \nu}^{(rad)}=\frac{2}{3}(\left[
\alpha_{1},\alpha_{4}\right]-2 \alpha_{1} ) \left( \ddot{
v}_{\text{$\mu $}} v_{\nu }-v_{\mu } \ddot{ v}_{\text{$\nu
$}}\right).
\end{equation}
Let us now analyze the second non-abelian case of interest, such
that $k$ in $B_{\mu}$ are both functions of velocity and
acceleration.
\subsection{The second non-abelian radiation damping
effect}\label{SUB2}

This case is more general than the previous one since it includes
divergences of order $\varepsilon^{-1}$, $\varepsilon^{-2}$ and
$\varepsilon^{-3}$ in the mass and of order $\varepsilon^{-1}$ in
the color's charge. If we take $\alpha_{2}=0$ and $\alpha_{4}=0$,
then equations of motion read
\begin{align}
\nonumber \dot{B}_{\mu}= &\left(-\frac{2 A_4^2 \dot{\textbf{v}}^2
}{3 \varepsilon ^2} -\frac{3}{8} A_4^2 \dot{\textbf{v}}^4
-\frac{1}{6} A_4^2 \ddot{\textbf{v}}^2 +\frac{3}{4} A_1 A_4
\dot{\textbf{v}} \ddot{\textbf{v}} -\frac{1}{6} A_4^2
\dot{\textbf{v}} \dddot{\textbf{v}} +\frac{5 A_1 A_4 \dot{\bf v}^2
}{6 \varepsilon }-\frac{5 A_4^2 \dot{\textbf{v}}\ddot{\textbf{v}}
}{6 \varepsilon }\right) v_{\mu }\\   \nonumber +&\left( \frac{3}{8}
A_1 A_4 \dot{\textbf{v}}^2+\frac{2}{3} A_4^2 \dot{\textbf{v}}
\ddot{\textbf{v}} +\frac{1}{6} A_4^2 \textbf{v}\dddot{\textbf{v}}
+\frac{A_4^2 }{\varepsilon ^3}+\frac{A_1 A_4}{2 \varepsilon
^2}-\frac{A_1^2 }{2 \varepsilon }\right) \dot{v}_{\mu}+\left(
\frac{A_4^2}{3 \varepsilon
^2}+\frac{A_1 A_4 }{3 \varepsilon }\right)\ddot{v}_{\mu} \\
-&A_1 f_{\mu \nu } v^{\nu }-\frac{A_4 f_{\mu \nu } v^{\nu
}}{\varepsilon }-A_4 f_{\mu \nu } \dot{v}^{\nu }.
\end{align}
The simplest expression for $\dot{B}_{\mu}$ is given by
\begin{equation}\label{bo}
\dot{B}_{\mu}=g_{0}v_{\mu}+k_{1}\dot{v}_{\mu}+k_{2}\ddot{v}_{\mu}+k_{3}f_{\mu
\nu } v^{\nu},
\end{equation}
for suitable coefficients $g_0$, $k_1$, $k_2$ and $k_3$. In this
case we can simplify the previous equation by defining the
non-abelian relations
\begin{equation}
k_{1}:=\varepsilon ^{-3}A_4^2 +\frac{\varepsilon^{-2}}{2}A_1
A_4-\frac{\varepsilon^{-1}}{2}A_1^2-m, \;\;\; k_{2}:=\frac{
\varepsilon ^{-2}}{3}A_4^2+\frac{\varepsilon^{-1}}{3}A_1
A_4-\mu_{3}, \;\;\; k_{3}:=-\varepsilon^{-1}A_{4}+q,
\end{equation}
where $m$, $\mu_3$ and $q$ are some suitable renormalization
parameters.

The first term in Eq. (\ref{bo}) depends on the function $g_{0}$
which is chosen as:
\begin{equation}
g_{0}=-\left(\frac{2 }{3}\varepsilon ^{-2} A_4^2-\frac{5}{6}
\varepsilon^{-1}  A_1 A_4 +\rho_{1} \right) \dot{\textbf{v}}^2
-\left( \frac{5}{6} \varepsilon^{-1} A_4^2+\rho_{2}\right)
\dot{\textbf{v}} \ddot{\textbf{v}},
\end{equation}
where now $\rho_1$ and $\rho_2$ are the renormalization parameters.

The last term in Eq. (\ref{bo}) is not consistent with Eq.
(\ref{sol}). However, it can be proved that this term satisfies the
condition $\dot{\textbf{B}}\widehat{\textbf{u}}_{i}=0$, but this is
possible only if $L_{\nu}\dot{v}^{\nu}=0$, where
$L_{\nu}=f_{\mu\nu}\widehat{u}_{4}^{\mu}$ is null, (see the Eq.
(\ref{UL}) in appendix C). Then it is allowed to add it in
(\ref{bo}). Thus the equations of motion can be expressed in the
form
\begin{align}
\nonumber&\left[ \rho_{1}\dot{\textbf{v}}^{2}+\left( \rho_{2}-\frac{3}{4} \alpha_1 [\alpha_{1},\alpha_{5}] \right) \dot{\textbf{v}}\ddot{\textbf{v}} -\frac{3}{8} [\alpha_{1},\alpha_{5}]^2 \dot{\textbf{v}}^4 -\frac{1}{6} [\alpha_{1},\alpha_{5}]^2 \ddot{\textbf{v}}^2 -\frac{1}{6} [\alpha_{1},\alpha_{5}]^2 \dot{\textbf{v}}\dddot{\textbf{v}}\right] v_{\mu}\\
&+m\dot{v}_{\mu}+(\mu_{1}\dot{\textbf{v}}^{2}
+\mu_{2}\dot{\textbf{v}}\ddot{\textbf{v}})\dot{v}_{\mu}+\mu_{3}\ddot{v}_{\mu}
=(-\alpha_{1}+q)f_{\mu \nu } v^{\nu }-[\alpha_{1},\alpha_{5}] f_{\mu
\nu } \dot{v}^{\nu}.
\end{align}
Of course, in order to find the equations of motion its is necessary
to take the traces in both sides of the previous equation. This
equation includes non-abelian corrections to the charge and a new
coupling of the field strength with the velocity of the particle,
where the tensor $f_{\mu\nu}$ is expressed through the radiated
field Eq. (\ref{Frad}) that now has the form
\begin{align}
\nonumber
F_{\mu \nu }^{(rad)}=&\frac{7}{12}[\alpha_{1},\alpha_{5}]\dot{v}^{2}\left( v_{\mu } \dot{v}_{\text{$\nu $}}-v_{\nu }\dot{v}_{\text{$\mu $}}\right)  -\frac{4}{3} \alpha_1  \left( \ddot{v}_{\text{$\mu $}} v_{\nu }-\ddot{v}_{\text{$\nu $}}v_{\mu} \right)\\
+&\frac{1}{4} [\alpha_{1},\alpha_{5}] \left( v_{\mu }
\dddot{v}_{\text{$\nu $}}+\frac{2}{3}\dot{v}_{\text{$\mu $}}
\ddot{v}_{\text{$\nu $}}-\frac{2}{3}\ddot{v}_{\text{$\mu $}}
\dot{v}_{\text{$\nu $}}-\dddot{v}_{\text{$\mu $}} v_{\nu } \right).
\end{align}
We have defined the constants $\mu_{1}:=-{3 \over 8}
\alpha_{1}[\alpha_{1},\alpha_{5}]$, $\mu_{2}:={2 \over
3}[\alpha_{1},\alpha_{5}]^{2}$ in order to simplify the expression.

\subsection{The third non-abelian radiation damping
effect}\label{SUB3}

In this case we choose the commutation relations to be
$[\alpha_{1},\alpha_{5}]= [\alpha_{2},\alpha_{4}]=
[\alpha_{2},\alpha_{5}]=0$. Then the equations of motion are
\begin{align}
\nonumber
\dot{B}_{\mu}=&\left( -\frac{1}{3} A_2^2 \dot{\textbf{v}}^4 -\frac{1}{2} A_3^2 \dot{\textbf{v}}^2 +\frac{2 A_1 A_2 \dot{\textbf{v}}^2 }{3 \varepsilon }-\frac{1}{2} A_1 A_3 \dot{\textbf{v}}^2 +\frac{A_2 A_3 \dot{\textbf{v}}^2 }{3 \varepsilon }-\frac{1}{3} A_2^2 \ddot{\textbf{v}}^2 +\frac{3}{4} A_1 A_2 \dot{\textbf{v}} \ddot{\textbf{v}}\right.\\ +&\left. \frac{17}{12} A_2 A_3 \dot{\textbf{v}} \ddot{\textbf{v}}-\frac{1}{3} A_2^2 \dot{\textbf{v}} \dddot{\textbf{v}}-\frac{2 A_2^2 \dot{\textbf{v}}^2 }{3 \varepsilon ^2}-\frac{A_2^2 \dot{\textbf{v}} \ddot{\textbf{v}} }{2 \varepsilon }\right)v_{\mu} \nonumber\\
+&\left( \frac{A_{3}^{2}}{2 \varepsilon}+\frac{5}{8} A_1 A_2 \dot{\textbf{v}}^2+\frac{3}{8} A_2 A_3 \dot{\textbf{v}}^{2}+\frac{3 A_1 A_2}{2 \varepsilon^{2}}+\frac{1}{6} A_2^2 \dot{\textbf{v}} \ddot{\textbf{v}} +\frac{1}{3} A_{2}^{2} \textbf{v}\dddot{\textbf{v}}-\frac{A_1^2 }{2 \varepsilon }-\frac{A_{2}^{2}\dot{\textbf{v}}^{2}}{2\varepsilon}\right. \nonumber\\
-&\left. \frac{A_2 A_3}{2 \varepsilon^{2}}-\frac{A_2^2}{\varepsilon ^3}\right) \dot{v}_{\text{$\mu $}} \nonumber\\
+&\left(-\frac{2 A_2 A_3}{3 \varepsilon }-\frac{1}{3} A_2^2
\dot{\textbf{v}}^2+\frac{2 A_1 A_2}{3 \varepsilon }-\frac{2 A_2^2}{3
\varepsilon ^2}\right)\ddot{v}_{\text{$\mu $}} +\left(
\frac{A_2}{\varepsilon }+A_3 - A_1\right) f_{\mu \nu } v^{\nu}.
\end{align}
The  vector $\dot{B}_{\mu}$ can be expressed in the form:
\begin{equation}\label{bo2}
\dot{B}_{\mu}=g_{0}v_{\mu}+g_{1}\dot{v}_{\mu}+k_{2}\ddot{v}_{\mu}+k_{3}f_{\mu
\nu } v^{\nu},
\end{equation}
for suitable parameters coefficients $g_0$, $g_1$, $k_2$ and $k_3$
that eliminate the divergences as $\varepsilon$ goes to zero.

The renormalization of the equations of motion is carried out
through the functions $g_{0}$ and $g_{1}$ depending on the
velocities and accelerations, and the constants $k_{2}$ and $k_{3}$
that appear in the solution (\ref{bo2})
\begin{equation}\label{28}
g_{0}:= \frac{2 A_1 A_2 \dot{\textbf{v}}^2 }{3 \varepsilon
}+\frac{A_2 A_3 \dot{\textbf{v}}^2 }{3 \varepsilon }-\frac{2 A_2^2
\dot{\textbf{v}}^2 }{3 \varepsilon ^2}-\frac{A_2^2 \dot{\textbf{v}}
\ddot{\textbf{v}} }{2 \varepsilon
}-\rho_{1}\dot{\textbf{v}}^{2}-\rho_{2}\dot{\textbf{v}}\ddot{\textbf{v}},
\end{equation}
\begin{equation}\label{29}
g_{1}:=\frac{A_{3}^{2}}{2 \varepsilon}+\frac{3 A_1 A_2}{2
\varepsilon^{2}}-\frac{A_1^2 }{2 \varepsilon
}-\frac{A_{2}^{2}}{2\varepsilon}\dot{\textbf{v}}^{2}-\frac{A_2
A_3}{2 \varepsilon^{2}}-\frac{A_2^2}{\varepsilon
^3}-\mu_{1}-\mu_{2}\dot{\textbf{v}}^{2},
\end{equation}
\begin{equation}\label{30}
k_{2}:=-\frac{2 A_2 A_3}{3 \varepsilon }+\frac{2 A_1 A_2}{3
\varepsilon }-\frac{2 A_2^2}{3 \varepsilon ^2}-\mu_{3}, \;\;\
k_{3}:=\left( \frac{A_2}{\varepsilon }+ q\right),
\end{equation}
where $\rho_1$, $\rho_2$, $\mu_1$, $\mu_2$, $\mu_3$ and $q$ are the
renormalization parameters.

The condition $\dot{\textbf{B}}\widehat{\textbf{u}}_{4}=0$ and Eq.
(\ref{sol}) to be fulfilled,  the term $k_{3}f_{\mu \nu}v^{\nu}$ in
(\ref{bo2}) must satisfy condition $f_{\mu \nu
}v^{\nu}\widehat{u}_{4}^{\mu}=0$. In order to fulfill the mentioned
conditions we should have $A_{4}, A_{5}=0$ (see Eq. (\ref{UL})). If
we use the Eqs. (\ref{bo2})-(\ref{30}), then the no-abelian
equations of motion are
\begin{align}
\nonumber&\left( -\frac{4}{3} \alpha_2^2 \dot{\textbf{v}}^4 -\frac{1}{2} \left( -2\dot{\alpha}_{2}+[\alpha_{1},\alpha_{4}]+[\alpha_{2},\alpha_{3}]\right) ^2 \dot{\textbf{v}}^2 +\frac{1}{2} \alpha_1 \left( -2\dot{\alpha}_{2}+[\alpha_{1},\alpha_{4}]+[\alpha_{2},\alpha_{3}]\right) \dot{\textbf{v}}^2\right. \nonumber\\
+&\left. \rho_{1}\dot{\textbf{v}}^{2}-\frac{4}{3} \alpha_2^2 \ddot{\textbf{v}}^2 +\frac{3}{2} \alpha_1 \alpha_2 \dot{\textbf{v}} \ddot{\textbf{v}}-\frac{17}{6} \alpha_2 \left( -2\dot{\alpha}_{2}+[\alpha_{1},\alpha_{4}]+[\alpha_{2},\alpha_{3}]\right)\dot{\textbf{v}}\ddot{\textbf{v}}-\frac{4}{3} \alpha_2^2 \dot{\textbf{v}}\dddot{\textbf{v}}+\rho_{2}\dot{\textbf{v}}\ddot{\textbf{v}}\right)v_{\mu} \nonumber\\
+&\left( \frac{5}{4} \alpha_1 \alpha_2 \dot{\textbf{v}}^2-\frac{3}{4} \alpha_2 \left( -2\dot{\alpha}_{2}+[\alpha_{1},\alpha_{4}]+[\alpha_{2},\alpha_{3}]\right)\dot{\textbf{v}}^{2}-\frac{10}{3} \alpha_2^2 \dot{\textbf{v}}\ddot{\textbf{v}}+\mu_{1}+\mu_{2}\dot{\textbf{v}}^{2}\right) \dot{v}_{\mu} \nonumber\\
&\left(-\frac{4}{3} \alpha_2^2
\dot{\textbf{v}}^2+\mu_{3}\right)\ddot{v}_{\mu} =\left( q-\alpha_1
+2\dot{\alpha}_{2}-[\alpha_{1},\alpha_{4}]-[\alpha_{2},\alpha_{3}]\right)
f_{\mu \nu } v^{\nu}.
\end{align}

In this expression the interaction term $f_{\mu \nu } v^{\nu}$
contains non-abelian corrections to the charge of the Yang-Mills
particle. Analogously to the previous cases we can determine the
radiated field $F_{\mu \nu }^{(rad)}$ through Eq. (\ref{Frad}). In
this case it becomes
\begin{align}
F_{\mu \nu }^{(rad)}=& \frac{11}{6} \alpha_2 \dot{\textbf{v}}^{2}\left( v_{\mu}\dot{v}_{\nu}-\dot{v}_{\mu}v_{\nu}\right) +\frac{2}{3} \left(-2\alpha_{1} -2\dot{\alpha}_{2}+[\alpha_{1},\alpha_{4}]+[\alpha_{2},\alpha_{3}]\right)  \left( \ddot{v}_{\mu}v_{\nu}-\ddot{v}_{\nu}v_{\mu}\right) \nonumber\\
-&\frac{3}{2} \alpha_2\left( v_{\mu }\dddot{v}_{\text{$\nu
$}}+\frac{2}{3}\dot{v}_{\text{$\mu $}} \ddot{v}_{\text{$\nu
$}}-\frac{2}{3}\ddot{v}_{\text{$\mu $}} \dot{v}_{\text{$\nu
$}}-\dddot{v}_{\text{$\mu $}} v_{\nu } \right).
\end{align}

\section{Final Remarks}

At the present time the classical problem of the dynamics of a
radiating particle is still a subject of intense work. In the
specific case of a color charged particle interacting with a
Yang-Mills field there is a huge interest in the problem of
non-abelian hydrodynamics and the study of the quark-gluon plasma.

In the present article we continue exploring the equations of motion
of a non-abelian charge in an external Yang-Mills field, where the
dynamics is described by Wong's equations \cite{Wong:1970fu}. There
has been much work in this subject as we mentioned in the
introduction section. However a self-consistent analysis of the
classical theory, as was given by Dirac \cite{Dirac:1938nz} for the
abelian case, has not been worked out in the literature for the
Yang-Mills case. This analysis has been carried out in the present
article.

The strategy we followed here was to adopt the ansatz (\ref{k10})
for the retarded L\'ienard-Wiechert potentials, given in Ref.
\cite{Sarioglu:2002qb}. This allowed to compute the retarded,
advanced and actual field strengths along the trajectory of the
particle. A summary and the results of this derivation were
collected in appendices A and B. Furthermore, using Dirac's method
of integration over the energy-momentum tensor, in appendix C is
outlined the computation of the equations of motion and its
corrections due the radiation damping. These results were used to
give precise formulas in Section 3 of the corrections due the
self-force to Wong's equations. Moreover it is shown that under
certain assumptions about the Lie algebra structure of the Dirac's
equation of motion \cite{Dirac:1938nz} can be recovered. We also
were able to find three non-trivial examples for the non-abelian
equations of motion with the properly cancelation of divergences.

Finally, as a possible perspective of this work, we point out that a
generalized Lorentz-Dirac equation was obtained in the context of
the AdS/CFT correspondence \cite{Chernicoff:2009re}. This derivation
In the near future we would like to address the problem of finding a
close correspondence between gravity dual in the AdS space and some
of the details that we found in the present article for the particle
Yang-Mills theory.

\vskip 1truecm
\centerline{\bf Acknowledgements}

It is a pleasure to thank G. Ares de Parga and A. Guijosa for very
useful discussions and suggestions. We thank A. Stern for pointing
out some important references and also some omissions and
imprecisions from the previous version. S. A. thanks to CONACyT for
a graduate fellowship. H. G.-C. was supported by a CONACyT project
128761. The work of R. C. was partially supported by COFAA-IPN and
by SIP-IPN grants 20161370, 20171100 and 20171168.


\break

\appendix

\section{Series expansions of fields near the trajectory of the particle}

In the following appendices we present some computations and results
we used in the body of the article. In this appendix we will give
some preliminary results that will be necessary in the subsequent
appendices.

In order to perform the calculation of the field strength $F_{\mu
\nu}$ in Eq. (\ref{efemunuret}) we need to compute first the terms
with different negative powers of $R$. $R^{-1}$ is a function of
$x_\mu$ and $z_\mu$ and therefore it can be expanded in powers of
$\sigma$ as is shown in Eqs. (\ref{k3}) and (\ref{k4}). Neglecting
the higher positive powers in $\sigma$ means that we are closer from
the trajectory of the particle. Then the expansion reads

\begin{align}
\nonumber R^{-1}=&\left[ \dot{z}^{\mu}\left(
x_{\mu}-z_{\mu}\right)\right]^{-1}\\ \nonumber =&\sigma^{-1}\left[
1+ \gamma {\bf f}(\sigma)\right]^{-1}\\
\times & \left\lbrace  1+\left[
-\frac{1}{6}\dot{\textbf{v}}^{2}\sigma^{2}
+\sum_{p=3}^{10}(-1)^{p}\sum_{n=0}^{p}\frac{\textbf{v}_{n}}{n!}\frac{\textbf{v}_{p-n}}{(p+1-n)!}
\sigma^{p}\right] \left[ 1+ \gamma {\bf f}(\sigma)
\right]^{-1}\right\rbrace^{-1}, \label{onevv}
\end{align}
where $\gamma{\bf f}(\sigma) =\sum_{k=1}^{10}(-1)^{k}
\gamma\textbf{v}_{k}\sigma^{k-1}$ and $
\gamma\textbf{v}_{k}=\gamma_{\mu} v^{\mu}_{k} $. The first factor
can in turns be expanded as power series of $\sigma$ and it can be
expressed as
\begin{align}
\nonumber \left[ 1+ \gamma {\bf f}(\sigma\right]^{-1} =&1-\gamma
{\bf f}(\sigma)+[\gamma
{\bf f}(\sigma)]^{2}-\cdots -[\gamma {\bf f}(\sigma)]^{11}\\
\nonumber
=&h_{10} \sigma ^{10}+h_9 \sigma ^9+h_8 \sigma ^8+h_7 \sigma ^7+h_6 \sigma ^6\\
+&h_5 \sigma ^5+h_4 \sigma ^4+h_3 \sigma ^3+h_2 \sigma ^2+h_1 \sigma
+h_0.
\label{expansioninhs}
\end{align}
In the process of expansion we have considered that the highest
possible value of the powers of $\sigma$ relevant in our computation
is up to 11. Moreover we have assumed that $\gamma$ is of the same
order in $\sigma$, thus for instance we have $[\gamma {\bf
f}(\sigma)]^{10}=(\gamma \dot{\textbf{v}})^{10}-5\sigma  (\gamma
\dot{\textbf{v}})^{9}\gamma \ddot{\textbf{v}}$ and $[\gamma
f(\sigma)]^{11}=-(\gamma \dot{\textbf{v}})^{11}$. The first
coefficients $h_i$ of expansion in Eq. (\ref{expansioninhs}) are
given by
\begin{align}
h_0=\text{$\gamma $}\dot{\textbf{v}}^{11}+\text{$\gamma
$}\dot{\textbf{v}}^{10}+\text{$\gamma
$}\dot{\textbf{v}}^9+\text{$\gamma
$}\dot{\textbf{v}}^8+\text{$\gamma
$}\dot{\textbf{v}}^7+\text{$\gamma
$}\dot{\textbf{v}}^6+\text{$\gamma
$}\dot{\textbf{v}}^5+\text{$\gamma
$}\dot{\textbf{v}}^4+\text{$\gamma
$}\dot{\textbf{v}}^3+\text{$\gamma
$}\dot{\textbf{v}}^2+\text{$\gamma $}\dot{\textbf{v}}+1,
\end{align}
\begin{align}
\nonumber h_1=&-5 \text{$\gamma $}\ddot{\textbf{v}} \text{$\gamma
$}\dot{\textbf{v}}^9-\frac{9}{2} \text{$\gamma $}\ddot{\textbf{v}}
\text{$\gamma $}\dot{\textbf{v}}^8-4 \text{$\gamma $}
\ddot{\textbf{v}} \text{$\gamma $}\dot{\textbf{v}}^7-\frac{7}{2}
\text{$\gamma $}\ddot{\textbf{v}} \text{$\gamma
$}\dot{\textbf{v}}^6-3 \text{$\gamma $}\ddot{\textbf{v}}
\text{$\gamma $}
\dot{\textbf{v}}^5-\frac{5}{2} \text{$\gamma $}\ddot{\textbf{v}} \text{$\gamma $}\dot{\textbf{v}}^4\\
-&2 \text{$\gamma $}\ddot{\textbf{v}} \text{$\gamma
$}\dot{\textbf{v}}^3-\frac{3}{2} \text{$\gamma $}\ddot{\textbf{v}}
\text{$\gamma $}\dot{\textbf{v}}^2-\text{$\gamma $}\ddot{\textbf{v}}
\text{$\gamma $}\dot{\textbf{v}}-\frac{\text{$\gamma
$}\ddot{\textbf{v}}}{2},
\end{align}
\begin{align}
\nonumber
h_2&=\frac{9}{6} \text{$\gamma $}\dddot{\textbf{v}} \text{$\gamma $}\dot{\textbf{v}}^8+9 \text{$\gamma $}\ddot{\textbf{v}}^2 \text{$\gamma $}\dot{\textbf{v}}^7+\frac{8}{6} \text{$\gamma $}\dddot{\textbf{v}} \text{$\gamma $}\dot{\textbf{v}}^7+7 \text{$\gamma $}\ddot{\textbf{v}}^2 \text{$\gamma $}\dot{\textbf{v}}^6+\frac{7}{6} \text{$\gamma $}\dddot{\textbf{v}} \text{$\gamma $}\dot{\textbf{v}}^6+\frac{21}{4} \text{$\gamma $}\ddot{\textbf{v}}^2 \text{$\gamma $}\dot{\textbf{v}}^5 \nonumber\\
&+\frac{6}{6} \text{$\gamma $}\dddot{\textbf{v}} \text{$\gamma $}\dot{\textbf{v}}^5+\frac{15}{4} \text{$\gamma $}\ddot{\textbf{v}}^2 \text{$\gamma $}\dot{\textbf{v}}^4+\frac{5}{6} \text{$\gamma $}\dddot{\textbf{v}} \text{$\gamma $}\dot{\textbf{v}}^4+\frac{5}{2} \text{$\gamma $}\ddot{\textbf{v}}^2 \text{$\gamma $}\dot{\textbf{v}}^3+\frac{4}{6} \text{$\gamma $}\dddot{\textbf{v}} \text{$\gamma $}\dot{\textbf{v}}^3+\frac{3}{2} \text{$\gamma $}\ddot{\textbf{v}}^2 \text{$\gamma $}\dot{\textbf{v}}^2 \nonumber\\
&+\frac{3}{6} \text{$\gamma $}\dddot{\textbf{v}} \text{$\gamma
$}\dot{\textbf{v}}^2+\frac{3}{4} \text{$\gamma $}\ddot{\textbf{v}}^2
\text{$\gamma $}\dot{\textbf{v}}+\frac{2 \text{$\gamma
$}\dddot{\textbf{v}} \text{$\gamma
$}\dot{\textbf{v}}}{6}+\frac{\text{$\gamma
$}\ddot{\textbf{v}}^2}{4}+\frac{\text{$\gamma
$}\dddot{\textbf{v}}}{6},
\end{align}
\begin{align}
\nonumber
h_3&=-\frac{1}{24} 8 \text{$\gamma $}\textbf{v}^{(4)} \text{$\gamma $}\dot{\textbf{v}}^7-\frac{28}{6} \text{$\gamma $}\ddot{\textbf{v}} \text{$\gamma $}\dddot{\textbf{v}} \text{$\gamma $}\dot{\textbf{v}}^6-\frac{7}{24} \text{$\gamma $}\textbf{v}^{(4)} \text{$\gamma $}\dot{\textbf{v}}^6-7 \text{$\gamma $}\ddot{\textbf{v}}^3 \text{$\gamma $}\dot{\textbf{v}}^5-\frac{21}{6} \text{$\gamma $}\ddot{\textbf{v}} \text{$\gamma $}\dddot{\textbf{v}} \text{$\gamma $}\dot{\textbf{v}}^5-\frac{6}{24} \text{$\gamma $}\textbf{v}^{(4)} \text{$\gamma $}\dot{\textbf{v}}^5 \nonumber\\
&-\frac{35}{8} \text{$\gamma $}\ddot{\textbf{v}}^3 \text{$\gamma $}\dot{\textbf{v}}^4-\frac{15}{6} \text{$\gamma $}\ddot{\textbf{v}} \text{$\gamma $}\dddot{\textbf{v}} \text{$\gamma $}\dot{\textbf{v}}^4-\frac{5}{24} \text{$\gamma $}\textbf{v}^{(4)} \text{$\gamma $}\dot{\textbf{v}}^4-\frac{5}{2} \text{$\gamma $}\ddot{\textbf{v}}^3 \text{$\gamma $}\dot{\textbf{v}}^3-\frac{10}{6} \text{$\gamma $}\ddot{\textbf{v}} \text{$\gamma $}\dddot{\textbf{v}} \text{$\gamma $}\dot{\textbf{v}}^3-\frac{4}{24} \text{$\gamma $}\textbf{v}^{(4)} \text{$\gamma $}\dot{\textbf{v}}^3 \nonumber\\
&-\frac{5}{4} \text{$\gamma $}\ddot{\textbf{v}}^3 \text{$\gamma $}\dot{\textbf{v}}^2-\frac{6}{6} \text{$\gamma $}\ddot{\textbf{v}} \text{$\gamma $}\dddot{\textbf{v}} \text{$\gamma $}\dot{\textbf{v}}^2-\frac{3}{24} \text{$\gamma $}\textbf{v}^{(4)} \text{$\gamma $}\dot{\textbf{v}}^2-\frac{1}{2} \text{$\gamma $}\ddot{\textbf{v}}^3 \text{$\gamma $}\dot{\textbf{v}}-\frac{3}{6} \text{$\gamma $}\ddot{\textbf{v}} \text{$\gamma $}\dddot{\textbf{v}} \text{$\gamma $}\dot{\textbf{v}}-\frac{2 \text{$\gamma $}\textbf{v}^{(4)} \text{$\gamma $}\dot{\textbf{v}}}{24} \nonumber\\
&-\frac{\text{$\gamma $}\ddot{\textbf{v}} \text{$\gamma
$}\dddot{\textbf{v}}}{6}-\frac{\text{$\gamma
$}\ddot{\textbf{v}}^3}{8}-\frac{\text{$\gamma
$}\textbf{v}^{(4)}}{24},
\end{align}
\begin{align}
\nonumber
h_4&=\frac{7}{120} \text{$\gamma $}\textbf{v}^{(5)} \text{$\gamma $}\dot{\textbf{v}}^6+\frac{21}{36} \text{$\gamma $}\dddot{\textbf{v}}^2 \text{$\gamma $}\dot{\textbf{v}}^5+\frac{21}{24} \text{$\gamma $}\ddot{\textbf{v}} \text{$\gamma $}\textbf{v}^{(4)} \text{$\gamma $}\dot{\textbf{v}}^5+\frac{6}{120} \text{$\gamma $}\textbf{v}^{(5)} \text{$\gamma $}\dot{\textbf{v}}^5+\frac{15}{36} \text{$\gamma $}\dddot{\textbf{v}}^2 \gamma \dot{\bf v}^4 \nonumber\\
&+\frac{105}{24} \text{$\gamma $}\ddot{\textbf{v}}^2 \text{$\gamma $}\dddot{\textbf{v}} \text{$\gamma $}\dot{\textbf{v}}^4+\frac{15}{24} \text{$\gamma $}\ddot{\textbf{v}} \text{$\gamma $}\textbf{v}^{(4)} \text{$\gamma $}\dot{\textbf{v}}^4+\frac{5}{120} \text{$\gamma $}\textbf{v} ^{(5)}\text{$\gamma $}\dot{\textbf{v}}^4+\frac{35}{16} \text{$\gamma $}\ddot{\textbf{v}}^4 \text{$\gamma $}\dot{\textbf{v}}^3+\frac{10}{36} \text{$\gamma $}\dddot{\textbf{v}}^2 \text{$\gamma $}\dot{\textbf{v}}^3 \nonumber\\
&+\frac{15}{6} \text{$\gamma $}\ddot{\textbf{v}}^2 \text{$\gamma $V}_3 \text{$\gamma $}\dot{\textbf{v}}^3+\frac{10}{24} \text{$\gamma $}\ddot{\textbf{v}} \text{$\gamma $}\textbf{v}^{(4)} \text{$\gamma $}\dot{\textbf{v}}^3+\frac{4}{120} \text{$\gamma $}\textbf{v}^{(5)} \text{$\gamma $}\dot{\textbf{v}}^3+\frac{15}{16} \text{$\gamma $}\ddot{\textbf{v}}^4 \text{$\gamma $}\dot{\textbf{v}}^2+\frac{6}{36} \text{$\gamma $}\dddot{\textbf{v}}^2 \text{$\gamma $}\dot{\textbf{v}}^2 \nonumber\\
&+\frac{15}{12} \text{$\gamma $}\ddot{\textbf{v}}^2 \text{$\gamma $}\dddot{\textbf{v}} \text{$\gamma $}\dot{\textbf{v}}^2+\frac{6}{24} \text{$\gamma $}\ddot{\textbf{v}} \text{$\gamma $}\textbf{v}^{(4)} \text{$\gamma $}\dot{\textbf{v}}^2+\frac{3}{120} \text{$\gamma $}\textbf{v} ^{(5)}\text{$\gamma $}\dot{\textbf{v}}^2+\frac{5}{16} \text{$\gamma $}\ddot{\textbf{v}}^4 \text{$\gamma $}\dot{\textbf{v}}+\frac{3}{36} \text{$\gamma $}\dddot{\textbf{v}}^2 \text{$\gamma $}\dot{\textbf{v}} \nonumber\\
&+\frac{3}{6} \text{$\gamma $}\ddot{\textbf{v}}^2 \text{$\gamma $}\dddot{\textbf{v}} \text{$\gamma $}\dot{\textbf{v}}+\frac{3}{24} \text{$\gamma $}\ddot{\textbf{v}} \text{$\gamma $}\textbf{v}^{(4)} \text{$\gamma $}\dot{\textbf{v}}+\frac{2 \text{$\gamma $}\textbf{v}^{(5)} \text{$\gamma $}\dot{\textbf{v}}}{120}+\frac{\text{$\gamma $}\ddot{\textbf{v}}^4}{16} \nonumber\\
&+\frac{\text{$\gamma $}\dddot{\textbf{v}}^2}{36}+\frac{3}{24}
\text{$\gamma $}\ddot{\textbf{v}}^2 \text{$\gamma
$}\dddot{\textbf{v}}+\frac{\text{$\gamma $}\ddot{\textbf{v}}
\text{$\gamma $}\textbf{v}^{(4)}}{24}+\frac{\text{$\gamma
$}\textbf{v}^{(5)}}{120},
\end{align}
where we adopted the notation $(\gamma \bf{v})^\alpha \equiv \gamma
\bf{v}^\alpha = (\gamma_\mu v^\mu)^\alpha$. Furthermore we have
written only the first coefficients of the expansion since that up
on computation one can see that the higher order terms will not
contribute to the equations of motion.

The first term of in the square brackets from the second factor of
the right-hand side of Eq. (\ref{onevv}) also can be expanded as
powers of $\sigma$ with coefficients given by certain functions
$p_{i}$
\begin{align}
\nonumber \left[
-\frac{1}{6}\dot{\textbf{v}}^{2}\sigma^{3}+\sum_{p=3}^{10}(-1)^{p}\sum_{n=0}^{p}
\frac{\textbf{v}_{n}}{n!}\frac{\textbf{v}_{p-n}}{(p+1-n)!}\sigma^{p+1}\right]=&p_3
\sigma ^3+p_4 \sigma ^4+p_5 \sigma ^5\\ \nonumber
+&p_6 \sigma ^6+p_7 \sigma ^7+p_8 \sigma ^8\\
+&p_9 \sigma ^9+p_{10} \sigma ^{10}+p_{11} \sigma ^{11},
\end{align}
where the important coefficients for our further computations are
\begin{equation}
p_3= -{1\over 6}\textbf{v}^2, \ \ \
p_{4}=\frac{5}{24}\dot{\textbf{v}}\ddot{\textbf{v}}, \ \ \
p_{5}=-\frac{1}{5!}\left(
8\ddot{\textbf{v}}+9\dot{\textbf{v}}\ddot{\textbf{v}}\right), \ \ \
p_{6}=-\frac{1}{6!}\left(
35\ddot{\textbf{v}}\dddot{\textbf{v}}+\dot{\textbf{v}}\textbf{v}^{(4)}\right).
\end{equation}

Moreover the expression $\dot{z}_{\mu}(x_{\nu}-z_{\nu}(s))$ is also
expanded in powers of $\sigma$ as follows
\begin{align}
\nonumber \dot{z}_{\mu}(x_{\nu}-z_{\nu}(s))=&\gamma _{\nu } v_{\mu
}+\sigma  \left(v_{\mu } v_{\nu }-\gamma _{\nu } \dot{v}_{\text{$\mu
$}}\right)+\sigma ^2 \left(k_{\text{$\mu \nu $2}} +\frac{\gamma
_{\nu } \ddot{v}_{\text{$\mu $}}}{2}\right)+\sigma ^3
\left(k_{\text{$\mu \nu $3}} -\frac{\gamma _{\nu }
\dddot{v}_{\text{$\mu $}}}{3!}\right)\\ \nonumber +&\sigma ^4
\left(\frac{\gamma _{\nu } v^{(4)}_{\text{$\mu
$}}}{4!}+k_{\text{$\mu \nu $4}}\right) +\sigma ^5
\left(k_{\text{$\mu \nu $5}}-\frac{\gamma _{\nu }
v^{(5)}_{\text{$\mu $}}}{5!}\right) +\sigma ^6 \left(k_{\text{$\mu
\nu $6}}+\frac{\gamma _{\nu } v^{(6)}_{\text{$\mu $}}}{6!}\right)\\
\nonumber
+&\sigma ^7 \left(k_{\text{$\mu \nu $7}}-\frac{\gamma _{\nu }
v^{(7)}_{\text{$\mu $}}}{7!}\right)+\sigma ^8 \left(k_{\text{$\mu \nu $8}}+\frac{\gamma _{\nu }
v^{(8)}_{\text{$\mu $}}}{8!}\right)+\sigma ^9 \left(k_{\text{$\mu \nu $9}}-\frac{\gamma _{\nu }
v^{(9)}_{\text{$\mu $}}}{9!}\right)\\
+&\sigma^{10} \left(\frac{\gamma _{\nu } v^{(10)}_{\text{$\mu
$}}}{10!}+k_{\text{$\mu \nu $10}}\right)+\sigma ^{11}
\left(k_{\text{$\mu \nu $11}}-\frac{\gamma _{\nu }
v^{(11)}_{\text{$\mu $}}}{11!}\right),
\end{align}
where $k_{\mu\nu i}$ are some functions depending on the velocities
and their derivatives. Actually we need to compute the
anti-symmetric version of the previous equation because this is the
case that arises in the computation of $F_{\mu \nu}^{(rad)}$, that
is: $\dot{z}_{\mu}(x_{\nu}-z_{\nu}(s)) -
\dot{z}_{\nu}(x_{\mu}-z_{\mu}(s))$. Thus, the relevant functions
$k_{\mu \nu i}$ are antisymmetric tensors and they are given by
\begin{equation}
k_{\mu\nu 2} =\frac{v_{\mu } \dot{v}_{\text{$\nu
$}}}{2}-\frac{\dot{v}_{\text{$\mu $}} v_{\nu }}{2}, \ \ \ k_{\mu\nu
3}=\frac{\ddot{v}_{\text{$\mu $}} v_{\nu }}{3}-\frac{v_{\mu }
\ddot{v}_{\text{$\nu $}}}{3},
\end{equation}
\begin{equation}
k_{\mu\nu 4} =\frac{v_{\mu } \dddot{v}_{\text{$\nu
$}}}{8}+\frac{\dot{v}_{\text{$\mu $}} \ddot{v}_{\text{$\nu
$}}}{12}-\frac{\ddot{v}_{\text{$\mu $}} \dot{v}_\text{$\nu
$}}{12}-\frac{1}{8} \dddot{v}_{\text{$\mu $}} v_{\nu},
\end{equation}
\begin{equation}
k_{\mu\nu 5} =-\frac{v_{\mu } v^{(4)}_{\text{$\nu
$}}}{30}-\frac{\dot{v}_{\text{$\mu $}} \dddot{v}_{\text{$\nu
$}}}{24}+\frac{\dddot{v}_{\text{$\mu $}} \dot{v}_{\text{$\nu
$}}}{24}+\frac{v^{(4)}_{\text{$\mu $}} v_{\nu }}{30}.
\end{equation}
Now we give some definitions that will be  useful later in order to
simplify the relevant equations in the following appendices are
\begin{align}
\nonumber
g_1=&h_2-h_0^2 p_3, \ \ \ g_{2}=-3 h_0^4 p_3+3 h_2 h_0^2+3 h_1^2 h_0, \nonumber\\
g_{3}=&3 g_1 h_0^2-2 h_0^4 p_3+2 h_2 h_0^2+5 h_1^2 h_0, \ \ \ g_{4}=g_1 h_0^2+2 h_0^4 p_3-2 h_2 h_0^2-h_1^2 h_0, \nonumber\\
g_{5}=&2 g_1 h_1 h_0+2 h_1 h_0^3 p_3-2 h_1 h_2 h_0, \ \  g_{6}=-4 h_0^5 p_3+4 h_2 h_0^3+6 h_1^2 h_0^2, \nonumber\\
g_{7}=&-4 h_0^5 p_3+4 h_2 h_0^3+6 h_1^2 h_0^2, \ \ g_{8}=-4 h_0^5 p_4-20 h_1 h_0^4 p_3+4 h_3 h_0^3+12 h_1 h_2 h_0^2+4 h_1^3 h_0, \nonumber\\
g_{9}=&-5 h_0^6 p_4-30 h_1 h_0^5 p_3+5 h_3 h_0^4+20 h_1 h_2 h_0^3+10 h_1^3 h_0^2, \ \ \ g_{10}=4 h_0^5 p_3-4 h_2 h_0^3-6 h_1^2 h_0^2, \nonumber\\
g_{11}=&-5 h_0^6 p_3+5 h_2 h_0^4+10 h_1^2 h_0^3, \ \ g_{12}=-5 h_0^6 p_3+5 h_2 h_0^4+10 h_1^2 h_0^3, \nonumber\\
g_{13}=&-5 h_0^6 p_4-30 h_1 h_0^5 p_3+5 h_3 h_0^4+20 h_1 h_2 h_0^3+10 h_1^3 h_0^2, \ \ \ g_{14}=5 h_0^6 p_3-5 h_2 h_0^4-10 h_1^2 h_0^3, \nonumber\\
g_{15}=&15 h_0^7 p_3^2-5 h_0^6 p_5-30 h_2 h_0^5 p_3-30 h_1 h_0^5 p_4-75 h_1^2 h_0^4 p_3+5 h_4 h_0^4, \nonumber\\
+&10 h_2^2 h_0^3+20 h_1 h_3 h_0^3+30 h_1^2 h_2 h_0^2+5 h_1^4 h_0, \nonumber\\
g_{16}=&5 h_0^6 p_4+30 h_1 h_0^5 p_3-5 h_3 h_0^4-20 h_1 h_2 h_0^3-10 h_1^3 h_0^2, \ \ \ g_{17}=-5 h_0^6 p_3+5 h_2 h_0^4+10 h_1^2 h_0^3, \nonumber \\
g_{18}=&-\frac{1}{2} 5 h_0^6 p_3+\frac{5}{2} h_2 h_0^4+5 h_1^2
h_0^3.
\end{align}


\section{Computation of $F_{\mu \nu}^{(rad)}$, $F_{\mu \nu}^{(act)}$
and $f_{\mu \nu}$}

In this appendix we will give some details of the computation of
$F_{\mu \nu }^{(ret)}$ and $F_{\mu \nu }^{(adv)}$, through which the
radiated field $F_{\mu \nu }^{(rad)}$ is also obtained. To carry out
the Taylor expansion we use Eqs. (\ref{k3}) and (\ref{k4}) and the
first term in the $F_{\mu \nu }^{(ret)}$ from Eq.
(\ref{efemunuret}), then it is written as
\begin{align}
\nonumber
-&R^{-1}  \dfrac{d}{ds}\bigg[ R^{-1}  \lbrace\dot{z}_{\mu}\left( x_{\nu}-z_{\nu}\right)-\dot{z}_{\nu}\left( x_{\mu}-z_{\mu}\right)\rbrace \bigg] \nonumber\\
=&g_1 h_1 \gamma _{\nu } v_{\mu }-2 g_1 h_0 \gamma _{\nu } \dot{v}_{\text{$\mu $}}+2 g_1 h_0 v_{\mu } v_{\nu }+3 h_1 h_0 k_{\text{$\mu \nu $2}} \nonumber\\
+&2 h_0^2 k_{\text{$\mu \nu $3}}+\frac{h_0^2 k_{\text{$\mu \nu $2}}-h_1 h_0 \gamma _{\nu } \dot{v}_{\text{$\mu $}}+\frac{1}{2} h_0^2 \gamma _{\nu } \ddot{v}_{\text{$\mu $}}
+h_1 h_0 v_{\mu } v_{\nu }}{\sigma } \nonumber\\
+&h_0^3 \left(-p_4\right) \gamma _{\nu } v_{\mu }-2 h_1 h_0^2 p_3 \gamma _{\nu } v_{\mu }-\frac{2 h_0^2 \gamma _{\nu } \dddot{v}_{\text{$\mu $}}}{6}
-\frac{h_0^2 \gamma _{\nu } v_{\mu }}{\sigma ^3} \nonumber\\
-&\frac{h_1 h_0 \gamma _{\nu } v_{\mu }}{\sigma ^2}+h_3 h_0 \gamma
_{\nu } v_{\mu }-h_1^2 \gamma _{\nu } \dot{v}_{\text{$\mu
$}}+\frac{3}{2} h_1 h_0 \gamma _{\nu } \ddot{v}_{\text{$\mu $}}
+h_1^2 v_{\mu } v_{\nu},
\end{align}
where the functions $k_{\mu \nu i}$ and $h_{i}$ were given
explicitly in appendix A. Similarly we can calculate the Taylor
expansion for the following terms in the retarded field
$F^{(ret)}_{\mu \nu}$. It is given by
\begin{align}
\nonumber
-&R^{-2}  \dfrac{d}{ds}\bigg[ R^{-1}  \lbrace\dot{z}_{\mu}\left( x_{\nu}-z_{\nu}\right)-\dot{z}_{\nu}\left( x_{\mu}-z_{\mu}\right)\rbrace \bigg] \nonumber\\
=&3 g_1 h_0^2 k_{\text{$\mu \nu $2}}+5 h_0 h_1^2 k_{\text{$\mu \nu $2}}+2 h_0^2 h_2 k_{\text{$\mu \nu $2}}+7 h_0^2 h_1 k_{\text{$\mu \nu $3}}+3 h_0^3 k_{\text{$\mu \nu $4}}-2 h_0^4 p_3 k_{\text{$\mu \nu $2}}+4 g_1 h_0 h_1 v_{\mu } v_{\nu } \nonumber\\
-&3 h_0^4 p_4 v_{\mu } v_{\nu }-8 h_1 h_0^3 p_3 v_{\mu } v_{\nu }-\frac{h_0^3 \gamma _{\nu } v_{\mu }}{\sigma ^4}+3 h_3 h_0^2 v_{\mu } v_{\nu }+2 h_1 h_2 h_0 v_{\mu } v_{\nu }+h_1^3 v_{\mu } v_{\nu } \nonumber\\
-&\frac{2 h_0^2 h_1 \gamma _{\nu } v_{\mu }}{\sigma ^3}+g_1 h_1^2 \gamma _{\nu } v_{\mu }+2 g_1 h_0 h_2 \gamma _{\nu } v_{\mu }-h_0 h_2^2 \gamma _{\nu } v_{\mu }+2 h_0 h_1 h_3 \gamma _{\nu } v_{\mu }+h_0^2 h_4 \gamma _{\nu } v_{\mu } \nonumber\\
-&2 g_1 h_0^3 p_3 \gamma _{\nu } v_{\mu }-5 h_0^2 h_1^2 p_3 \gamma _{\nu } v_{\mu }-4 h_0^3 h_1 p_4 \gamma _{\nu } v_{\mu }-h_0^4 p_5 \gamma _{\nu } v_{\mu }-4 g_1 h_0 h_1 \gamma _{\nu } \dot{v}_{\text{$\mu $}} \nonumber\\
-&3 h_0^4 p_4 \gamma _{\nu } \dot{v}_{\text{$\mu $}}+8 h_1 h_0^3 p_3 \gamma _{\nu } \dot{v}_{\text{$\mu $}}-3 h_3 h_0^2 \gamma _{\nu } \dot{v}_{\text{$\mu $}}-2 h_1 h_2 h_0 \gamma _{\nu } \dot{v}_{\text{$\mu $}}+h_1^3 \gamma _{\nu } \dot{v}_{\text{$\mu $}} \nonumber\\
+&\frac{3}{2} g_1 h_0^2 \gamma _{\nu } \ddot{v}_{\text{$\mu $}}+\frac{5}{2} h_0 h_1^2 \gamma _{\nu } \ddot{v}_{\text{$\mu $}}+h_0^2 h_2 \gamma _{\nu } \ddot{v}_{\text{$\mu $}}-h_0^4 p_3 \gamma _{\nu } \ddot{v}_{\text{$\mu $}}-\frac{7 h_0^2 h_1 \gamma _{\nu } \dddot{v}_{\text{$\mu $}}}{6}+\frac{3 h_0^3 \gamma _{\nu } v^{(4)}_{\text{$\mu $}}}{24} \nonumber\\
+&\frac{g_1 h_0^2 \gamma _{\nu } v_{\mu }+h_0^3 k_{\text{$\mu \nu $2}}+2 h_0^4 p_3 \gamma _{\nu } v_{\mu }-2 h_2 h_0^2 \gamma _{\nu } v_{\mu }-h_1^2 h_0 \gamma _{\nu } v_{\mu }-h_1 h_0^2 \gamma _{\nu } \dot{v}_{\text{$\mu $}}+\frac{1}{2} h_0^3 \gamma _{\nu } \ddot{v}_{\text{$\mu $}}+h_1 h_0^2 v_{\mu } v_{\nu }}{\sigma ^2} \nonumber\\
+&\frac{1}{\sigma}\left[ 2 g_1 h_1 h_0 \gamma _{\nu } v_{\mu }+2 g_1 h_0^2 v_{\mu } v_{\nu }+4 h_1 h_0^2 k_{\text{$\mu \nu $2}}+2 h_0^3 k_{\text{$\mu \nu $3}}-2 h_1 h_2 h_0 \gamma _{\nu } v_{\mu }+2 h_1^2 h_0 v_{\mu } v_{\nu }\right. \nonumber\\
-&\left. 2 g_1 h_0^2 \gamma _{\nu } \dot{v}_{\text{$\mu $}}+2 h_1
h_0^3 p_3 \gamma _{\nu } v_{\mu }-\frac{2 h_0^3 \gamma _{\nu }
\dddot{v}_{\text{$\mu $}}}{6}-2 h_1^2 h_0 \gamma _{\nu }
\dot{v}_{\text{$\mu $}}+2 h_1 h_0^2 \gamma _{\nu }
\ddot{v}_{\text{$\mu $}}\right],
\end{align}
where $h_{i}$, $p_i$, $k_{\mu \nu i}$ and $g_i$ are given in
appendix A.

Further terms of higher negative powers of $R$ are given by
\begin{align}
\nonumber
&R^{-3}  \bigg[\dot{z}_{\mu}\left( x_{\nu}-z_{\nu}\right)-\dot{z}_{\nu}\left( x_{\mu}-z_{\mu}\right) \bigg] \nonumber\\
=& 3 h_1 h_0^2 k_{\text{$\mu \nu $2}}+h_0^3 k_{\text{$\mu \nu $3}} \nonumber \\
+&\left. \frac{1}{\sigma }\bigg[  h_0^3 k_{\text{$\mu \nu $2}}-3 h_0^4 p_3 \gamma _{\nu } v_{\mu }+3 h_2 h_0^2 \gamma _{\nu } v_{\mu }+3 h_1^2 h_0 \gamma _{\nu } v_{\mu }-3 h_1 h_0^2 \gamma _{\nu } \dot{v}_{\text{$\mu $}} \right. \nonumber\\
+&\left. \frac{1}{2} h_0^3 \gamma _{\nu } \ddot{v}_{\text{$\mu $}}+3 h_1 h_0^2 v_{\mu } v_{\nu }\bigg] -3 h_0^4 p_4 \gamma _{\nu } v_{\mu }-12 h_1 h_0^3 p_3 \gamma _{\nu } v_{\mu } \right. \nonumber\\
+&3 h_0^4 p_3 \gamma _{\nu } \dot{v}_{\text{$\mu $}}-3 h_0^4 p_3 v_{\mu } v_{\nu }-\frac{h_0^3 \gamma _{\nu } \dddot{v}_{\text{$\mu $}}}{6} \nonumber\\
+&\frac{3 h_1 h_0^2 \gamma _{\nu } v_{\mu }-h_0^3 \gamma _{\nu } \dot{v}_{\text{$\mu $}}+h_0^3 v_{\mu } v_{\nu }}{\sigma ^2}+\frac{h_0^3 \gamma _{\nu } v_{\mu }}{\sigma ^3} \nonumber\\
+&3 h_3 h_0^2 \gamma _{\nu } v_{\mu }+6 h_1 h_2 h_0 \gamma _{\nu } v_{\mu }+h_1^3 \gamma _{\nu } v_{\mu }-3 h_2 h_0^2 \gamma _{\nu } \dot{v}_{\text{$\mu $}} \nonumber\\
-&3 h_1^2 h_0 \gamma _{\nu } \dot{v}_{\text{$\mu $}}+\frac{3}{2} h_1
h_0^2 \gamma _{\nu } \ddot{v}_{\text{$\mu $}}+3 h_2 h_0^2 v_{\mu }
v_{\nu }+3 h_1^2 h_0 v_{\mu } v_{\nu }.
\end{align}
The following term is written as
\begin{align}
\nonumber
&R^{-4}  \bigg[\dot{z}_{\mu}\left( x_{\nu}-z_{\nu}\right)-\dot{z}_{\nu}\left( x_{\mu}-z_{\mu}\right)\bigg] \\
=&10 p_3^2 v_{\mu } \gamma _{\nu } h_0^6-4 k_{\text{$\mu \nu $2}} p_3 h_0^5-4 p_4 v_{\mu } v_{\nu } h_0^5-4 p_5 v_{\mu } \gamma _{\nu } h_0^5 \nonumber\\
+&4 p_4 \dot{v}_{\text{$\mu $}} \gamma _{\nu } h_0^5-2 p_3 \ddot{v}_{\text{$\mu $}} \gamma _{\nu } h_0^5+k_{\text{$\mu \nu $4}} h_0^4-20 h_1 p_3 v_{\mu } v_{\nu } h_0^4 \nonumber\\
-&20 h_2 p_3 v_{\mu } \gamma _{\nu } h_0^4-20 h_1 p_4 v_{\mu } \gamma _{\nu } h_0^4+\frac{v_{\mu } \gamma _{\nu } h_0^4}{\sigma ^4}+20 h_1 p_3 \dot{v}_{\text{$\mu $}} \gamma _{\nu } h_0^4 \nonumber\\
+&\frac{v^{(4)}_{\text{$\mu $}} \gamma _{\nu } h_0^4}{24}+4 h_2 k_{\text{$\mu \nu $2}} h_0^3+4 h_1 k_{\text{$\mu \nu $3}} h_0^3+4 h_3 v_{\mu } v_{\nu } h_0^3+4 h_4 v_{\mu } \gamma _{\nu } h_0^3 \nonumber\\
-&40 h_1^2 p_3 v_{\mu } \gamma _{\nu } h_0^3-4 h_3 \dot{v}_{\text{$\mu $}} \gamma _{\nu } h_0^3+2 h_2 \ddot{v}_{\text{$\mu $}} \gamma _{\nu } h_0^3-\frac{4 h_1 \dddot{v}_{\text{$\mu $}} \gamma _{\nu } h_0^3}{6} \nonumber\\
+&6 h_1^2 k_{\text{$\mu \nu $2}} h_0^2+12 h_1 h_2 v_{\mu } v_{\nu } h_0^2+6 h_2^2 v_{\mu } \gamma _{\nu } h_0^2+12 h_1 h_3 v_{\mu } \gamma _{\nu } h_0^2 \nonumber\\
-&12 h_1 h_2 \dot{v}_{\text{$\mu $}} \gamma _{\nu } h_0^2+3 h_1^2 \ddot{v}_{\text{$\mu $}} \gamma _{\nu } h_0^2+4 h_1^3 v_{\mu } v_{\nu } h_0+12 h_1^2 h_2 v_{\mu } \gamma _{\nu } h_0 \nonumber\\
-&4 h_1^3 \dot{v}_{\text{$\mu $}} \gamma _{\nu } h_0 \nonumber\\
+&\frac{1}{\sigma }\bigg[ -4 p_3 v_{\mu } v_{\nu } h_0^5-4 p_4
v_{\mu } \gamma _{\nu } h_0^5+4 p_3 \dot{v}_{\text{$\mu $}} \gamma
_{\nu } h_0^5
+k_{\text{$\mu \nu $3}} h_0^4 \nonumber\\
-&20 h_1 p_3 v_{\mu } \gamma _{\nu } h_0^4-\frac{\dddot{v}_{\text{$\mu $}} \gamma _{\nu } h_0^4}{6}+4 h_1 k_{\text{$\mu \nu $2}} h_0^3+4 h_2 v_{\mu } v_{\nu } h_0^3 \nonumber\\
+&4 h_3 v_{\mu } \gamma _{\nu } h_0^3-4 h_2 \dot{v}_{\text{$\mu $}} \gamma _{\nu } h_0^3+2 h_1 \ddot{v}_{\text{$\mu $}} \gamma _{\nu } h_0^3 \nonumber\\
+& 6 h_1^2 v_{\mu } v_{\nu } h_0^2+12 h_1 h_2 v_{\mu } \gamma _{\nu } h_0^2-6 h_1^2 \dot{v}_{\text{$\mu $}} \gamma _{\nu } h_0^2+4 h_1^3 v_{\mu } \gamma _{\nu } h_0\bigg] \nonumber\\
+&\frac{1}{\sigma ^2}\left[ -4 p_3 v_{\mu } \gamma _{\nu } h_0^5+k_{\text{$\mu \nu $2}} h_0^4+\frac{1}{2} \ddot{v}_{\text{$\mu $}} \gamma _{\nu } h_0^4+4 h_1 v_{\mu } v_{\nu } h_0^3 \right] \nonumber \\
+&\left. 4 h_2 v_{\mu } \gamma _{\nu } h_0^3-4 h_1 \dot{v}_{\text{$\mu $}} \gamma _{\nu } h_0^3+6 h_1^2 v_{\mu } \gamma _{\nu } h_0^2 \right] \nonumber\\
+&\frac{v_{\mu } v_{\nu } h_0^4-\dot{v}_{\text{$\mu $}} \gamma _{\nu
} h_0^4+4 h_1 v_{\mu } \gamma _{\nu } h_0^3}{\sigma ^3}+h_1^4 v_{\mu
} \gamma _{\nu }.
\end{align}
Finally the last expansion reads
\begin{align}
\nonumber
R^{-5} & \bigg[\dot{z}_{\mu}\left( x_{\nu}-z_{\nu}\right)-\dot{z}_{\nu}\left( x_{\mu}-z_{\mu}\right)\bigg]  \nonumber\\
=&15 p_3^2 v_{\mu } v_{\nu } h_0^7+30 p_3 p_4 v_{\mu } \gamma _{\nu } h_0^7-15 p_3^2 \dot{v}_{\text{$\mu $}} \gamma _{\nu } h_0^7-5 k_{\text{$\mu \nu $3}} p_3 h_0^6-5 k_{\text{$\mu \nu $2}} p_4 h_0^6-5 p_5 v_{\mu } v_{\nu } h_0^6 \nonumber\\
+&105 h_1 p_3^2 v_{\mu } \gamma _{\nu } h_0^6
-5 p_6 v_{\mu } \gamma _{\nu } h_0^6+5 p_5 \dot{v}_{\text{$\mu $}} \gamma _{\nu } h_0^6-\frac{5}{2} p_4 \ddot{v}_{\text{$\mu $}} \gamma _{\nu } h_0^6+\frac{5 p_3 \dddot{v}_{\text{$\mu $}} \gamma _{\nu } h_0^6}{6}+k_{\text{$\mu \nu $5}} h_0^5 \nonumber\\
-&30 h_1 k_{\text{$\mu \nu $2}} p_3 h_0^5-30 h_2 p_3 v_{\mu } v_{\nu } h_0^5-30 h_1 p_4 v_{\mu } v_{\nu } h_0^5-30 h_3 p_3 v_{\mu } \gamma _{\nu } h_0^5-30 h_2 p_4 v_{\mu } \gamma _{\nu } h_0^5 \nonumber\\
-&30 h_1 p_5 v_{\mu } \gamma _{\nu } h_0^5+\frac{v_{\mu } \gamma _{\nu } h_0^5}{\sigma ^5}+30 h_2 p_3 \dot{v}_{\text{$\mu $}} \gamma _{\nu } h_0^5+30 h_1 p_4 \dot{v}_{\text{$\mu $}} \gamma _{\nu } h_0^5-15 h_1 p_3 \ddot{v}_{\text{$\mu $}} \gamma _{\nu } h_0^5 \nonumber\\
-&\frac{v^{(5)}_{\text{$\mu $}} \gamma _{\nu } h_0^5}{120}+5 h_3 k_{\text{$\mu \nu $2}} h_0^4+5 h_2 k_{\text{$\mu \nu $3}} h_0^4+5 h_1 k_{\text{$\mu \nu $4}} h_0^4+5 h_4 v_{\mu } v_{\nu } h_0^4-75 h_1^2 p_3 v_{\mu } v_{\nu } h_0^4 \nonumber\\
+&5 h_5 v_{\mu } \gamma _{\nu } h_0^4-150 h_1 h_2 p_3 v_{\mu } \gamma _{\nu } h_0^4-75 h_1^2 p_4 v_{\mu } \gamma _{\nu } h_0^4-5 h_4 \dot{v}_{\text{$\mu $}} \gamma _{\nu } h_0^4+75 h_1^2 p_3 \dot{v}_{\text{$\mu $}} \gamma _{\nu } h_0^4+\frac{5}{2} h_3 \ddot{v}_{\text{$\mu $}} \gamma _{\nu } h_0^4 \nonumber\\
+&\frac{5 h_1 v^{(4)}_{\text{$\mu $}} \gamma _{\nu } h_0^4}{24}-\frac{5 h_2 \dddot{v}_{\text{$\mu $}} \gamma _{\nu } h_0^4}{6}+20 h_1 h_2 k_{\text{$\mu \nu $2}} h_0^3+10 h_1^2 k_{\text{$\mu \nu $3}} h_0^3+10 h_2^2 v_{\mu } v_{\nu } h_0^3+20 h_1 h_3 v_{\mu } v_{\nu } h_0^3 \nonumber\\
+&20 h_2 h_3 v_{\mu } \gamma _{\nu } h_0^3+20 h_1 h_4 v_{\mu } \gamma _{\nu } h_0^3-100 h_1^3 p_3 v_{\mu } \gamma _{\nu } h_0^3-10 h_2^2 \dot{v}_{\text{$\mu $}} \gamma _{\nu } h_0^3-20 h_1 h_3 \dot{v}_{\text{$\mu $}} \gamma _{\nu } h_0^3 \nonumber\\
+&10 h_1 h_2 \ddot{v}_{\text{$\mu $}} \gamma _{\nu } h_0^3-\frac{10 h_1^2 \dddot{v}_{\text{$\mu $}} \gamma _{\nu } h_0^3}{6}+10 h_1^3 k_{\text{$\mu \nu $2}} h_0^2+30 h_1^2 h_2 v_{\mu } v_{\nu } h_0^2+30 h_1 h_2^2 v_{\mu } \gamma _{\nu } h_0^2+30 h_1^2 h_3 v_{\mu } \gamma _{\nu } h_0^2 \nonumber\\
-&30 h_1^2 h_2 \dot{v}_{\text{$\mu $}} \gamma _{\nu } h_0^2+5 h_1^3 \ddot{v}_{\text{$\mu $}} \gamma _{\nu } h_0^2+5 h_1^4 v_{\mu } v_{\nu } h_0+20 h_1^3 h_2 v_{\mu } \gamma _{\nu } h_0-5 h_1^4 \dot{v}_{\text{$\mu $}} \gamma _{\nu } h_0 \nonumber\\
+&\frac{1}{\sigma }\bigg[ 15 p_3^2 v_{\mu } \gamma _{\nu } h_0^7-5 k_{\text{$\mu \nu $2}} p_3 h_0^6-5 p_4 v_{\mu } v_{\nu } h_0^6-5 p_5 v_{\mu } \gamma _{\nu } h_0^6+5 p_4 \dot{v}_{\text{$\mu $}} \gamma _{\nu } h_0^6 \nonumber\\
-& \frac{5}{2} p_3 \ddot{v}_{\text{$\mu $}} \gamma _{\nu } h_0^6+k_{\text{$\mu \nu $4}} h_0^5-30 h_1 p_3 v_{\mu } v_{\nu } h_0^5-30 h_2 p_3 v_{\mu } \gamma _{\nu } h_0^5-30 h_1 p_4 v_{\mu } \gamma _{\nu } h_0^5 \nonumber\\
+& 30 h_1 p_3 \dot{v}_{\text{$\mu $}} \gamma _{\nu } h_0^5+\frac{v^{(4)}_{\text{$\mu $}} \gamma _{\nu } h_0^5}{24}+5 h_2 k_{\text{$\mu \nu $2}} h_0^4+5 h_1 k_{\text{$\mu \nu $3}} h_0^4+5 h_3 v_{\mu } v_{\nu } h_0^4 \nonumber\\
+&5 h_4 v_{\mu } \gamma _{\nu } h_0^4-75 h_1^2 p_3 v_{\mu } \gamma _{\nu } h_0^4-5 h_3 \dot{v}_{\text{$\mu $}} \gamma _{\nu } h_0^4+\frac{5}{2} h_2 \ddot{v}_{\text{$\mu $}} \gamma _{\nu } h_0^4-\frac{5 h_1 \dddot{v}_{\text{$\mu $}} \gamma _{\nu } h_0^4}{6}+10 h_1^2 k_{\text{$\mu \nu $2}} h_0^3 \nonumber\\
+&20 h_1 h_2 v_{\mu } v_{\nu } h_0^3+10 h_2^2 v_{\mu } \gamma _{\nu } h_0^3+20 h_1 h_3 v_{\mu } \gamma _{\nu } h_0^3-20 h_1 h_2 \dot{v}_{\text{$\mu $}} \gamma _{\nu } h_0^3 \nonumber\\
+& 5 h_1^2 \ddot{v}_{\text{$\mu $}} \gamma _{\nu } h_0^3+10 h_1^3 v_{\mu } v_{\nu } h_0^2+30 h_1^2 h_2 v_{\mu } \gamma _{\nu } h_0^2-10 h_1^3 \dot{v}_{\text{$\mu $}} \gamma _{\nu } h_0^2+5 h_1^4 v_{\mu } \gamma _{\nu } h_0\bigg] \nonumber\\
+&\frac{1}{\sigma ^2}\bigg[ -5 p_3 v_{\mu } v_{\nu } h_0^6-5 p_4 v_{\mu } \gamma _{\nu } h_0^6+5 p_3 \dot{v}_{\text{$\mu $}} \gamma _{\nu } h_0^6+k_{\text{$\mu \nu $3}} h_0^5-30 h_1 p_3 v_{\mu } \gamma _{\nu } h_0^5-\frac{\dddot{v}_{\text{$\mu $}} \gamma _{\nu } h_0^5}{6} \nonumber\\
+&5 h_1 k_{\text{$\mu \nu $2}} h_0^4+5 h_2 v_{\mu } v_{\nu } h_0^4+5 h_3 v_{\mu } \gamma _{\nu } h_0^4-5 h_2 \dot{v}_{\text{$\mu $}} \gamma _{\nu } h_0^4+\frac{5}{2} h_1 \ddot{v}_{\text{$\mu $}} \gamma _{\nu } h_0^4 \nonumber\\
+& 10 h_1^2 v_{\mu } v_{\nu } h_0^3+20 h_1 h_2 v_{\mu } \gamma _{\nu } h_0^3-10 h_1^2 \dot{v}_{\text{$\mu $}} \gamma _{\nu } h_0^3+10 h_1^3 v_{\mu } \gamma _{\nu } h_0^2\bigg]+h_1^5 v_{\mu } \gamma _{\nu } \nonumber\\
+&\frac{1}{\sigma ^3}\bigg[ -5 p_3 v_{\mu } \gamma _{\nu } h_0^6+k_{\text{$\mu \nu $2}} h_0^5+\frac{1}{2} \ddot{v}_{\text{$\mu $}} \gamma _{\nu } h_0^5+5 h_1 v_{\mu } v_{\nu } h_0^4+5 h_2 v_{\mu } \gamma _{\nu } h_0^4 \nonumber\\
-& 5 h_1 \dot{v}_{\text{$\mu $}} \gamma _{\nu } h_0^4+10 h_1^2 v_{\mu } \gamma _{\nu } h_0^3\bigg]  \nonumber\\
+&\frac{v_{\mu } v_{\nu } h_0^5-\dot{v}_{\text{$\mu $}} \gamma_{\nu}
h_0^5+5 h_1 v_{\mu } \gamma _{\nu } h_0^4}{\sigma ^4}.
\end{align}
Using these expansions obtained in this appendix we can calculate
the field $F_{\mu \nu }^{(ret)}$ in a neighborhood near the
trajectory of the particle. To obtain this field in the points on
the trajectories of the non-abelian particle
\begin{align}
\nonumber
F_{\mu \nu}^{(ret)}=&\left(2 k_{\text{$\mu \nu $3}} h_0^2-\frac{v_{\mu } \gamma _{\nu } h_0^2}{\sigma ^3}+3 h_1 k_{\text{$\mu \nu $2}} h_0-\frac{h_1 v_{\mu } \gamma _{\nu } h_0}{\sigma ^2}+\frac{k_{\text{$\mu \nu $2}} h_0^2+\frac{1}{2} \ddot{v}_{\text{$\mu $}} \gamma _{\nu } h_0^2-h_1 \dot{v}_{\text{$\mu $}} \gamma _{\nu } h_0}{\sigma }\right) A_1 \nonumber\\
+&\left(k_{\text{$\mu \nu $3}} h_0^3+\frac{v_{\mu } \gamma _{\nu } h_0^3}{\sigma ^3}+3 h_1 k_{\text{$\mu \nu $2}} h_0^2+\frac{k_{\text{$\mu \nu $2}} h_0^3+\frac{1}{2} \ddot{v}_{\text{$\mu $}} \gamma _{\nu } h_0^3-3 h_1 \dot{v}_{\text{$\mu $}} \gamma _{\nu } h_0^2+g_{2} v_{\mu } \gamma _{\nu }}{\sigma }\right. \nonumber\\
+&\left. \frac{3 h_0^2 h_1 v_{\mu } \gamma _{\nu }-h_0^3 \dot{v}_{\text{$\mu $}} \gamma _{\nu }}{\sigma ^2}\right) A_3 \nonumber\\
+&A_5 \left(k_{\text{$\mu \nu $5}} h_0^5+\frac{v_{\mu } \gamma _{\nu } h_0^5}{\sigma ^5}+5 h_1 k_{\text{$\mu \nu $4}} h_0^4+g_{9} k_{\text{$\mu \nu $2}}+k_{\text{$\mu \nu $3}} g_{11}\right. \nonumber\\
+&\left. \frac{k_{\text{$\mu \nu $2}} h_0^5+\frac{1}{2} \ddot{v}_{\text{$\mu $}} \gamma _{\nu } h_0^5-5 h_1 \dot{v}_{\text{$\mu $}} \gamma _{\nu } h_0^4+g_{12} v_{\mu } \gamma _{\nu }}{\sigma ^3}+\frac{5 h_0^4 h_1 v_{\mu } \gamma _{\nu }-h_0^5 \dot{v}_{\text{$\mu $}} \gamma _{\nu }}{\sigma ^4}\right. \nonumber\\
+& \frac{1}{\sigma }\left( k_{\text{$\mu \nu $4}} h_0^5+\frac{1}{24} v^{(4)}_{\text{$\mu $}} \gamma _{\nu } h_0^5+5 h_1 k_{\text{$\mu \nu $3}} h_0^4-\frac{5}{6} h_1 \dddot{v}_{\text{$\mu $}} \gamma _{\nu } h_0^4+k_{\text{$\mu \nu $2}} g_{17}+g_{15} v_{\mu } \gamma _{\nu }\right. \nonumber\\
+&\left. \dot{v}_{\text{$\mu $}} g_{16} \gamma _{\nu }+\ddot{v}_{\text{$\mu $}} \gamma _{\nu } g_{18}\right)  \nonumber\\
+&\left.  \frac{k_{\text{$\mu \nu $3}} h_0^5-\frac{1}{6} \dddot{v}_{\text{$\mu $}} \gamma _{\nu } h_0^5+5 h_1 k_{\text{$\mu \nu $2}} h_0^4+\frac{5}{2} h_1 \ddot{v}_{\text{$\mu $}} \gamma _{\nu } h_0^4+g_{13} v_{\mu } \gamma _{\nu }-g_{14} \dot{v}_{\text{$\mu $}} \gamma _{\nu }}{\sigma ^2}\right) \nonumber\\
+&A_2 \left(3 k_{\text{$\mu \nu $4}} h_0^3-\frac{v_{\mu } \gamma _{\nu } h_0^3}{\sigma ^4}+7 h_1 k_{\text{$\mu \nu $3}} h_0^2-\frac{2 h_1 v_{\mu } \gamma _{\nu } h_0^2}{\sigma ^3}\right. \nonumber\\
+&\left. g_{3} k_{\text{$\mu \nu $2}}+\frac{k_{\text{$\mu \nu $2}} h_0^3+\frac{1}{2} \ddot{v}_{\text{$\mu $}} \gamma _{\nu } h_0^3-h_1 \dot{v}_{\text{$\mu $}} \gamma _{\nu } h_0^2+g_{4} v_{\mu } \gamma _{\nu }}{\sigma ^2}\right. \nonumber\\
+&\left. \frac{2 k_{\text{$\mu \nu $3}} h_0^3-\frac{1}{3} \dddot{v}_{\text{$\mu $}} \gamma _{\nu } h_0^3+4 h_1 k_{\text{$\mu \nu $2}} h_0^2-2 g_1 \dot{v}_{\text{$\mu $}} \gamma _{\nu } h_0^2+2 h_1 \ddot{v}_{\text{$\mu $}} \gamma _{\nu } h_0^2-2 h_1^2 \dot{v}_{\text{$\mu $}} \gamma _{\nu } h_0+g_{5} v_{\mu } \gamma _{\nu }}{\sigma }\right) \nonumber\\
+&A_4 \left(k_{\text{$\mu \nu $4}} h_0^4+\frac{v_{\mu } \gamma _{\nu } h_0^4}{\sigma ^4}+4 h_1 k_{\text{$\mu \nu $3}} h_0^3+g_{6} k_{\text{$\mu \nu $2}}\right. \nonumber\\
+&\left. \frac{k_{\text{$\mu \nu $2}} h_0^4+\frac{1}{2} \ddot{v}_{\text{$\mu $}} \gamma _{\nu } h_0^4-4 h_1 \dot{v}_{\text{$\mu $}} \gamma _{\nu } h_0^3+g_{7} v_{\mu } \gamma _{\nu }}{\sigma ^2}+\frac{4 h_0^3 h_1 v_{\mu } \gamma _{\nu }-h_0^4 \dot{v}_{\text{$\mu $}} \gamma _{\nu }}{\sigma ^3}\right. \nonumber\\
+&\left. \frac{k_{\text{$\mu \nu $3}} h_0^4-\frac{1}{6}
\dddot{v}_{\text{$\mu $}} \gamma _{\nu } h_0^4+4 h_1 k_{\text{$\mu
\nu $2}} h_0^3+2 h_1 \ddot{v}_{\text{$\mu $}} \gamma _{\nu }
h_0^3+g_{8} v_{\mu } \gamma _{\nu }+\dot{v}_{\text{$\mu $}} g_{10}
\gamma _{\nu }}{\sigma }\right),
\end{align}
where $A_1$, $A_2$, $A_3$, $A_4$ and $A_5$ are those given after Eq.
(\ref{a}).

Now to calculate the actual field we use the Eq. (\ref{Frad}) and
solve the equation for $\sigma$ in terms of the parameter
$\varepsilon$. These arguments were already presented in
\cite{Dirac:1938nz}, to calculate $F_{\mu \nu }^{(act)}$, it is
necessary to determine first $\sigma$ as a function of $\varepsilon$
for which we have
\begin{align}
\nonumber
\sigma ^2=&\frac{\varepsilon ^2}{1-\gamma  \dot{\textbf{v}}}\left[  1-\frac{\gamma\ddot{\textbf{v}} \varepsilon }{3}-\varepsilon ^9 \left(\frac{\gamma  \textbf{v}^{(10)}}{19958400}+\frac{\textbf{v}^{(4)} \textbf{v}^{(5)}}{21600}+\frac{\dddot{\textbf{v}} \textbf{v}^{(6)}}{33600}+\frac{\ddot{\textbf{v}} \textbf{v}^{(7)}}{86400}+\frac{\dot{\textbf{v}} \textbf{v}^{(8)}}{453600}\right)\right. \nonumber\\
+&\varepsilon ^8 \left(\frac{\gamma  \textbf{v}^{(9)}}{1814400}+\frac{\textbf{v}^{(4)2}}{8100}+\frac{\dddot{\textbf{v}} \textbf{v}^{(5)}}{5184}+\frac{\ddot{\textbf{v}} \textbf{v}^{(6)}}{11340}+\frac{\dot{\textbf{v}} \textbf{v}^{(7)}}{51840}\right) \nonumber\\
-&\varepsilon ^7 \left(\frac{\gamma  \textbf{v}^{(8)}}{181440}+\frac{\dddot{\textbf{v}} \textbf{v}^{(4)}}{960}+\frac{\ddot{\textbf{v}} \textbf{v}^{(5)}}{1728}+\frac{\dot{\textbf{v}} \textbf{v}^{(6)}}{6720}\right)+\varepsilon ^6 \left(\frac{\gamma  \textbf{v}^{(7)}}{20160}+\frac{\dddot{\textbf{v}}^{2}}{448}+\frac{\ddot{\textbf{v}} \textbf{v}^{(4)}}{315}+\frac{\dot{\textbf{v}} \textbf{v}^{(5)}}{1008}\right) \nonumber\\
-&\varepsilon ^5 \left(\frac{\gamma  \textbf{v}^{(6)}}{2520}+\frac{\ddot{\textbf{v}} \dddot{\textbf{v}}}{72}+\frac{\dot{\textbf{v}} \textbf{v}^{(4)}}{180}\right)+\varepsilon ^4 \left(\frac{\gamma  \textbf{v}^{(5)}}{360}+\frac{\ddot{\textbf{v}}^{2}}{45}+\frac{\dot{\textbf{v}} \dddot{\textbf{v}}}{40}\right)-\varepsilon ^3 \left(\frac{\gamma  \textbf{v}^{(4)}}{60}+\frac{ \dot{\textbf{v}}\ddot{\textbf{v}}}{12}\right) \nonumber\\
+&\left. \varepsilon ^2 \left(\frac{\gamma
\dddot{\textbf{v}}}{12}+\frac{\dot{\textbf{v}}^{2}}{12}\right)\right].
\end{align}
This expression is obtained from expansion (\ref{k3}) and the
condition $ (x_{\mu}-z_{\mu}(s))(x^{\mu}-z^{\mu}(s))=0 $. The
substitution of these equations, in $F_{\mu \nu }^{(act)}$ gives the
field strength very close to the trajectory of the Yang-Mills
particle
\begin{align*}
\nonumber
F_{\mu \nu }^{(act)}=&f_{\mu \nu }-\frac{A_5 k_{\text{$\mu \nu $4}} h_0^5}{\varepsilon  U_1}-\frac{A_5 h_0^5}{24 \varepsilon  U_1}\left[ v_{\mu}^{(4)} \gamma _{\nu }-v_{\nu}^{(4)} \gamma _{\mu }\right]  +\frac{A_4 h_0^4}{6 \varepsilon  U_1}\left[  \dddot{v}_{\text{$\mu $}} \gamma _{\nu }- \dddot{v}_{\text{$\nu $}} \gamma _{\mu }\right] \nonumber\\
 -&\frac{A_4 k_{\text{$\mu \nu $3}} h_0^4}{\varepsilon  U_1}+\frac{A_2 h_0^3}{3 \varepsilon  U_1}\left[ \dddot{v}_{\text{$\mu $}} \gamma _{\nu }- \dddot{v}_{\text{$\nu $}} \gamma _{\mu }  \right] \nonumber\\
-&\frac{A_3 k_{\text{$\mu \nu $2}} h_0^3}{\varepsilon  U_1}-\frac{2 A_2 k_{\text{$\mu \nu $3}} h_0^3}{\varepsilon  U_1}-\frac{A_3 h_0^3}{2 \varepsilon  U_1}\left[ \ddot{v}_{\text{$\mu $}} \gamma _{\nu }- \ddot{v}_{\text{$\nu $}} \gamma _{\mu } \right]-\frac{A_1 k_{\text{$\mu \nu $2}} h_0^2}{\varepsilon  U_1} \nonumber\\
-&\frac{A_1 h_0^2}{2 \varepsilon  U_1}\left[ \ddot{v}_{\text{$\mu $}} \gamma _{\nu }- \ddot{v}_{\text{$\nu $}} \gamma _{\mu }  \right]+\frac{A_1 h_1 h_0}{\varepsilon  U_1}\left[ \dot{v}_{\text{$\mu $}} \gamma _{\nu }- \dot{v}_{\text{$\nu $}} \gamma _{\mu } \right] \nonumber\\
+&A_2 k_{\text{$\mu \nu $2}} \left(\frac{2 a_1 h_0^3}{\varepsilon  U_1^2}-\frac{4 h_0^2 h_1}{\varepsilon  U_1}\right)+A_4 k_{\text{$\mu \nu $2}} \left(\frac{2 a_1 h_0^4}{\varepsilon  U_1^2}-\frac{4 h_0^3 h_1}{\varepsilon  U_1}\right)+A_5 k_{\text{$\mu \nu $3}} \left(\frac{2 a_1 h_0^5}{\varepsilon  U_1^2}-\frac{5 h_0^4 h_1}{\varepsilon  U_1}\right) \nonumber\\
+&A_5 k_{\text{$\mu \nu $2}} \left(-\frac{6 a_1^2 h_0^5}{\varepsilon  U_1^3}+\frac{3 a_2 h_0^5}{\varepsilon  U_1^3}-\frac{h_0^5}{\varepsilon ^3 U_1^3}+\frac{10 a_1 h_1 h_0^4}{\varepsilon  U_1^2}-\frac{g_{17}}{\varepsilon  U_1}\right) \nonumber\\
+&A_2 \left(-\frac{20 a_1^3 h_0^3}{\varepsilon  U_1^4}+\frac{20 a_1 a_2 h_0^3}{\varepsilon  U_1^4}-\frac{4 a_3 h_0^3}{\varepsilon  U_1^4}-\frac{4 a_1 h_0^3}{\varepsilon ^3 U_1^4}+\frac{12 a_1^2 h_1 h_0^2}{\varepsilon  U_1^3}\right. \nonumber\\
-&\left. \frac{6 a_2 h_1 h_0^2}{\varepsilon  U_1^3}+\frac{2 h_1 h_0^2}{\varepsilon ^3 U_1^3}-\frac{g_{5}}{\varepsilon  U_1}+\frac{2 a_1 g_{4}}{\varepsilon  U_1^2}\right) \left[ v_{\text{$\mu $}} \gamma _{\nu }-v_{\text{$\nu $}} \gamma _{\mu }  \right] \nonumber\\
+&A_3 \left(-\frac{6 a_1^2 h_0^3}{\varepsilon  U_1^3}+\frac{3 a_2 h_0^3}{\varepsilon  U_1^3}-\frac{h_0^3}{\varepsilon ^3 U_1^3}+\frac{6 a_1 h_1 h_0^2}{\varepsilon  U_1^2}-\frac{g_{2}}{\varepsilon  U_1}\right) \left[ v_{\text{$\mu $}} \gamma _{\nu }-v_{\text{$\nu $}} \gamma _{\mu }  \right] \nonumber\\
+&A_4 \left(\frac{20 a_1^3 h_0^4}{\varepsilon  U_1^4}-\frac{20 a_1 a_2 h_0^4}{\varepsilon  U_1^4}+\frac{4 a_3 h_0^4}{\varepsilon  U_1^4}+\frac{4 a_1 h_0^4}{\varepsilon ^3 U_1^4}-\frac{24 a_1^2 h_1 h_0^3}{\varepsilon  U_1^3}\right. \nonumber\\
+&\left. \frac{12 a_2 h_1 h_0^3}{\varepsilon  U_1^3}-\frac{4 h_1 h_0^3}{\varepsilon ^3 U_1^3}-\frac{g_{8}}{\varepsilon  U_1}+\frac{2 a_1 g_{7}}{\varepsilon  U_1^2}\right) \left[ v_{\text{$\mu $}} \gamma _{\nu }-v_{\text{$\nu $}} \gamma _{\mu }  \right] \nonumber\\
\end{align*}
\begin{align}
\nonumber
+&A_5 \left(-\frac{70 a_1^4 h_0^5}{\varepsilon  U_1^5}-\frac{15 a_2^2 h_0^5}{\varepsilon  U_1^5}+\frac{105 a_1^2 a_2 h_0^5}{\varepsilon  U_1^5}-\frac{30 a_1 a_3 h_0^5}{\varepsilon  U_1^5}+\frac{5 a_4 h_0^5}{\varepsilon  U_1^5}-\frac{15 a_1^2 h_0^5}{\varepsilon ^3 U_1^5}+\frac{5 a_2 h_0^5}{\varepsilon ^3 U_1^5}\right. \nonumber\\
-& \frac{h_0^5}{\varepsilon ^5 U_1^5}+\frac{100 a_1^3 h_1 h_0^4}{\varepsilon  U_1^4}-\frac{100 a_1 a_2 h_1 h_0^4}{\varepsilon  U_1^4}+\frac{20 a_3 h_1 h_0^4}{\varepsilon  U_1^4}+\frac{20 a_1 h_1 h_0^4}{\varepsilon ^3 U_1^4}-\frac{g_{15}}{\varepsilon  U_1}+\frac{2 a_1 g_{12}}{\varepsilon  U_1^2} \nonumber\\
-&\left. \frac{6 a_1^2 g_{12}}{\varepsilon  U_1^3}+\frac{3 a_2 g_{12}}{\varepsilon  U_1^3}-\frac{g_{12}}{\varepsilon ^3 U_1^3}\right) \left[ v_{\text{$\mu $}} \gamma _{\nu }-v_{\text{$\nu $}} \gamma _{\mu }  \right] \nonumber\\
+&A_1 \left(\frac{6 a_1^2 h_0^2}{\varepsilon  U_1^3}-\frac{3 a_2 h_0^2}{\varepsilon  U_1^3}+\frac{h_0^2}{\varepsilon ^3 U_1^3}-\frac{2 a_1 h_1 h_0}{\varepsilon  U_1^2}\right) \left[ v_{\text{$\mu $}} \gamma _{\nu }-v_{\text{$\nu $}} \gamma _{\mu }  \right] \nonumber\\
+&A_3 \left(\frac{3 h_0^2 h_1}{\varepsilon  U_1}-\frac{2 a_1
h_0^3}{\varepsilon  U_1^2}\right) \left[\dot{v}_{\text{$\mu $}}
\gamma _{\nu }- \dot{v}_{\text{$\nu $}} \gamma _{\mu } \right]
+A_2 \left(\frac{2 g_1 h_0^2}{\varepsilon  U_1}-\frac{2 a_1 h_1 h_0^2}{\varepsilon  U_1^2}+\frac{2 h_1^2 h_0}{\varepsilon  U_1}\right) \left[ \dot{v}_{\text{$\mu $}} \gamma _{\nu }- \dot{v}_{\text{$\nu $}} \gamma _{\mu }  \right] \nonumber\\
+&A_4 \left(\frac{6 a_1^2 h_0^4}{\varepsilon  U_1^3}-\frac{3 a_2 h_0^4}{\varepsilon  U_1^3}+\frac{h_0^4}{\varepsilon ^3 U_1^3}-\frac{8 a_1 h_1 h_0^3}{\varepsilon  U_1^2}-\frac{g_{10}}{\varepsilon  U_1}\right) \left[ \dot{v}_{\text{$\mu $}} \gamma _{\nu }- \dot{v}_{\text{$\nu $}} \gamma _{\mu }  \right] \nonumber\\
+&A_5 \left(-\frac{20 a_1^3 h_0^5}{\varepsilon  U_1^4}+\frac{20 a_1 a_2 h_0^5}{\varepsilon  U_1^4}-\frac{4 a_3 h_0^5}{\varepsilon  U_1^4}-\frac{4 a_1 h_0^5}{\varepsilon ^3 U_1^4}+\frac{30 a_1^2 h_1 h_0^4}{\varepsilon  U_1^3}-\frac{15 a_2 h_1 h_0^4}{\varepsilon  U_1^3}+\frac{5 h_1 h_0^4}{\varepsilon ^3 U_1^3}\right. \nonumber\\
-&\left. \frac{g_{16}}{\varepsilon  U_1}-\frac{2 a_1 g_{14}}{\varepsilon  U_1^2}\right) \left[ \dot{v}_{\text{$\mu $}} \gamma _{\nu }- \dot{v}_{\text{$\nu $}} \gamma _{\mu }  \right] \nonumber\\
+&A_2 \left(\frac{a_1 h_0^3}{\varepsilon  U_1^2}-\frac{2 h_0^2 h_1}{\varepsilon  U_1}\right)\left[ \ddot{v}_{\text{$\mu $}} \gamma _{\nu }- \ddot{v}_{\text{$\nu $}} \gamma _{\mu }  \right]+A_4 \left(\frac{a_1 h_0^4}{\varepsilon  U_1^2}-\frac{2 h_0^3 h_1}{\varepsilon  U_1}\right) \left[ \ddot{v}_{\text{$\mu $}} \gamma _{\nu }- \ddot{v}_{\text{$\nu $}} \gamma _{\mu }  \right] \nonumber\\
+&A_5 \left(-\frac{3 a_1^2 h_0^5}{\varepsilon  U_1^3}+\frac{3 a_2 h_0^5}{2 \varepsilon  U_1^3}-\frac{h_0^5}{2 \varepsilon ^3 U_1^3}+\frac{5 a_1 h_1 h_0^4}{\varepsilon  U_1^2}-\frac{g_{18}}{\varepsilon  U_1}\right) \left[ \ddot{v}_{\text{$\mu $}} \gamma _{\nu }- \ddot{v}_{\text{$\nu $}} \gamma _{\mu }  \right] \nonumber\\
+&A_5 \left(\frac{5 h_0^4 h_1}{6 \varepsilon  U_1}-\frac{a_1
h_0^5}{3 \varepsilon  U_1^2}\right) \left[ \dddot{v}_{\text{$\mu $}}
\gamma _{\nu }- \dddot{v}_{\text{$\nu $}} \gamma _{\mu } \right],
\label{VGAlarge}
\end{align}
where $U_{1}\equiv (1-\gamma \dot{\textbf{v}})^{-1/2}$, and $a_1$,
$a_2$, $a_3$ and $a_4$ are given by
\begin{align}
\nonumber
a_1&=-\frac{\text{$\gamma $}\ddot{\textbf{v}}}{6};\;\;\;a_2=\frac{1}{72} \left(-\text{$\gamma $}\ddot{\textbf{v}}^2+3 \gamma  \dddot{\textbf{v}}+3 \dot{\textbf{v}}^2\right) \nonumber\\
a_{3}&=\frac{1}{432} \left(-\text{$\gamma $}\ddot{\textbf{v}}^3-\frac{18 \text{$\gamma $}\textbf{v}^{(4)}}{5}+3 \text{$\gamma $}\ddot{\textbf{v}} \left(\text{$\gamma $}\dddot{\textbf{v}}+\dot{\textbf{v}}^2\right)-18 \dot{\textbf{v}} \ddot{\textbf{v}}\right) \nonumber\\
a_{4}&=-\frac{1}{128} 5 \left( \frac{\text{$\gamma $}\ddot{\textbf{v}}}{3}\right)^4+\frac{3}{16} \left(\frac{\gamma  \dddot{\textbf{v}}}{12}+\frac{\dot{\textbf{v}}^2}{12}\right)   \left( \frac{\text{$\gamma $}\ddot{\textbf{v}}}{3}\right)^2-\frac{1}{4}\left( \frac{\gamma  \textbf{v}^{(4)}}{60}+\frac{\dot{\textbf{v}} \ddot{\textbf{v}}}{12}\right) \left( \frac{\text{$\gamma $}\ddot{\textbf{v}}}{3}\right) \nonumber\\
&-\frac{1}{2}\left( \frac{\gamma
\textbf{v}^{(5)}}{360}+\frac{\ddot{\textbf{v}}^2}{45}+\frac{\dot{\textbf{v}}
\dddot{\textbf{v}}}{40}\right) +\frac{1}{8}\left(\frac{\gamma
\dddot{\textbf{v}}}{12}+\frac{\dot{\textbf{v}}^2}{12}\right)^{2}.
\end{align}

Now we can calculate the radiating field, in order to do that we
calculate first the advanced field strength $F_{\mu \nu }^{(adv)}$.
Similarly as in the abelian case, it can be done by changing
$\varepsilon$ by $-\varepsilon$. Thus the radiated field strength
$F_{\mu \nu }^{(rad)} = F_{\mu \nu }^{(ret)} - F_{\mu \nu }^{(adv)}$
is written as
\begin{align}
F_{\mu \nu }^{(rad)}=&A_5 k_{\text{$\mu \nu $2}} \left(-\frac{20 a_1^3 h_0^5}{U_1^3}+\frac{24 a_1 a_2 h_0^5}{U_1^3}-\frac{6 a_3 h_0^5}{U_1^3}+\frac{30 a_1^2 h_1 h_0^4}{U_1^2}-\frac{20 a_2 h_1 h_0^4}{U_1^2}-\frac{2 a_1 g_{17}}{U_1}+2 g_{9}\right) \nonumber \\
+&A_2 k_{\text{$\mu \nu $2}} \left(\frac{6 a_1^2 h_0^3}{U_1^2}-\frac{4 a_2 h_0^3}{U_1^2}-\frac{8 a_1 h_1 h_0^2}{U_1}+2 g_{3}\right) \nonumber\\
+&A_4 k_{\text{$\mu \nu $2}} \left(\frac{6 a_1^2 h_0^4}{U_1^2}-\frac{4 a_2 h_0^4}{U_1^2}-\frac{8 a_1 h_1 h_0^3}{U_1}+2 g_{6}\right) \nonumber\\
+&A_5 k_{\text{$\mu \nu $3}} \left(\frac{6 a_1^2 h_0^5}{U_1^2}-\frac{4 a_2 h_0^5}{U_1^2}-\frac{10 a_1 h_1 h_0^4}{U_1}+2 g_{11}\right) \nonumber\\
+&A_1 k_{\text{$\mu \nu $2}} \left(6 h_0 h_1-\frac{2 a_1 h_0^2}{U_1}\right)+A_3 k_{\text{$\mu \nu $2}} \left(6 h_0^2 h_1-\frac{2 a_1 h_0^3}{U_1}\right)+A_2 k_{\text{$\mu \nu $3}} \left(14 h_0^2 h_1-\frac{4 a_1 h_0^3}{U_1}\right) \nonumber\\
+&A_4 k_{\text{$\mu \nu $3}} \left(8 h_0^3 h_1-\frac{2 a_1 h_0^4}{U_1}\right)+A_5 k_{\text{$\mu \nu $4}} \left(10 h_0^4 h_1-\frac{2 a_1 h_0^5}{U_1}\right) \nonumber\\
+&2 A_3 h_0^3 k_{\text{$\mu \nu $3}}+4 A_1 h_0^2 k_{\text{$\mu \nu
$3}}+2 A_4 h_0^4 k_{\text{$\mu \nu $4}}+6 A_2 h_0^3 k_{\text{$\mu
\nu $4}}+2 A_5 h_0^5 k_{\text{$\mu \nu $5}}.
\end{align}
If we take all the $A_{i}=0$, $ i=2,3,4,5 $, except $A_{1} \not=0$,
then we can calculate the radiated field by an abelian particle.
Therefore the Yang-Mills field is reduced to the one of Dirac's
radiation damping \cite{Dirac:1938nz}
\begin{equation}
F_{\mu \nu }^{(rad) D}=4 A_1 h_0^2 k_{\text{$\mu \nu $3}}=4 A_1
h_0^2 \left( \frac{\ddot{v}_{\text{$\mu $}} v_{\nu
}}{3}-\frac{v_{\mu } \ddot{v}_{\text{$\nu $}}}{3}\right).
\end{equation}
Notice that in $F_{\mu \nu }^{(rad)}$ there are two terms depending
on $A_{1}$. Moreover the terms depending on $h_{1}$ and $a_{1}$ will
not contribute to the equations of motion, since they are linear in
$\gamma$ and they will be eliminated in the integral when the flow
of the particle be computed.


\section{Computation of equations of motion}

In order to find the equations of motion we first should calculate
the energy-momentum tensor (\ref{T1}) in terms of the radiation
field at points near the trajectory of the particle. For this we
make use of $F_{\mu\nu}^{(act)}$ on the world-line. When we use
these relations in (\ref{T1}), we can find the divergent terms in
$T_{\mu\nu}$. In the process of computing and simplifying the
energy-momentum tensor (\ref{T1}) it is useful to consider the
following relations
\begin{align}
\nonumber
&\left[
\gamma_{\mu}\upsilon_{\nu}-\gamma_{\nu}\upsilon_{\mu}\right]\left[
\gamma^{\nu}\upsilon_{\rho}-\gamma_{\rho}\upsilon^{\nu}\right]=
-\gamma_{\mu}\gamma_{\rho}+\varepsilon^{2}\upsilon_{\mu}\upsilon_{\rho} \nonumber\\
&\left[ \gamma_{\mu}\upsilon_{\nu}-\gamma_{\nu}\upsilon_{\mu}\right]\left[ \gamma^{\nu}\upsilon^{(i)}_{\rho}-\gamma_{\rho}\upsilon^{(i)\nu}\right]=\text{$\gamma $}\textbf{v}^{(i)} \gamma _{\rho } v_{\mu }+\varepsilon ^2 v_{\mu } v^{(i)}_{\text{$\rho $}}-\textbf{v}\textbf{v}^{(i)} \gamma _{\mu } \gamma _{\rho }, \nonumber\\
&\left[
\gamma_{\mu}\upsilon_{\nu}-\gamma_{\nu}\upsilon_{\mu}\right]\left[
\upsilon^{\nu}\dot{\upsilon}_{\rho}-\upsilon_{\rho}\dot{\upsilon}^{\nu}\right]=
\gamma_{\mu}\dot{\upsilon}_{\rho}+(\gamma\dot{\textbf{v}})\upsilon_{\mu}\upsilon_{\rho} \nonumber\\
&\left[
\gamma_{\mu}\upsilon_{\nu}-\gamma_{\nu}\upsilon_{\mu}\right]\left[
\ddot{v}^{\nu}\gamma_{\rho}-\ddot{v}_{\rho}\gamma^{\nu}\right]=
-\dot{\textbf{v}}^{2}\gamma_{\mu}\gamma_{\rho}
-(\gamma\ddot{\textbf{v}})\upsilon_{\mu}\gamma_{\rho}
-\varepsilon^{2}\upsilon_{\mu}\ddot{\upsilon}_{\rho} \nonumber\\
&\left[
\upsilon_{\mu}\dot{\upsilon}_{\nu}-\upsilon_{\nu}\dot{\upsilon}_{\mu}\right]\left[
\upsilon^{\nu}\dot{\upsilon}_{\rho}-\upsilon_{\rho}\dot{\upsilon}^{\nu}\right]=
-\dot{\textbf{v}}^{2}\upsilon_{\mu}\upsilon_{\rho}-
\dot{\upsilon}_{\mu}\dot{\upsilon}_{\rho} \nonumber\\
&\left[ \gamma_{\mu}\upsilon^{(i)}_{\nu}-\gamma_{\nu}\upsilon^{(i)}_{\mu}\right]\left[ \gamma^{\nu}\upsilon_{\rho}-\gamma_{\rho}\upsilon^{\nu}\right]=\text{$\gamma $}\textbf{v}^{(i)} \gamma _{\mu } v_{\rho }+\varepsilon ^2 v_{\text{$\mu $}}^{(i)} v_{\rho }-\textbf{v}\textbf{v}^{(i)} \gamma _{\mu } \gamma _{\rho } \nonumber\\
&\left[
\gamma_{\mu}\upsilon^{(i)}_{\nu}-\gamma_{\nu}\upsilon^{(i)}_{\mu}\right]\left[
\gamma^{\nu}\upsilon^{(j)}_{\rho}-\gamma_{\rho}\upsilon^{(j)\nu}\right]=\text{$\gamma
$}\textbf{v}^{(i)} \gamma _{\mu } v^{(j)}_{\text{$\rho
$}}+\text{$\gamma $}\textbf{v}^{(j)} \gamma _{\rho } v_{\text{$\mu
$}}^{(i)}-\textbf{v}^{(i)} \textbf{v}^{(j)} \gamma _{\mu } \gamma
_{\rho }+\varepsilon ^2 v^{(i)}_{\text{$\mu $}} v^{(j)}_{\text{$\rho
$}}.
\end{align}

Integrate out in a hyper-surface of radius $\varepsilon$ and using
the Stokes theorem, with the normal vector to the surface being in
the direction of $\gamma^{\rho}$, which comes from the fact that the
equation $\gamma dx=0$ that is obtained from the variation of
$(x_{\mu}-z_{\mu})^{2}=\varepsilon^{2}$ and $R=0$ is satisfied in
the same way as in \cite{Dirac:1938nz} if we perform a splitting
$dx\to dx_{\perp}dx_{\parallel}$ we get
\begin{align}
\nonumber
-&\int \varepsilon^{-1}T_{\mu\rho}\gamma^{\rho}ds\left|dx_{\parallel}\right| =-\int 4\pi \varepsilon T_{\mu\rho}\gamma^{\rho}U_{1}^{-2}ds \nonumber\\
=&\int {\rm Tr}\left\lbrace \frac{A_5 f_{\mu \nu } v^{\nu }+Q_{14} v_{\mu }+Q_{11} \dot{v}_{\mu}+Q_{17} \ddot{v}_{\mu}+Q_{16} \dddot{v}_{\mu}}{\varepsilon ^2}\right. \nonumber\\
+&\varepsilon  \left(A_3 f_{\mu \nu } \dot{v}^{\nu}+Q_{27} f_{\mu \nu } v^{\nu }+Q_{31} f_{\mu \nu } \dot{v}^{\nu}+Q_{34} f_{\mu \nu } \ddot{v}^{\nu }+Q_{29} f_{\mu \nu } \dddot{v}^{\nu }+Q_7 v_{\mu }+Q_1 \dot{v}_{\mu}\right. \nonumber\\
+&\left. Q_5 \ddot{v}_{\mu}+Q_4 \dddot{v}_{\mu}\right)+\frac{A_5 f_{\mu \nu } \dot{v}^{\nu }+Q_{36} f_{\mu \nu } v^{\nu }+Q_9 v_{\mu }+Q_3 \dot{v}_{\mu}+Q_{10} \ddot{v}_{\mu}+Q_{15} \dddot{v}_{\mu}}{\varepsilon } \nonumber\\
-&A_4 f_{\mu \nu } \dot{v}^{\nu }+Q_{35} f_{\mu \nu } v^{\nu } \nonumber\\
+&\varepsilon ^2 \left(Q_{25} f_{\mu \nu } v^{\nu }+Q_{26} f_{\mu \nu } \dot{v}^{\nu }+Q_{28} f_{\mu \nu } \ddot{v}^{\nu }+Q_{32} f_{\mu \nu } \dddot{v}^{\nu }+Q_{30} f_{\mu \nu } v^{\nu (4) }\right) \nonumber\\
+&Q_{33} f_{\mu \nu } \ddot{v}^{\nu }+Q_8 v_{\mu }+Q_6 \dot{v}_{\mu}+Q_2 \ddot{v}_{\mu}+Q_{12} \dddot{v}_{\mu}+\frac{Q_{24} \dot{v}_{\mu}}{\varepsilon ^5} \nonumber\\
+&\left. \frac{Q_{23} v_{\mu }+Q_{22} \dot{v}_{\mu}+Q_{20}
\ddot{v}_{\mu}}{\varepsilon ^4}+\frac{Q_{13} v_{\mu }+Q_{21}
\dot{v}_{\mu}+Q_{19} \ddot{v}_{\mu}+Q_{18}
\dddot{v}_{\mu}}{\varepsilon ^3}\right\rbrace  ds,
\end{align}
where we have integrated on a hyper-sphere of area
$4\pi\varepsilon^{2}$ and used the line element
$\left|dx_{\parallel}\right|=(1-\gamma
\dot{\textbf{v}})ds=U_{1}^{-2}ds$. If we remember that the linear
terms in $\gamma$ do not contribute to the equations of motion and
we neglect all terms that depend on positive powers of $\varepsilon$
(in the limit $\varepsilon \to 0$) we get Eq. (\ref{eq0}). The
functions $Q_{i}$ are given by
\begin{align}
\nonumber
Q_{1}=&-\frac{35}{16} a_4 A_5^2 \dot{\textbf{v}}^2-\frac{5}{2} a_4 A_1 A_5-\frac{5}{2} a_4 A_3 A_5-\frac{17}{144} A_2 A_5 \dot{\textbf{v}}^2 \textbf{v}\dddot{\textbf{v}}-\frac{17}{288} A_4 A_5 \dot{\textbf{v}}^2 \textbf{v}\dddot{\textbf{v}} \nonumber\\
+&\frac{17 A_5^2 \dot{\textbf{v}}^2 \textbf{v}\textbf{v}^{(4)}}{1152}+\frac{209 A_5^2 \dot{\textbf{v}}^6}{9216}-\frac{25}{192} A_1 A_5 \dot{\textbf{v}}^4-\frac{1}{48} A_3 A_5 \dot{\textbf{v}}^4+\frac{21}{128} A_5^2 \ddot{\textbf{v}}\dot{\textbf{v}} \dot{\textbf{v}}^2 \nonumber\\
-&\frac{13}{288} A_2 A_5 \ddot{\textbf{v}}\dot{\textbf{v}} \dot{\textbf{v}}^2-\frac{17}{72} A_4 A_5 \ddot{\textbf{v}} \dot{\textbf{v}}\dot{\textbf{v}}^2-\frac{3}{16} A_1^2 \dot{\textbf{v}}^2-\frac{1}{16} A_3^2 \dot{\textbf{v}}^2+\frac{17}{576} A_5^2 \ddot{\textbf{v}}^2 \dot{\textbf{v}}^2 \nonumber\\
-&\frac{1}{4} A_1 A_3 \dot{\textbf{v}}^2-\frac{1}{96} A_4 A_5 \dot{\textbf{v}}\ddot{\textbf{v}} \dot{\textbf{v}}^2-\frac{1}{12} A_1 A_2 \ddot{\textbf{v}} \dot{\textbf{v}}-\frac{1}{12} A_2 A_3 V_2 \dot{\textbf{v}}-\frac{1}{3} A_1 A_4 \ddot{\textbf{v}} \dot{\textbf{v}} \nonumber \\
-&\frac{1}{3} A_3 A_4 \ddot{\textbf{v}}\dot{\textbf{v}}+\frac{3}{16} A_1 A_5 \ddot{\textbf{v}} \dot{\textbf{v}}+\frac{3}{16} A_3 A_5 \ddot{\textbf{v}} \dot{\textbf{v}}+\frac{1}{8} A_1 A_5 \ddot{\textbf{v}}^2+\frac{1}{8} A_3 A_5 \ddot{\textbf{v}}^2 \nonumber\\
+&\frac{1}{72} A_5^2 (\dot{\textbf{v}}\ddot{\textbf{v}})^2+\frac{1}{36} A_2 A_5 \ddot{\textbf{v}}\dddot{\textbf{v}}+\frac{1}{72} A_4 A_5 \ddot{\textbf{v}}\dddot{\textbf{v}}-\frac{1}{288} A_5^2 \ddot{\textbf{v}}\textbf{v}^{(4)}-\frac{1}{6} A_1 A_2 \textbf{v}\dddot{\textbf{v}}-\frac{1}{6} A_2 A_3 \textbf{v}\dddot{\textbf{v}} \nonumber\\
-&\frac{1}{12} A_1 A_4 \textbf{v}\dddot{\textbf{v}}-\frac{1}{12} A_3 A_4
\textbf{v}\dddot{\textbf{v}}+\frac{1}{48} A_1 A_5 \textbf{v}\textbf{v}^{(4)}+\frac{1}{48} A_3 A_5
\textbf{v}\textbf{v}^{(4)},
\end{align}
\begin{align}
\nonumber
Q_{2}=&-\frac{10}{3} a_4 A_5^2-\frac{1}{32} 3 A_5^2 \dot{\textbf{v}}^4-\frac{1}{3} A_2^2 \dot{\textbf{v}}^2-\frac{1}{6} A_2 A_4 \dot{\textbf{v}}^2-\frac{1}{4} A_1 A_5 \dot{\textbf{v}}^2+\frac{1}{4} A_5^2 \ddot{\textbf{v}} \dot{\textbf{v}}-\frac{17}{24} A_2 A_5 \ddot{\textbf{v}}\dot{\textbf{v}} \nonumber \\
-&\frac{55}{72} A_4 A_5 \ddot{\textbf{v}} \dot{\textbf{v}}+\frac{1}{72} A_5^2 \dddot{\textbf{v}}
\dot{\textbf{v}}+\frac{2}{9} A_5^2 \ddot{\textbf{v}}^2-\frac{1}{3} A_2 A_5
\textbf{v}\dddot{\textbf{v}}-\frac{1}{6} A_4 A_5 \textbf{v}\dddot{\textbf{v}}+\frac{1}{36} A_5^2
\textbf{v}\textbf{v}^{(4)},
\end{align}
\begin{align}
\nonumber
Q_{3}=&-\frac{5}{2} a_4 A_5^2+\frac{17}{192} A_5^2 \dot{\textbf{v}}^4-\frac{1}{2} A_2^2 \dot{\textbf{v}}^2-\frac{1}{2} A_2 A_4 \dot{\textbf{v}}^2-\frac{9}{16} A_1 A_5 \dot{\textbf{v}}^2 \nonumber\\
+&\frac{7}{16} A_3 A_5 \dot{\textbf{v}}^2+\frac{3}{16} A_5^2 \ddot{\textbf{v}} \dot{\textbf{v}}-\frac{11}{12} A_2 A_5 \ddot{\textbf{v}} \dot{\textbf{v}}-\frac{13}{12} A_4 A_5 \ddot{\textbf{v}} \dot{\textbf{v}}+\frac{1}{8} A_5^2 \ddot{\textbf{v}}^2 \nonumber\\
-&\frac{1}{3} A_2 A_5 \textbf{v}\dddot{\textbf{v}}-\frac{1}{4} A_4 A_5
\textbf{v}\dddot{\textbf{v}}+\frac{1}{48} A_5^2
\textbf{v}\textbf{v}^{(4)}+\frac{A_3^2}{2}-\frac{A_1^2}{2},
\end{align}
\begin{align}
\nonumber
Q_{4}=&-\frac{5}{8} a_4 A_5^2-\frac{1}{512} A_5^2 \dot{\textbf{v}}^4-\frac{3}{64} A_1 A_5 \dot{\textbf{v}}^2-\frac{1}{64} A_3 A_5 \dot{\textbf{v}}^2+\frac{3}{64} A_5^2 \ddot{\textbf{v}} \dot{\textbf{v}}-\frac{1}{48} A_2 A_5 \ddot{\textbf{v}} \dot{\textbf{v}} \nonumber\\
-&\frac{1}{12} A_4 A_5 \ddot{\textbf{v}} \dot{\textbf{v}}+\frac{1}{24} A_5^2 \ddot{\textbf{v}}^2-\frac{1}{24}
A_2 A_5 \textbf{v}\dddot{\textbf{v}}-\frac{1}{48} A_4 A_5 \textbf{v}\dddot{\textbf{v}}+\frac{1}{192}
A_5^2 \textbf{v}\textbf{v}^{(4)},
\end{align}
\begin{align}
\nonumber
Q_{5}=&\frac{10}{3} a_4 A_2 A_5+\frac{5}{3} a_4 A_4 A_5-\frac{1}{288} 5 A_2 A_5 \dot{\textbf{v}}^4-\frac{23}{576} A_4 A_5 \dot{\textbf{v}}^4+\frac{59}{576} A_5^2 \ddot{\textbf{v}} \dot{\textbf{v}}^3 \nonumber\\
+&\frac{1}{4} A_1 A_2 \dot{\textbf{v}}^2+\frac{1}{12} A_2 A_3 \dot{\textbf{v}}^2+\frac{1}{8} A_1 A_4 \dot{\textbf{v}}^2+\frac{1}{24} A_3 A_4 \dot{\textbf{v}}^2+\frac{1}{9} A_2^2 \ddot{\textbf{v}} \dot{\textbf{v}} \nonumber\\
+&\frac{2}{9} A_4^2 \ddot{\textbf{v}} \dot{\textbf{v}}+\frac{1}{2} A_2 A_4 \ddot{\textbf{v}} \dot{\textbf{v}}+\frac{1}{24} A_1 A_5 \ddot{\textbf{v}} \dot{\textbf{v}}-\frac{1}{4} A_2 A_5 \ddot{\textbf{v}} \dot{\textbf{v}}+\frac{1}{24} A_3 A_5 \ddot{\textbf{v}} \dot{\textbf{v}} \nonumber\\
-&\frac{1}{8} A_4 A_5 \ddot{\textbf{v}} \dot{\textbf{v}}-\frac{1}{36} A_2 A_5 \dddot{\textbf{v}} \dot{\textbf{v}}-\frac{1}{72} A_4 A_5 \dddot{\textbf{v}} \dot{\textbf{v}}+\frac{1}{288} A_5^2 \textbf{v}^{(4)} \dot{\textbf{v}}-\frac{2}{9} A_2 A_5 \ddot{\textbf{v}}^2 \nonumber\\
-&\frac{1}{9} A_4 A_5 \ddot{\textbf{v}}^2+\frac{2}{9} A_2^2
\textbf{v}\dddot{\textbf{v}}+\frac{1}{18} A_4^2\textbf{v}\dddot{\textbf{v}}+\frac{2}{9} A_2 A_4
\textbf{v}\dddot{\textbf{v}}-\frac{1}{36} A_2 A_5 \textbf{v}\textbf{v}^{(4)}-\frac{1}{72} A_4 A_5
\textbf{v}\textbf{v}^{(4)},
\end{align}
\begin{align}
\nonumber
Q_{6}=&5 a_4 A_2 A_5+5 a_4 A_4 A_5+\frac{17}{288} A_5^2 \dot{\textbf{v}}^2 \textbf{v}\dddot{\textbf{v}}+\frac{37}{192} A_2 A_5 \dot{\textbf{v}}^4-\frac{5}{192} A_4 A_5 \dot{\textbf{v}}^4 \nonumber\\
+&\frac{125}{288} A_5^2 \ddot{\textbf{v}} \dot{\textbf{v}}^3+\frac{5}{8} A_1 A_2 \dot{\textbf{v}}^2+\frac{3}{8} A_2 A_3 \dot{\textbf{v}}^2+\frac{3}{8} A_1 A_4 \dot{\textbf{v}}^2+\frac{1}{8} A_3 A_4 \dot{\textbf{v}}^2 \nonumber\\
+&\frac{1}{72} A_5^2 \dot{\textbf{v}}\ddot{\textbf{v}} \dot{\textbf{v}}^2+\frac{1}{6} A_2^2 \ddot{\textbf{v}} \dot{\textbf{v}}+\frac{2}{3} A_4^2 \ddot{\textbf{v}} \dot{\textbf{v}}+\frac{5}{6} A_2 A_4 \ddot{\textbf{v}} \dot{\textbf{v}}+\frac{5}{12} A_1 A_5 \ddot{\textbf{v}} \dot{\textbf{v}} \nonumber\\
-&\frac{3}{8} A_2 A_5 \ddot{\textbf{v}} \dot{\textbf{v}}+\frac{1}{3} A_3 A_5 \ddot{\textbf{v}} \dot{\textbf{v}}-\frac{3}{8} A_4 A_5 \ddot{\textbf{v}} \dot{\textbf{v}}-\frac{7}{24} A_2 A_5 \ddot{\textbf{v}}^2-\frac{7}{24} A_4 A_5 \ddot{\textbf{v}}^2 \nonumber\\
-&\frac{1}{72} A_5^2 \ddot{\textbf{v}}\dddot{\textbf{v}}+\frac{1}{3} A_2^2 \textbf{v}\dddot{\textbf{v}}+\frac{1}{6} A_4^2 \textbf{v}\dddot{\textbf{v}}+\frac{1}{2} A_2 A_4 \textbf{v}\dddot{\textbf{v}}+\frac{1}{12} A_1 A_5 \textbf{v}\dddot{\textbf{v}} \nonumber\\
+&\frac{1}{12} A_3 A_5 \textbf{v}\dddot{\textbf{v}}-\frac{1}{24} A_2
A_5 \textbf{v}\textbf{v}^{(4)}-\frac{1}{24} A_4 A_5
\textbf{v}\textbf{v}^{(4)},
\end{align}
\begin{align}
\nonumber
Q_{7}=&-\frac{1}{288} 31 A_2 A_5 \dot{\textbf{v}}^6+\frac{41}{576} A_4 A_5 \dot{\textbf{v}}^6+\frac{1}{12} A_1 A_2 \dot{\textbf{v}}^4-\frac{1}{12} A_2 A_3 \dot{\textbf{v}}^4+\frac{11}{48} A_1 A_4 \dot{\textbf{v}}^4\\
+&\frac{7}{48} A_3 A_4 \dot{\textbf{v}}^4+\frac{1}{128} A_4 A_5 \dot{\textbf{v}}^{2} \dot{\textbf{v}}^4-\frac{65 A_5^2 \textbf{v}\dddot{\textbf{v}} \dot{\textbf{v}}^4}{1536}-\frac{979 A_5^2 \dot{\textbf{v}}\ddot{\textbf{v}} \dot{\textbf{v}}^4}{2304}-\frac{1}{9} A_2^2 \ddot{\textbf{v}}\dot{\textbf{v}} \dot{\textbf{v}}^2 \nonumber\\
-&\frac{1}{18} A_2 A_4 \ddot{\textbf{v}}\dot{\textbf{v}} \dot{\textbf{v}}^2-\frac{19}{24} A_1 A_5 \ddot{\textbf{v}}\dot{\textbf{v}} \dot{\textbf{v}}^2-\frac{1}{4} A_2 A_5 \ddot{\textbf{v}}\dot{\textbf{v}} \dot{\textbf{v}}^2-\frac{19}{24} A_3 A_5 \ddot{\textbf{v}}\dot{\textbf{v}} \dot{\textbf{v}}^2-\frac{1}{8} A_4 A_5 \ddot{\textbf{v}}\dot{\textbf{v}} \dot{\textbf{v}}^2 \nonumber\\
+&\frac{23}{144} A_2 A_5 \dddot{\textbf{v}}\dot{\textbf{v}} \dot{\textbf{v}}^2+\frac{41}{288} A_4 A_5 \dddot{\textbf{v}}\dot{\textbf{v}} \dot{\textbf{v}}^2-\frac{17 A_5^2 \textbf{v}^{(4)}\dot{\textbf{v}} \dot{\textbf{v}}^2}{1152}+\frac{17}{24} A_2 A_5 \ddot{\textbf{v}}^2 \dot{\textbf{v}}^2+\frac{17}{48} A_4 A_5 \ddot{\textbf{v}}^2 \dot{\textbf{v}}^2 \nonumber\\
-&\frac{67}{384} A_5^2 \ddot{\textbf{v}} \dddot{\textbf{v}} \dot{\textbf{v}}^2+\frac{1}{24} A_4^2 \dot{\textbf{v}}\ddot{\textbf{v}}\dot{\textbf{v}}^2+\frac{1}{12} A_2 A_4 \dot{\textbf{v}}\ddot{\textbf{v}} \dot{\textbf{v}}^2-\frac{1}{64} A_4 A_5 \dot{\textbf{v}}\dddot{\textbf{v}} \dot{\textbf{v}}^2-\frac{1}{64} A_1 A_5 \textbf{v}\dddot{\textbf{v}} \dot{\textbf{v}}^2 \nonumber\\
-&\frac{3}{64} A_3 A_5 \textbf{v}\dddot{\textbf{v}} \dot{\textbf{v}}^2+\frac{1}{6} A_1 A_2 \dddot{\textbf{v}} \dot{\textbf{v}}+\frac{1}{6} A_2 A_3 \dddot{\textbf{v}} \dot{\textbf{v}}+\frac{1}{12} A_1 A_4 \dddot{\textbf{v}} \dot{\textbf{v}}+\frac{1}{12} A_3 A_4 \dddot{\textbf{v}} \dot{\textbf{v}} \nonumber\\
-&\frac{1}{48} A_1 A_5 \textbf{v}^{(4)} \dot{\textbf{v}}-\frac{1}{48} A_3 A_5 \textbf{v}^{(4)} \dot{\textbf{v}}-\frac{3}{64} A_5^2 \ddot{\textbf{v}}\dot{\textbf{v}} \textbf{v}\dddot{\textbf{v}}+\frac{1}{48} A_2 A_5 \ddot{\textbf{v}}\dot{\textbf{v}}\textbf{v}\dddot{\textbf{v}}+\frac{1}{12} A_4 A_5\ddot{\textbf{v}}\dot{\textbf{v}} \textbf{v}\dddot{\textbf{v}}  \nonumber\\
+&\frac{1}{3} A_1 A_2 \ddot{\textbf{v}}^2+\frac{1}{3} A_2 A_3 \ddot{\textbf{v}}^2+\frac{1}{6} A_1 A_4 \ddot{\textbf{v}}^2+\frac{1}{6} A_3 A_4 \ddot{\textbf{v}}^2+\frac{1}{24} A_2 A_5 \dddot{\textbf{v}}^2 \nonumber\\
+&\frac{1}{48} A_4 A_5 \dddot{\textbf{v}}^2-\frac{1}{9} A_2 A_5 \dot{\textbf{v}}\ddot{\textbf{v}}^2-\frac{1}{18} A_4 A_5 \dot{\textbf{v}}\ddot{\textbf{v}}^2-\frac{1}{16} A_1 A_5 \ddot{\textbf{v}} \dddot{\textbf{v}}-\frac{1}{16} A_3 A_5 \ddot{\textbf{v}} \dddot{\textbf{v}} \nonumber\\
+&\frac{10}{3} a_4 A_2 A_5 \dot{\textbf{v}}^{2}+\frac{5}{3} a_4 A_4 A_5 \dot{\textbf{v}}^{2}-\frac{1}{4} A_1^2 \dot{\textbf{v}}\ddot{\textbf{v}}-\frac{1}{4} A_3^2 \dot{\textbf{v}}\ddot{\textbf{v}}-\frac{1}{96} A_5^2 \dot{\textbf{v}}^4 \dot{\textbf{v}}\ddot{\textbf{v}}-\frac{1}{2} A_1 A_3 \dot{\textbf{v}}\ddot{\textbf{v}} \nonumber\\
+&\frac{1}{48} A_5^2 \dot{\textbf{v}}\ddot{\textbf{v}}\dot{\textbf{v}}\dddot{\textbf{v}}-\frac{2}{9} A_2^2 \ddot{\textbf{v}}\dddot{\textbf{v}}-\frac{1}{18} A_4^2 \ddot{\textbf{v}}\dddot{\textbf{v}}-\frac{2}{9} A_2 A_4 V_{23}+\frac{1}{36} A_2 A_5 \ddot{\textbf{v}}\textbf{v}^{(4)}+\frac{1}{72} A_4 A_5 \ddot{\textbf{v}}\textbf{v}^{(4)} \nonumber\\
-&\frac{1}{192} A_5^2 \dddot{\textbf{v}}\textbf{v}^{(4)}+\frac{5}{8}
a_4 A_5^2 \textbf{v}\dddot{\textbf{v}}-\frac{1}{24} A_5^2 V_2^2
\textbf{v}\dddot{\textbf{v}},
\end{align}
\begin{align}
\nonumber
Q_{8}=&-\frac{10}{3} a_4 A_5^2 \dot{\textbf{v}}^{2}+\frac{1}{16} A_4 A_5 \dot{\textbf{v}}^2 \textbf{v}\dddot{\textbf{v}}-\frac{5}{48} A_5^2 \ddot{\textbf{v}} \dot{\textbf{v}} \textbf{v}\dddot{\textbf{v}}+\frac{85}{144} A_5^2 \dot{\textbf{v}}^6-\frac{1}{3} A_2^2 \dot{\textbf{v}}^4 \nonumber\\
-&\frac{3}{8} A_4^2 \dot{\textbf{v}}^4-\frac{13}{24} A_2 A_4 \dot{\textbf{v}}^4+\frac{7}{12} A_1 A_5 \dot{\textbf{v}}^4+\frac{19}{48} A_3 A_5 \dot{\textbf{v}}^4-\frac{1}{96} A_5^2 \dot{\textbf{v}}^{2} \dot{\textbf{v}}^4 \nonumber\\
+&\frac{1}{4} A_5^2 \ddot{\textbf{v}} \dot{\textbf{v}}^3+\frac{13}{32} A_2 A_5 \ddot{\textbf{v}} \dot{\textbf{v}}^3+\frac{245}{288} A_4 A_5 \ddot{\textbf{v}}\dot{\textbf{v}} \dot{\textbf{v}}^2+\frac{13}{288} A_5^2 \dddot{\textbf{v}} \dot{\textbf{v}}^3-\frac{17}{24} A_5^2 V_2^2 \dot{\textbf{v}}^2 \nonumber\\
-&\frac{1}{9} A_2 A_5 \dot{\textbf{v}}\ddot{\textbf{v}} \dot{\textbf{v}}^2-\frac{5}{36} A_4 A_5 \dot{\textbf{v}}\ddot{\textbf{v}}\dot{\textbf{v}}^2+\frac{1}{48} A_5^2 \dot{\textbf{v}}\dddot{\textbf{v}} \dot{\textbf{v}}^2-\frac{1}{3} A_2^2 \dddot{\textbf{v}} \dot{\textbf{v}}-\frac{1}{6} A_4^2 \dddot{\textbf{v}} \dot{\textbf{v}} \nonumber\\
-&\frac{1}{2} A_2 A_4 \dddot{\textbf{v}}\dot{\textbf{v}}-\frac{1}{12} A_1 A_5 \dddot{\textbf{v}}\dot{\textbf{v}}-\frac{5}{24} A_3 A_5 \dddot{\textbf{v}} \dot{\textbf{v}}+\frac{1}{24} A_2 A_5 \textbf{v}^{(4)}  \dot{\textbf{v}}+\frac{1}{24} A_4 A_5  \textbf{v}^{(4)}  \dot{\textbf{v}}-\frac{1}{3} A_2^2  \ddot{\textbf{v}}^2 \nonumber\\
-&\frac{1}{6} A_4^2  \ddot{\textbf{v}}^2-\frac{1}{2} A_2 A_4  \ddot{\textbf{v}}^2-\frac{1}{3} A_1 A_5  \ddot{\textbf{v}}^2-\frac{1}{3} A_3 A_5  \ddot{\textbf{v}}^2-\frac{1}{48} A_5^2  \dddot{\textbf{v}}^2+\frac{1}{9} A_5^2 ( \dot{\textbf{v}} \ddot{\textbf{v}})^2 \nonumber\\
+&\frac{1}{16} A_2 A_5  \ddot{\textbf{v}}  \dddot{\textbf{v}}+\frac{1}{16} A_4 A_5  \ddot{\textbf{v}}  \dddot{\textbf{v}}-\frac{1}{2} A_3^2  \dot{\textbf{v}}^{2}-\frac{1}{2} A_1 A_3  \dot{\textbf{v}}^{2}+\frac{3}{4} A_1 A_2  \dot{\textbf{v}} \ddot{\textbf{v}}+\frac{17}{12} A_2 A_3  \dot{\textbf{v}} \ddot{\textbf{v}} \nonumber\\
+&\frac{3}{4} A_1 A_4  \dot{\textbf{v}}
\ddot{\textbf{v}}+\frac{13}{12} A_3 A_4
 \dot{\textbf{v}} \ddot{\textbf{v}}+\frac{1}{3} A_2 A_5  \ddot{\textbf{v}} \dddot{\textbf{v}}+\frac{1}{6} A_4 A_5
 \ddot{\textbf{v}} \dddot{\textbf{v}}-\frac{1}{36} A_5^2  \ddot{\textbf{v}}\textbf{v}^{(4)},
\end{align}
\begin{align}
\nonumber
Q_{9}=&-\frac{5}{64} A_5^2  \dot{\textbf{v}}^2 \textbf{v}\dddot{\textbf{v}}+\frac{1}{4} (-5) A_2 A_5 \dot{\textbf{v}}^4-\frac{11}{12} A_4 A_5 \dot{\textbf{v}}^4+\frac{23}{72} A_5^2 \ddot{\textbf{v}} \dot{\textbf{v}}^3+\frac{1}{9} A_5^2 \dot{\textbf{v}}\ddot{\textbf{v}} \dot{\textbf{v}}^2 \nonumber\\
+&\frac{1}{3} A_2 A_5 \dddot{\textbf{v}} \dot{\textbf{v}}+\frac{3}{8} A_4 A_5 \dddot{\textbf{v}} \dot{\textbf{v}}-\frac{1}{48} A_5^2 \textbf{v}^{(4)}\dot{\textbf{v}}+\frac{2}{3} A_2 A_5 \ddot{\textbf{v}}^2+\frac{1}{2} A_4 A_5 \ddot{\textbf{v}}^2 \nonumber\\
-&\frac{1}{16} A_5^2 \ddot{\textbf{v}} \dddot{\textbf{v}}+\frac{2}{3} A_1 A_2 \dot{\textbf{v}}^{2}+\frac{1}{3} A_2 A_3 \dot{\textbf{v}}^{2}+\frac{5}{6} A_1 A_4 \dot{\textbf{v}}^{2}+\frac{7}{6} A_3 A_4 \dot{\textbf{v}}^{2}-\frac{1}{2} A_2^2 \dot{\textbf{v}}\ddot{\textbf{v}}-\frac{5}{6} A_4^2 \dot{\textbf{v}}\ddot{\textbf{v}} \nonumber\\
-&\frac{5}{3} A_2 A_4 \dot{\textbf{v}}\ddot{\textbf{v}}-\frac{1}{2}
A_1 A_5 \dot{\textbf{v}}\ddot{\textbf{v}}-\frac{7}{6} A_3 A_5
\dot{\textbf{v}}\ddot{\textbf{v}}-\frac{1}{9} A_5^2
\ddot{\textbf{v}}\dddot{\textbf{v}}+\frac{1}{8} A_1 A_5
\textbf{v}\dddot{\textbf{v}}-\frac{1}{8} A_3 A_5
\textbf{v}\dddot{\textbf{v}},
\end{align}
\begin{equation}
Q_{10}=\frac{43}{72} A_5^2 \dot{\textbf{v}}
\ddot{\textbf{v}}+\frac{1}{4} A_2 A_5 \dot{\textbf{v}}^2-\frac{1}{8}
A_4 A_5 \dot{\textbf{v}}^2+\frac{1}{9} A_5^2
\textbf{v}\dddot{\textbf{v}}+\frac{2 A_1 A_2}{3}+\frac{A_1
A_4}{3}-\frac{2 A_2 A_3}{3}-\frac{A_3 A_4}{3},
\end{equation}
\begin{align}
Q_{11}=&\frac{1}{3} A_5^2 \dot{\textbf{v}} \ddot{\textbf{v}}+\frac{9}{16} A_2 A_5 \dot{\textbf{v}}^2-\frac{9}{16} A_4 A_5 \dot{\textbf{v}}^2+\frac{1}{12} A_5^2 \textbf{v}\dddot{\textbf{v}}+\frac{3 A_1 A_2}{2} \nonumber\\
+&\frac{A_1 A_4}{2}-\frac{A_2 A_3}{2}-\frac{3 A_3 A_4}{2},
\end{align}
\begin{align}
Q_{12}=\frac{5}{48} A_5^2 \dot{\textbf{v}} \ddot{\textbf{v}}+\frac{1}{16} A_2 A_5
\dot{\textbf{v}}^2+\frac{1}{48} A_5^2 \textbf{v}\dddot{\textbf{v}},
\end{align}
\begin{align}
Q_{13}=-\frac{1}{12} 11 A_5^2 \dot{\textbf{v}}\ddot{\textbf{v}}+\frac{1}{2} A_4 A_5 \dot{\textbf{v}}^2+A_2
A_5 \dot{\textbf{v}}^{2}-\frac{1}{8} A_5^2 \textbf{v}\dddot{\textbf{v}},
\end{align}
\begin{align}
Q_{14}=&\frac{9}{16} A_5^2 \dot{\textbf{v}}^4-\frac{2}{3} A_4^2 \dot{\textbf{v}}^2-\frac{5}{24} A_5^2 \dddot{\textbf{v}} \dot{\textbf{v}}-\frac{1}{3} A_5^2 \ddot{\textbf{v}}^2-\frac{2}{3} A_2^2 \dot{\textbf{v}}^{2}-\frac{2}{3} A_2 A_4 \dot{\textbf{v}}^{2} \nonumber\\
-&\frac{7}{6} A_1 A_5 \dot{\textbf{v}}^{2}-\frac{1}{3} A_3 A_5
\dot{\textbf{v}}^{2}+\frac{17}{12} A_2 A_5
\dot{\textbf{v}}\ddot{\textbf{v}}+\frac{7}{4} A_4 A_5
\dot{\textbf{v}}\ddot{\textbf{v}}-\frac{1}{8} A_2 A_5
\textbf{v}\dddot{\textbf{v}}+\frac{1}{8} A_4 A_5
\textbf{v}\dddot{\textbf{v}},
\end{align}
\begin{align}
Q_{15}=\frac{1}{64} A_5^2 \dot{\textbf{v}}^2-\frac{A_1
A_5}{8}+\frac{A_3 A_5}{8}, \ \ \ \ \ \ \ \ \ \ Q_{16}=\frac{A_2
A_5}{8}-\frac{A_4 A_5}{8},
\end{align}
\begin{align}
Q_{17}=\frac{1}{6} A_5^2 \dot{\textbf{v}}^2-\frac{1}{3} 2 A_2^2+\frac{A_4
A_2}{3}+\frac{A_4^2}{3}+\frac{2 A_3 A_5}{3}-\frac{2 A_1 A_5}{3},
\end{align}
\begin{align}
Q_{18}=\frac{A_5^2}{8}, \  \ \ \ \ \ \ Q_{19}=-A_4 A_5, \ \ \ \ \ \
\ Q_{20}=\frac{2 A_5^2}{3},
\end{align}
\begin{align}
Q_{21}=\frac{1}{2} A_5^2 \dot{\textbf{v}}^2-A_2^2+A_4^2+A_3 A_5, \ \
\ \ \ \ \  Q_{22}=-\frac{1}{2} A_2 A_5-\frac{3 A_4 A_5}{2},
\end{align}
\begin{align}
Q_{23}=\frac{1}{6} A_5^2 \dot{\textbf{v}}^{2}, \ \ \ \ \ \ \ \
Q_{24}=\frac{A_5^2}{2},
\end{align}
\begin{align}
Q_{25}&=-5 a_4 A_5+\frac{65}{192} A_5 \dot{\textbf{v}}^4+\frac{1}{8}
A_1 \dot{\textbf{v}}^2+\frac{3}{8} A_3 \dot{\textbf{v}}^2 +\frac{3
A_5 \dot{\textbf{v}}\ddot{\textbf{v}}}{8}+\frac{A_5
\ddot{\textbf{v}}^{2}}{3}-\frac{2 A_4
\dot{\textbf{v}}\ddot{\textbf{v}}}{3}-\frac{A_2
\dot{\textbf{v}}\ddot{\textbf{v}}}{6},
\end{align}
\begin{align}
Q_{26}=-\frac{1}{3} A_2 \dot{\textbf{v}}^{2}-\frac{13 A_4
\dot{\textbf{v}}^{2}}{24}+\frac{7 A_5
\dot{\textbf{v}}\ddot{\textbf{v}}}{8},\;\;Q_{27}=\frac{5 A_5
\dot{\textbf{v}}\ddot{\textbf{v}}}{6}-\frac{1}{2} A_4
\dot{\textbf{v}}^2,\;\;Q_{28}=\frac{17 A_5
\dot{\textbf{v}}^{2}}{48}+\frac{A_1}{2}+\frac{A_3}{2},
\end{align}
\begin{align}
Q_{29}=\frac{A_5}{6}, \ \ Q_{30}=\frac{A_5}{24}, \ \ Q_{31}=-A_5
\dot{\textbf{v}}^{2}, \ \ Q_{32}=-\frac{A_2}{3}-\frac{A_4}{6}, \ \
Q_{33}=\frac{A_5}{2},
\end{align}
\begin{align}
Q_{34}=-\frac{A_2}{2}-\frac{A_4}{2}, \ \ \ \ \ Q_{35}=\frac{5}{8}
A_5 \dot{\textbf{v}}^2-A_1+A_3, \ \ \ \ \ Q_{36}=A_2-A_4.
\end{align}
Now the equations of motion (\ref{eq1}) need to be renormalized,
thus as we will need in Section \ref{Ka}, a set of four-vectors can
be built in an orthonormal basis from the set $\{\dot{B}_{\mu},
v_{\mu}, \dot{v}_{\mu }, \ddot{v}_{\mu},  \dddot{v}_{\mu}\}$, where
these vectors have the following form
\begin{equation}
\dot{\widehat{B}}_{\mu}=\frac{\dot{B}_{\mu}}{\dot{B}}.
\end{equation}
Here $\dot{\bf B}$ is the norm  of $\dot{B}_\mu$. The first
orthonormal vector is given by
\begin{equation}\label{u1}
\widehat{u}_{1 \mu
}=\frac{v_{\mu}-(\textbf{v}\dot{\widehat{\textbf{B}}})\dot{\widehat{B}}_{\mu}}{u_{1}},
\end{equation}
with contraction of the form
$\left|u_{1}\right|^{2}=u_{1}^{2}=1-\frac{\textbf{v}\dot{\textbf{B}}}{\dot{B}^{2}}$.
For the second normalized vector in terms of $\dot{v}_{\mu}$ we find
\begin{equation}\label{u2}
\widehat{u}_{2 \mu }=\frac{1}{u_{2}}\bigg[
\dot{v}_{\mu}-(\dot{\textbf{v}}\dot{\widehat{\textbf{B}}})
\dot{\widehat{B}}_{\mu}+\frac{(\textbf{v}\dot{\widehat{\textbf{B}}})(\dot{\textbf{v}}\dot{\widehat{\textbf{B}}})}{u_{1}^{2}}\left(
v_{\mu}-(\textbf{v}\dot{\widehat{\textbf{B}}})\dot{\widehat{B}}_{\mu}\right)\bigg],
\end{equation}
whose magnitude can be expressed in the simple form
\begin{equation}
u_{2}^{2}=\dot{\textbf{v}}^{2}-(\dot{\textbf{v}}\dot{\widehat{\textbf{B}}})^{2}-
\frac{(\textbf{v}\dot{\widehat{\textbf{B}}})^{2}(\dot{\textbf{v}}\dot{\widehat{\textbf{B}}})^{2}}{u_{1}^{2}}.
\end{equation}
The vector $u_{3\mu}$ is expressed in terms of $\ddot{v}_{\mu}$ and
it is given as follows
\begin{align}\label{u3}
\nonumber
u_{3 \mu }=&\ddot{v}_{\mu }-\frac{\textbf{B}\ddot{\textbf{v}} B_{\mu }}{B^2}-\frac{\textbf{v}\ddot{\textbf{v}}-(\textbf{v}\dot{\widehat{\textbf{B}}})(\ddot{\textbf{v}}\dot{\widehat{\textbf{B}}})}{u_{1}^{2}}\left( v_{\mu}-(\textbf{v}\dot{\widehat{\textbf{B}}})\dot{\widehat{B}}_{\mu}\right) \nonumber\\
-&\frac{1}{u_{2}^{2}}\left[ \dot{\textbf{v}}\ddot{\textbf{v}}-(\dot{\textbf{v}}\dot{\widehat{\textbf{B}}})(\ddot{\textbf{v}}\dot{\widehat{\textbf{B}}})+\frac{(\textbf{v}
\dot{\widehat{\textbf{B}}})(\dot{\textbf{v}}\dot{\widehat{\textbf{B}}})}{u_{1}^{2}}\left( \textbf{v}\ddot{\textbf{v}}-(\textbf{v}\dot{\widehat{\textbf{B}}})
\ddot{\textbf{v}}\dot{\widehat{\textbf{B}}}\right) \right] \nonumber\\
\times &\left[
\dot{v}_{\mu}-(\dot{\textbf{v}}\dot{\widehat{\textbf{B}}})
\dot{\widehat{B}}_{\mu}+\frac{(\textbf{v}\dot{\widehat{\textbf{B}}})(\dot{\textbf{v}}\dot{\widehat{\textbf{B}}})}{u_{1}^{2}}\left(
v_{\mu}-(\textbf{v}\dot{\widehat{\textbf{B}}})\dot{\widehat{B}}_{\mu}\right)
\right].
\end{align}
Consequently we have
\begin{align}
\nonumber
u_{3}^{2}=&\ddot{\textbf{v}}^{2}-(\ddot{\textbf{v}}\dot{\widehat{\textbf{B}}})^{2}-\frac{1}{u_{1}^{2}}\left[  \textbf{v}\ddot{\textbf{v}}-(\textbf{v}\dot{\widehat{\textbf{B}}})(\ddot{\textbf{v}}\dot{\widehat{\textbf{B}}})\right]^{2} \nonumber\\
-&\frac{1}{u_{2}^{2}}\left[
\dot{\textbf{v}}\ddot{\textbf{v}}-(\dot{\textbf{v}}\dot{\widehat{\textbf{B}}})(\ddot{\textbf{v}}\dot{\widehat{\textbf{B}}})
+\frac{(\textbf{v}\dot{\widehat{\textbf{B}}})(\dot{\textbf{v}}\dot{\widehat{\textbf{B}}})}{u_{1}^{2}}\left(
\textbf{v}\ddot{\textbf{v}}
-(\textbf{v}\dot{\widehat{\textbf{B}}})(\ddot{\textbf{v}}\dot{\widehat{\textbf{B}}})\right)
\right]^{2}.
\end{align}
All these vectors are constructed in such a way that the conditions
of orthnormalization are fulfilled in all the expressions of vectors
$u_{i}$. Here the index $i=1,2,3,4 $ stands for a label of the
basis. The index $j=1,2,3,4$ in $v^{(j)}$ stands to the derivative
with respect to the proper-time of the particle. Analogously the
last more general orthonormal vector that we build in this basis is
\begin{align}\label{u4}
\nonumber
u_{4 \mu }=&\dddot{v}_{ \mu
}-(\dddot{\textbf{v}}\dot{\widehat{\textbf{B}}})\dot{\widehat{B}}_{\mu}-\frac{\left[
\textbf{v}\dddot{\textbf{v}}-(\textbf{v}\dot{\widehat{\textbf{B}}})(\dddot{\textbf{v}}\dot{\widehat{\textbf{B}}})\right]
}{u_{1}^{2}}\left[v_{\mu}-(\textbf{V}\dot{\widehat{\textbf{B}}})
\dot{\widehat{B}}_{\mu} \right] \nonumber \\
-&\frac{1}{u_{2}^{2}}\left[ \dot{\textbf{v}}\dddot{\textbf{v}}-(\dot{\textbf{v}}\dot{\widehat{\textbf{B}}})(\dddot{\textbf{v}}\dot{\widehat{\textbf{B}}})
+\frac{(\textbf{v}\dot{\widehat{\textbf{B}}})(\dddot{\textbf{v}}\dot{\widehat{\textbf{B}}})}{u_{1}^{2}}\left( \textbf{v}\dddot{\textbf{v}}-(\textbf{v}\dot{\widehat{\textbf{B}}})(\dddot{\textbf{v}}\dot{\widehat{\textbf{B}}})\right) \right] \nonumber\\
\times &\left[ \dot{v}_{\mu}-(\dot{\textbf{v}}\dot{\widehat{\textbf{B}}})
\dot{\widehat{B}}_{\mu}+\frac{(\textbf{v}\dot{\widehat{\textbf{B}}})(\dot{\textbf{v}}\dot{\widehat{\textbf{B}}})}{u_{1}^{2}}\left( v_{\mu}-(\textbf{v}\dot{\widehat{\textbf{B}}})\dot{\widehat{B}}_{\mu}\right) \right]  \nonumber\\
-&\frac{1}{u_{3}^{2}}\left\lbrace
\ddot{\textbf{v}}\dddot{\textbf{v}}-(\ddot{\textbf{v}}\dot{\widehat{\textbf{B}}})(\dddot{\textbf{v}}\dot{\widehat{\textbf{B}}})
-\frac{\textbf{v}\ddot{\textbf{v}}-(\textbf{v}\dot{\widehat{\textbf{B}}})(\ddot{\textbf{v}}\dot{\widehat{\textbf{B}}})}{u_{1}^{2}}\left( \textbf{v}\dddot{\textbf{v}}-(\textbf{v}\dot{\widehat{\textbf{B}}})(\dddot{\textbf{v}}\dot{\widehat{\textbf{B}}})\right) \right.  \nonumber\\
-&\frac{1}{u_{2}^{2}}\left[
\dot{\textbf{v}}\ddot{\textbf{v}}-(\dot{\textbf{v}}\dot{\widehat{\textbf{B}}})(\ddot{\textbf{v}}\dot{\widehat{\textbf{B}}})
+\frac{(\textbf{v}\dot{\widehat{\textbf{B}}})(\dot{\textbf{v}}\dot{\widehat{\textbf{B}}})}{u_{1}^{2}}\left(
\textbf{v}\ddot{\textbf{v}}-(\textbf{v}\dot{\widehat{\textbf{B}}})
\ddot{\textbf{v}}\dot{\widehat{\textbf{B}}}\right) \right] \nonumber \\
\times &\left. \left[ \dot{\textbf{v}}\dddot{\textbf{v}}-(\dot{\textbf{v}}\dot{\widehat{\textbf{B}}})(\dddot{\textbf{v}}\dot{\widehat{\textbf{B}}})
+\frac{(\textbf{v}\dot{\widehat{\textbf{B}}})(\dot{\textbf{v}}\dot{\widehat{\textbf{B}}})}{u_{1}^{2}}\left( \textbf{v}\dddot{\textbf{v}}-(\textbf{v}\dot{\widehat{\textbf{B}}})(\dddot{\textbf{v}}\dot{\widehat{\textbf{B}}})\right) \right] \right\rbrace \nonumber\\
\times &\left\lbrace  \ddot{v}_{\mu}-(\ddot{\textbf{v}}\dot{\widehat{\textbf{B}}})
\dot{\widehat{B}}_{\mu}-\frac{\textbf{v}\ddot{\textbf{v}}
-(\textbf{v}\dot{\widehat{\textbf{B}}})(\ddot{\textbf{v}}\dot{\widehat{\textbf{B}}})}{u_{1}^{2}}\left( v_{\mu}-(\textbf{v}\dot{\widehat{\textbf{B}}})\dot{\widehat{B}}_{\mu}\right) \right.  \nonumber\\
-&\frac{1}{u_{2}^{2}}\left[
\dot{\textbf{v}}\ddot{\textbf{v}}-(\dot{\textbf{v}}\dot{\widehat{\textbf{B}}})(\ddot{\textbf{v}}\dot{\widehat{\textbf{B}}})
+\frac{(\textbf{v}\dot{\widehat{\textbf{B}}})(\dot{\textbf{v}}\dot{\widehat{\textbf{B}}})}{u_{1}^{2}}\left(
\textbf{v}\ddot{\textbf{v}}
-(\textbf{v}\dot{\widehat{\textbf{B}}})\ddot{\textbf{v}}\dot{\widehat{\textbf{B}}}\right) \right]  \nonumber\\
\times &\left. \left[
\dot{v}_{\mu}-(\dot{\textbf{v}}\dot{\widehat{\textbf{B}}})
\dot{\widehat{B}}_{\mu}+\frac{(\textbf{v}\dot{\widehat{\textbf{B}}})(\dot{\textbf{v}}\dot{\widehat{\textbf{B}}})}{u_{1}^{2}}\left(
v_{\mu}-(\textbf{v}\dot{\widehat{\textbf{B}}})\dot{\widehat{B}}_{\mu}\right)
\right] \right\rbrace.
\end{align}
Now  with Eqs. (\ref{u1}), (\ref{u2}), (\ref{u3}), (\ref{u4}) and
the conditions of orthonormalization, it can be proved that
$v_{\mu}^{(i)}\widehat{u}_{4}^{\mu}=v^{(i)}\widehat{u}_{4}=0$ with
$i=0,1,2,3$ and $v^{(0)}_{\mu}=v_{\mu}$. With Eqs. (\ref{u4}) it is
easy to show that the product $\dot{\textbf{B}}\widehat{u}_{4}=0$,
where $\dot{B}_\mu$ is taken from  Eq. (\ref{eq1}) and this give
rises to
\begin{equation}\label{UL}
\begin{split}
\left(
\varepsilon^{-1}Q_{36}+Q_{35}\right)L_{\nu}v^{\nu}-A_{4}L_{\nu}\dot{v}^{\nu}=0,
\end{split}
\end{equation}
where  $L_{\nu}:=f_{\mu\nu}\widehat{u}_{4}^{\mu}$.  This expression
is used in the subsections (\ref{SUB2}) and (\ref{SUB3}) to obtain
the equations of motion.

\bibliography{octaviostrings}

\begin{thebibliography}{99}

\bibitem{Abraham} M. Abraham. Ann. Phys. (Leypzig) \textbf{10} (1903) 105; H.A. Lorentz
second edition (Dover, N. Y. 1952).

\bibitem{Dirac:1938nz}
  P.~A.~M.~Dirac,
  ``Classical theory of radiating electrons,''
  Proc.\ Roy.\ Soc.\ Lond.\ A {\bf 167}, 148 (1938);
  doi:10.1098/rspa.1938.0124.

\bibitem{Teitelboim} C. Teitelboim, ``Splitting of the Maxwell Tensor:
Radiation Reaction without Advanced Fields,'' Phys. Rev. D {\bf 1}
(1970) 1572; Erratum Phys. Phys. Rev. D {\bf 2} (1970) 1763.

\bibitem{bonnor}  W.B. Bonnor, ``A new equation of motion for a radiating charged particle,''
Proc. R. Soc. Lond.  A \textbf{337} (1974) 591-598.

\bibitem{AresdeParga:1998ti}
  G.~Ares de Parga and R.~Mares,
  ``A generalized equation of motion for a charged point particle,''
  Nuovo Cim.\ B {\bf 113}, 1469 (1998).

\bibitem{Jackson:1998nia}
  J.~D.~Jackson,
  {\it Classical Electrodynamics} (Wiley, New York, 1999), Third
  edition.

\bibitem{FRohrlich}
  F.~Rohrlich,
  {\it Classical Charged Particles} (Addison Wesley, Rodwood City, California, 1990),
  Second edition.

\bibitem{RohrlichAJP2000}
  F.~Rohrlich,
  ``The Self-force and radiation reaction,''
  Am.\ J.\ Phys. {\bf 68} (12), 1109 (2000).

\bibitem{Rohrlich:1999gd}
  F.~Rohrlich,
  ``Classical self-force,''
  Phys.\ Rev.\ D {\bf 60}, 084017 (1999);
  doi:10.1103/PhysRevD.60.084017.


\bibitem{Wong:1970fu}
  S.~K.~Wong,
  ``Field and particle equations for the classical Yang-Mills field and particles with isotopic spin,''
  Nuovo Cim.\ A {\bf 65}, 689 (1970);
  doi:10.1007/BF02892134.

\bibitem{Balachandran:1976ya}
  A.~P.~Balachandran, P.~Salomonson, B.~S.~Skagerstam and J.~O.~Winnberg,
  ``Classical Description of Particle Interacting with Nonabelian Gauge Field,''
  Phys.\ Rev.\ D {\bf 15}, 2308 (1977).
  doi:10.1103/PhysRevD.15.2308

\bibitem{Balachandran:1977ub}
  A.~P.~Balachandran, S.~Borchardt and A.~Stern,
  ``Lagrangian and Hamiltonian Descriptions of Yang-Mills Particles,''
  Phys.\ Rev.\ D {\bf 17}, 3247 (1978).
  doi:10.1103/PhysRevD.17.3247

\bibitem{Arodz:1982fm}
  H.~Arodz,
  ``Colored, Spinning Classical Particle in an External Nonabelian Gauge Field,''
  Phys.\ Lett.\  {\bf 116B}, 251 (1982).
  doi:10.1016/0370-2693(82)90336-7

\bibitem{Bastianelli:2013pta}
  F.~Bastianelli, R.~Bonezzi, O.~Corradini and E.~Latini,
  ``Particles with non abelian charges,''
  JHEP {\bf 1310}, 098 (2013)
  doi:10.1007/JHEP10(2013)098
  [arXiv:1309.1608 [hep-th]].



\bibitem{Heinz:1983nx}
  U.~W.~Heinz,
  ``Kinetic Theory for Nonabelian Plasmas,''
  Phys.\ Rev.\ Lett.\  {\bf 51}, 351 (1983).
  doi:10.1103/PhysRevLett.51.351

\bibitem{Heinz:1985vf}
  U.~W.~Heinz,
  ``Quark - Gluon Transport Theory,''
  Nucl.\ Phys.\ A {\bf 418}, 603C (1984).
  doi:10.1016/0375-9474(84)90579-7

\bibitem{Kelly:1994dh}
  P.~F.~Kelly, Q.~Liu, C.~Lucchesi and C.~Manuel,
  ``Classical transport theory and hard thermal loops in the quark - gluon plasma,''
  Phys.\ Rev.\ D {\bf 50}, 4209 (1994)
  doi:10.1103/PhysRevD.50.4209
  [hep-ph/9406285].

\bibitem{Gyulassy:1994ug}
  M.~Gyulassy and A.~V.~Selikhov,
  ``Developments in QCD transport theory,''
  Nucl.\ Phys.\ A {\bf 566}, 133C (1994).
  doi:10.1016/0375-9474(94)90617-3

\bibitem{Nayak:1996ex}
  G.~C.~Nayak and V.~Ravishankar,
  ``Preequilibrium evolution of nonAbelian plasma,''
  Phys.\ Rev.\ D {\bf 55}, 6877 (1997)
  doi:10.1103/PhysRevD.55.6877
  [hep-th/9610215].

\bibitem{Litim:1999id}
  D.~F.~Litim and C.~Manuel,
  ``Effective transport equations for nonAbelian plasmas,''
  Nucl.\ Phys.\ B {\bf 562}, 237 (1999)
  doi:10.1016/S0550-3213(99)00531-3
  [hep-ph/9906210].

\bibitem{Bistrovic:2002jx}
  B.~Bistrovic, R.~Jackiw, H.~Li, V.~P.~Nair and S.~Y.~Pi,
  ``NonAbelian fluid dynamics in Lagrangian formulation,''
  Phys.\ Rev.\ D {\bf 67}, 025013 (2003)
  doi:10.1103/PhysRevD.67.025013
  [hep-th/0210143].

\bibitem{Arnold:2005vb}
  P.~B.~Arnold, G.~D.~Moore and L.~G.~Yaffe,
  ``The Fate of non-Abelian plasma instabilities in 3+1 dimensions,''
  Phys.\ Rev.\ D {\bf 72}, 054003 (2005)
  doi:10.1103/PhysRevD.72.054003
  [hep-ph/0505212].

\bibitem{Dumitru:2007rp}
  A.~Dumitru, Y.~Nara, B.~Schenke and M.~Strickland,
  ``Jet broadening in unstable non-Abelian plasmas,''
  Phys.\ Rev.\ C {\bf 78}, 024909 (2008)
  doi:10.1103/PhysRevC.78.024909
  [arXiv:0710.1223 [hep-ph]].

\bibitem{PeraltaRamos:2012er}
  J.~Peralta-Ramos and E.~Calzetta,
  ``Effective dynamics of a nonabelian plasma out of equilibrium,''
  Phys.\ Rev.\ D {\bf 86}, 125024 (2012)
  doi:10.1103/PhysRevD.86.125024
  [arXiv:1208.2715 [hep-ph]].

\bibitem{Fernandez-Melgarejo:2016xiv}
  J.~J.~Fernandez-Melgarejo, S.~J.~Rey and P.~Surówka,
  ``A New Approach to Non-Abelian Hydrodynamics,''
  JHEP {\bf 1702}, 122 (2017)
  doi:10.1007/JHEP02(2017)122
  [arXiv:1605.06080 [hep-th]].



\bibitem{Poschl:1998fr}
  W.~Poschl and B.~Muller,
  ``Gluon field dynamics in ultrarelativistic heavy ion collisions: Time evolution on a gauge lattice in (3+1)-dimensions,''
  nucl-th/9808031.

\bibitem{JalilianMarian:2000ad}
  J.~Jalilian-Marian, S.~Jeon and R.~Venugopalan,
  ``Wong's equations and the small x effective action in QCD,''
  Phys.\ Rev.\ D {\bf 63}, 036004 (2001)
  doi:10.1103/PhysRevD.63.036004
  [hep-ph/0003070].

\bibitem{Voronyuk:2015ita}
  V.~Voronyuk, V.~V.~Goloviznin, G.~M.~Zinovjev, W.~Cassing, S.~V.~Molodtsov, A.~M.~Snigirev and V.~D.~Toneev,
  ``Classical Gluon Fields and Collective Dynamics of Color-Charge Systems,''
  Phys.\ Atom.\ Nucl.\  {\bf 78}, no. 2, 312 (2015).
  doi:10.1134/S1063778815010202

\bibitem{Mrowczynski:2016etf}
  S.~Mrowczynski, B.~Schenke and M.~Strickland,
  ``Color instabilities in the quarkgluon plasma,''
  Phys.\ Rept.\  {\bf 682}, 1 (2017)
  doi:10.1016/j.physrep.2017.03.003
  [arXiv:1603.08946 [hep-ph]].

\bibitem{Dumitru:2018vpr}
  A.~Dumitru, G.~A.~Miller and R.~Venugopalan,
  ``Extracting many-body color charge correlators in the proton from exclusive DIS at large Bjorken x,''
  arXiv:1808.02501 [hep-ph].

\bibitem{Goldberger:2016iau}
  W.~D.~Goldberger and A.~K.~Ridgway,
  ``Radiation and the classical double copy for color charges,''
  Phys.\ Rev.\ D {\bf 95}, no. 12, 125010 (2017)
  doi:10.1103/PhysRevD.95.125010
  [arXiv:1611.03493 [hep-th]].

\bibitem{Dzhunushaliev:2018hui}
  V.~Dzhunushaliev, V.~Folomeev and N.~Protsenko,
  ``The motion of color-charged particles as a means of testing the non-Abelian dark matter model,''
  arXiv:1805.11440 [gr-qc].

\bibitem{Drechsler:1981nc}
  W.~Drechsler and A.~Rosenblum,
  ``Equations of Motion and Iteration of Lienard-wiechert Type Solutions in Classical {Yang-Mills} Theory,''
  Phys.\ Lett.\  {\bf 106B}, 81 (1981).
  doi:10.1016/0370-2693(81)91085-6

\bibitem{Kates:1984cn}
  R.~E.~Kates and A.~Rosenblum,
  ``Radiation Damping Of Color In Classical Su(2) Yang-mills Theory,''
  Phys.\ Rev.\ D {\bf 28}, 3066 (1983).
  doi:10.1103/PhysRevD.28.3066

\bibitem{Trautman:1981qd}
  A.~Trautman,
  ``Radiation of Energy and Change in Color of a Point Source of the Yang-Mills Field,''
  Phys.\ Rev.\ Lett.\  {\bf 46}, 875 (1981);
  doi:10.1103/PhysRevLett.46.875

\bibitem{Oh:1985gj}
  C.~H.~Oh, C.~H.~Lai and R.~Teh,
  ``Color Radiation In The Classical Yang-Mills Theory,''
  Phys.\ Rev.\ D {\bf 33}, 1133 (1986).
  doi:10.1103/PhysRevD.33.1133

\bibitem{Sarioglu:2002qb}
  O.~Sarioglu,
  ``Lienard-Wiechert potentials of a nonAbelian Yang-Mills charge,''
  Phys.\ Rev.\ D {\bf 66}, 085005 (2002);
  doi:10.1103/PhysRevD.66.085005
  [hep-th/0207227].

\bibitem{Chernicoff:2009re}
  M.~Chernicoff, J.~A.~Garcia and A.~Guijosa,
  ``Generalized Lorentz-Dirac Equation for a Strongly-Coupled Gauge Theory,''
  Phys.\ Rev.\ Lett.\  {\bf 102}, 241601 (2009)
  doi:10.1103/PhysRevLett.102.241601
  [arXiv:0903.2047 [hep-th]].





\end{thebibliography}
\addcontentsline{toc}{section}{Bibliography}
\bibliographystyle{TitleAndArxiv}


\end{document}